\documentclass[prd, twocolumn, superscriptaddress, nofootinbib, notitlepage, floatfix, preprintnumbers]{revtex4-1}
\usepackage{xcolor}
\usepackage{graphicx}
\usepackage{wrapfig}
\usepackage{amsmath}
\usepackage{amsfonts}
\usepackage{amssymb}
\usepackage{multirow}
\usepackage{slashed}
\usepackage{physics}
\usepackage{soul}

\usepackage{ulem}
\usepackage[colorlinks=true, linkcolor=blue, citecolor=blue, urlcolor=blue]{hyperref}
\newcommand\befs{\begin{figure*}}
\newcommand\eefs[1]{\label{fig:#1}\end{figure*}}
\newcommand\bef{\begin{figure}}
\newcommand\eef[1]{\label{fig:#1}\end{figure}}
\newcommand\beq{\begin{equation}}
\newcommand\eeq[1]{\label{#1}\end{equation}}
\newcommand\beqa{\begin{eqnarray}}
\newcommand\eeqa[1]{\label{#1}\end{eqnarray}}
\newcommand\bet{\begin{table}}
\newcommand\eet[1]{\label{tb:#1}\end{table}}
\newcommand\bets{\begin{table*}}
\newcommand\eets[1]{\label{tb:#1}\end{table*}}
\newcommand\fgn[1]{Fig.\ \ref{fig:#1}}
\newcommand\eqn[1]{Eq.\ (\ref{#1})}
\newcommand\scn[1]{Section \ref{sec:#1}}
\newcommand\apx[1]{Appendix \ref{sec:#1}}

\newcommand{\msbar}{{\overline{\mathrm{MS}}}}
\newcommand\reM{{\rm Re} \: {\cal M}}
\newcommand\reMtw{{\rm Re} \: {\cal M}^{\rm tw2,corr}}
\newcommand\MS{\overline{\rm MS}}
\newcommand\nn{\nonumber}

\begin{document}

\widetext
\setstcolor{red}

\title{Pion distribution amplitude at the physical point using the leading-twist expansion of the quasi-distribution-amplitude matrix element}

\author{Xiang Gao}
\email{gaox@anl.gov}
\affiliation{Physics Division, Argonne National Laboratory, Lemont, IL 60439, USA}

\author{Andrew D. Hanlon}
\affiliation{Physics Department, Brookhaven National Laboratory, Bldg. 510A, Upton, New York 11973, USA}

\author{Nikhil Karthik}
\email{nkarthik.work@gmail.com}
\affiliation{Department of Physics, College of William \& Mary, Williamsburg, VA 23185, USA}
\affiliation{Thomas Jefferson National Accelerator Facility, Newport News, VA 23606, USA}

\author{Swagato Mukherjee}
\affiliation{Physics Department, Brookhaven National Laboratory, Bldg. 510A, Upton, New York 11973, USA}

\author{Peter Petreczky}
\affiliation{Physics Department, Brookhaven National Laboratory, Bldg. 510A, Upton, New York 11973, USA}

\author{Philipp Scior}
\affiliation{Physics Department, Brookhaven National Laboratory, Bldg. 510A, Upton, New York 11973, USA}

\author{Sergey Syritsyn}
\affiliation{RIKEN-BNL Research Center, Brookhaven National Laboratory, Upton, New York 11973}
\affiliation{Department of Physics and Astronomy, Stony Brook University, Stony Brook, New York 11790}

\author{Yong Zhao}
\email{yong.zhao@anl.gov}
\affiliation{Physics Division, Argonne National Laboratory, Lemont, IL 60439, USA}

\begin{abstract} 

We present a lattice QCD determination of the distribution amplitude
(DA) of the pion and the first few Mellin moments from an analysis
of the quasi-DA matrix element within the leading-twist framework.
We perform our study on a HISQ ensemble with $a=0.076$ fm lattice
spacing with the Wilson-clover valence quark mass tuned to the
physical point. We analyze the ratios of pion quasi-DA matrix
elements at short distances using the leading-twist Mellin operator
product expansion (OPE) at the next-to-leading order and the
conformal OPE at the leading-logarithmic order. We find a robust result
for the first non-vanishing Mellin moment $\langle x^2 \rangle =
0.287(6)(6)$ at a factorization scale $\mu=2$ GeV. We also present
different Ans\"atze-based reconstructions of the $x$-dependent DA,
from which we determine the perturbative leading-twist expectations
for the pion electromagnetic and gravitational form-factors at large
momentum transfers.

\end{abstract} 
\date{\today} 
\maketitle

\section{Introduction}\label{sec:intro}
The study of inclusive deep-inelastic processes by describing them
using process-independent parton distribution functions (PDFs) has
resulted in a good understanding of collinear internal structures
of hadrons.  The next generation of experimental facilities
(e.g.,~\cite{Aguilar:2019teb,Arrington:2021biu,
Denisov:2018unj,Adams:2018pwt}) will focus on observables that
characterize hard semi-inclusive and exclusive processes to relate
intrinsic properties of hadrons to those of the partons.  The
generalized parton distribution functions (GPDs) and the meson
distribution amplitudes (DAs) will play crucial roles as universal
soft-functions in the descriptions of exclusive reactions in hard
kinematical regimes through factorization. Thus, non-perturbative
determination of such parton distributions and amplitudes are
currently essential.

Concretely, the pion DA, $\phi(x)$, captures the overlap of the
pion with a state with two collinear valence quarks carrying fractions
$x$ and $(1-x)$ of the pion light-front momentum
$P^+$~\cite{Lepage:1980fj,Lepage:1979zb,Efremov:1979qk}.  Hence,
the pion DA is of theoretical interest due to its proximity to being
the light-front wave-function~\cite{Lepage:1979zb} of the Nambu-Goldstone
boson of chiral symmetry breaking; by comparison with DAs of
non-Goldstone pseudoscalar mesons, one could learn about how the
fundamental quark-gluon interaction leads to the special properties
of the pion (see review~\cite{Roberts:2021nhw}, and Ref.~\cite{Gao:2021hvs}
for our lattice calculation in a related direction).  Phenomenologically,
the properties of the pion DA, such as its shape and its Mellin
moments, are still not precisely known and the main experimental
input has been from the factorization of the pion-photon transition
form
factor~\cite{CELLO:1990klc,CLEO:1997fho,BaBar:2009rrj,Belle:2012wwz,Melic:2002ij,
Gao:2021iqq}, and from
past~\cite{Dally:1981ur,Dally:1982zk,Amendolia:1984nz,NA7:1986vav,JeffersonLabFpi:2000nlc,JeffersonLabFpi:2007vir,JeffersonLab:2008jve,JeffersonLab:2008gyl}
(and ongoing~\cite{Dudek:2012vr}) investigations of the large-$Q^2$
behavior of the pion electromagnetic form
factor~\cite{Efremov:1979qk,Farrar:1979aw,Lepage:1980fj}.  From a
field-theoretic standpoint, the determination of the pion DA involves
the light-front
correlation~\cite{Radyushkin:1977gp,Efremov:1979qk,Lepage:1980fj},
\beq
\phi(x,\mu) = \int\frac{d\lambda}{2\pi} e^{-i\frac{x}{2}\lambda} {\cal I}(\lambda,\mu),\ {\rm with}\ \lambda=P^+ z^-,
\eeq{phidef}
with a dimensionless invariant amplitude ${\cal I}$ renormalized in the $\msbar$ scheme at scale $\mu$ defined as,
\beq
i f_\pi P^+ {\cal I}(\lambda,\mu) = \mel**{0}{\overline{d}(-z^-/2)\gamma^+\gamma_5 W_+ u(z^-/2)}{\pi^+; P},
\eeq{mdef}
where $W_+$ is a straight Wilson line from $-z^-/2$ to $z^-/2$ along
the light-cone.  Various model-based determinations
(e.g.,~\cite{Chernyak:1981zz,Bakulev:2001pa,Agaev:2010aq,Chang:2013pq,RuizArriola:2006jge})
of $\phi$ have given key insights into the full $x$-dependence and
the moments.  For a rigorous QCD-based non-perturbative calculation,
one needs to rely on lattice QCD computation. However, the unequal
time-separation in the above light-front correlator has prevented
a direct computation of DA.

The first few Mellin and Gegenbauer moments of the pion DA have
been computed from leading-twist local operators on the
lattice~\cite{Kronfeld:1984zv,Martinelli:1987si,DeGrand:1987vy,Daniel:1990ah,DelDebbio:1999mq,DelDebbio:2002mq,Braun:2006dg,Arthur:2010xf,Braun:2015axa,Bali:2017ude,RQCD:2019osh}.
The difficulty of this approach is the nontrivial renormalization
of these operators on the lattice, which limits the number of lowest calculable moments. One way to avoid this challenge
is through a leading-twist expansion of the pion-to-vacuum transition
matrix element of certain spatially separated
operators~\cite{Braun:2007wv}.  The short-distance logarithmic
divergences are absorbed as part of perturbatively computed Wilson
coefficients, and most importantly, the expansion coefficients are
proportional to the moments of the DA.  Such a leading-twist expansion
approach was first applied using the pion-to-vacuum transition
matrix element using two current operator insertions~\cite{Braun:2007wv},
which has been been utilized in further studies of DA and the
moments~\cite{Bali:2017gfr,Bali:2018spj}.  Analogous real-space
analysis using current-current correlators was also applied to the
pion-to-pion forward matrix element to determine the pion
PDF~\cite{Ma:2014jla,Sufian:2019bol,Sufian:2020vzb}. Besides, there are also approaches based on the leading-twist expansion of current-current correlators in the Fourier space~\cite{Chambers:2017dov}, including the method that uses an intermediate heavy quark~\cite{Detmold:2005gg,Detmold:2021uru,Detmold:2021qln}, to calculate the higher moments of DAs and PDFs.

Recently, there have been new advancements in the determination of
parton distribution functions using multiplicatively renormalized,
spatially-extended quark-antiquark operators based on the
LaMET~\cite{Ji:2013dva,Ji:2014gla,Ji:2020ect} and the
pseudodistribution~\cite{Radyushkin:2017cyf,Orginos:2017kos,Radyushkin:2019mye}
approaches. For example,
Refs.~\cite{Izubuchi:2019lyk,Gao:2020ito,Lin:2020ssv,Gao:2021dbh}
are recent computations of the pion PDF using the bilocal quark-antiquark
operator. For DA, one can construct the pion-to-vacuum transition
matrix element of such a bilocal quark-antiquark
operator~\cite{Ji:2015qla}, which we simply refer to as the quasi-DA
matrix element.  The aim of this paper is to apply the leading-twist
expansion of the quasi-DA matrix element in a manner similar to
Ref~\cite{Braun:2007wv}, to extract the Mellin moments of the pion
DA, and attempt to reconstruct the shape of the pion DA based on
strategies utilized in the case of PDFs.  Previously, such quasi-DA
matrix elements of both the pion and kaon have been investigated
using the $x$-space LaMET
matching~\cite{Zhang:2017bzy,Zhang:2017zfe,Liu:2018tox,Zhang:2020gaj,Hua:2022kcm}.
Due to the differences in the renormalization method and the analysis
methodology for a leading-twist expansion approach, we expect this
work to shed new light into the quasi-DA method as a way to obtain
the pion DA and its moments.
Apart from the above leading-twist expansion or effective theory matching approaches, it has also been proposed to directly obtain the structure functions from the hadronic tensor on the lattice through a nontrivial inversion of Euclidean correlators~\cite{Liu:1993cv}, which has not yet been applied to the extraction of DAs.

The plan of the paper is as follows. First, we 
give an overall description of our methodology in 
\scn{method}. Then, in \scn{matching}, we 
present the perturbative results pertaining to conformal and 
Mellin OPEs. In \scn{setup}, we give the details of 
our lattice calculation. In \scn{extrapol}, we present the details of 
the determinations of the bare quasi-DA matrix element 
and the renormalized ratios thereof. 
In \scn{results}, we present the results on the
pion DA; here, we first present the analysis specifications, then we present the 
Mellin moments for fixed spatial distances, after which we present our 
model-independent determination of the first two Mellin 
moments, and finally, we describe the model-dependent 
reconstructions of the shape of the pion DA. 
From such reconstructed $x$-dependent DA, we present 
the perturbative expectations for pion form-factors at large 
momentum transfers.
In \scn{concl},
we summarize our findings.

\section{Method}\label{sec:method}

We specify four-vectors as $v_\rho$, whose components are
$(v_0,v_1,v_2,v_3)$ with $v_0$ being the temporal component, and
$\mathbf{v}=(v_1,v_2,v_3)$ as the spatial component. We use the
$\rho=3$ direction for spatial separations and as the direction of
the pion momentum. The metric convention is $v\cdot w = v_0 w_0 -
\mathbf{v}\cdot \mathbf{w}$.  We specify the Dirac $\gamma$-matrices
as $\gamma_\rho$, and we use the Minkowskian convention for them.
For the ease of understanding, we specify them as
$(\gamma_t,\gamma_x,\gamma_y,\gamma_z)$ respectively, when explicitly
mentioning a matrix. We specify the bare and renormalized quantities
using superscript ``B" and ``R" respectively.

We use the short-distance behavior of the quasi-DA
matrix element of a boosted pion, $\pi^+ (u \bar d)$, to determine
its leading-twist DA. The quasi-DA operator is the equal-time bilocal
quark bilinear operator,
\beq
O^B_{\rho}(z) = \overline{d}(-z/2)\gamma_\rho\gamma_5 W_{-z/2,z/2} u(z/2),
\eeq{quasidaop}
with the straight Wilson-line $W_{-z/2,z/2}$ connecting the quark
and antiquark that are separated spatially as $z=(0,0,0,z_3)$.  At
non-zero $z$, the operator suffers from a linear divergence in the
self-energy of the Wilson-line, $e^{-c|z|}$, and also from the
end-point logarithmic divergence, and therefore, it needs to be
renormalized.  Thus the operator above is bare, and hence, the
superscript $B$ to specify this.  Let $O_\rho^R(z,\mu)$ be the
renormalized operator in the $\msbar$ scheme at scale $\mu$.  The
quasi-DA matrix element for the pion is,
\beq
i P_\rho h^R\left(z\cdot P, z^2, \mu\right) \equiv \mel**{0}{O^R_\rho(z,\mu)}{\pi^+; P},
\eeq{quasidadef}
with the on-shell pion momentum $P=(E(P_3),0,0,P_3)$. The Lorentz
invariant $\lambda = -z\cdot P = z_3 P_3$ is called the Ioffe-time
or light-cone distance in the literature. We note that the
left-hand side of the above equation is not a Lorentz decomposition,
instead we have defined $h$ above in a form convenient for the
leading-twist expansion that is proportional to $P_\rho$~\footnote{It must be implicitly
understood that $h$ is also labeled by $\hat n_\rho = z_\rho/\sqrt{-z^2}$,
as will be reflected in the $\rho$-dependent Wilson coefficients (c.f.~\cite{Izubuchi:2018srq})
in the operator product expansion (OPE) of $O^R_\rho(z)$, and hence,
of the leading-twist expansion of $h^R(z)$.}.  As a specific case, $z_3=0$
and for all momentum $P_3$, the local operator $O^R_\rho(0)$ is
the axial-current operator, and $h(z_3=0, P_3) = f_\pi$, the pion decay
constant.

The idea used in this paper is that for quark-antiquark separations $z_3$
that are small in QCD scales and for momenta $P_3 > 0$, one can
describe the $\lambda$ and $z^2$ dependencies of $h^R(\lambda,z^2,\mu)$
within a leading-twist OPE framework valid up to higher-twist contributions.
This lets us relate the lattice-calculable equal-time
quantity $h^R(\lambda,z^2)$ to the light-cone distribution amplitude
$\phi(x,\mu)$ at a factorization scale $\mu$ and its Mellin moments,
\beq
\langle x^n \rangle = \int_{-1}^1 \phi(x,\mu) x^n dx. 
\eeq{momentsdef}
The framework is similar to the one used in the determination of
the parton distribution functions (PDFs), however, the key difference
for the case of DA is that the matrix element in \eqn{quasidadef}
is between the boosted pion state and vacuum state.  This results
in a different leading-twist expansion than in the case of the
forward matrix element for PDFs. As we will show in the paper, the
leading-twist expansion of $h^R(\lambda,z^2,\mu)$ for DA is
\beq
h^{\rm tw2}(\lambda,z^2,\mu) = \sum_{n=0} \frac{(-i \lambda/2)^n}{n!} \sum_{m=0}^n C_{n,m}(z^2 \mu^2) \langle x^m \rangle ,
\eeq{mope}
where $C_{n,m}(\mu^2 z^2)$ are the Wilson coefficients calculable
in perturbation theory that relates $h$ to
the DA via its Mellin moments at scale $\mu$. By the superscript
``tw2" we mean that the expansion ignores all terms with twists
bigger than two. Henceforth, we will refer to \eqn{mope} as the
Mellin OPE (M-OPE). In \scn{matching}, we provide the NLO results
for $C_{n,m}$ for $O^R_3$.
Under an ERBL~\cite{Lepage:1980fj,Lepage:1979zb,Efremov:1979qk} evolution in $\mu$,  the different Mellin
moments mix, which is also reflected in the non-vanishing off-diagonal
nature of $C_{n,m}$.  At the level of leading logarithms and up to
finite ${\cal O}(\alpha_s)$ corrections, massless QCD is conformal
and this helps in diagonalizing the leading-twist expansion (see review~\cite{Braun:2003rp})
with respect to evolution. Such an expansion in terms of the conformal
partial waves ${\cal F}_n(\lambda/2, \mu^2 z^2; \alpha_s)$ is given
as
\beq
h^{\rm tw2}(\lambda,z^2,\mu) = \sum_{n=0} a_n(\mu) {\cal F}_n(\lambda/2, z^2 \mu^2; \alpha_s),
\eeq{cope}
where $a_n$ are the Gegenbauer moments at scale $\mu$, and they
satisfy the simpler LO DGLAP evolution.  In  \scn{matching}, we provide
the expressions for ${\cal F}_n$. We will refer to \eqn{cope} as
the conformal OPE (C-OPE). 

At next-to-leading order, the Mellin and Conformal OPEs differ in
the finite $\alpha_s$ terms and are the same up to $\alpha_s^0$ and
$\alpha_s \log(z^2 \mu^2)$ terms.  At the level of practical
implementation, we have to truncate \eqn{mope} and \eqn{cope} ---
for C-OPE, it is a truncation in conformal partial waves that each
have infinite order in $\lambda$, whereas for M-OPE, it is a
truncation in the order of $\lambda$.  We will use M-OPE primarily
in this paper to implement the matching at NLO, and compare the
results with that obtained with C-OPE to cross-check that the results
are approximately the same. The expressions in \eqn{mope} and
\eqn{cope} are general for any pseudoscalar mesons, but it gets
further simplified for the pion; due to isosopin symmetry, the
Mellin and Gegenbauer moments for odd $n$ vanish, and therefore,
at leading-twist the matrix elements are purely real. We will use
this fact in our analysis and set odd $n$ moments to zero.  Since
the Gegenbauer moments for all $n>0$ approach zero under evolution
to $\mu\to\infty$, the C-OPE expression simply approaches ${\cal
F}_0(\lambda/2)$ asymptotically.

In the above discussion, we assumed that the operator is renormalized
in the $\msbar$ scheme, which cannot be directly implemented on the
lattice.  Furthermore, it was shown in
Refs.~\cite{Constantinou:2017sej,Chen:2017mzz}, that the operator
$O^B_3(z)$, that has the $\gamma_z\gamma_5$ structure, is
multiplicatively renormalizable, whereas the choice $O^B_0(z)$ that
has the $\gamma_t\gamma_5$ structure, mixes with the
$\bar{u}\gamma_5\gamma_z\gamma_t d$ operator when lattice-regulated
fermions that break chiral-symmetry are used. Therefore, in this
work, we only work with the $O^B_3(z)$ operator to avoid the mixing.
We adapt the renormalization group invariant (RGI)
ratios~\cite{Orginos:2017kos,Fan:2020nzz,Gao:2020ito} of hadronic
matrix elements for the renormalization.  Since the renormalization
of $O^B_3(z)$ is purely multiplicative, the ratio,
\beq
{\cal M}(\lambda,z^2, P^0) \equiv \frac{h^B(\lambda,z_3^2)}{h^B(\lambda_0, z_3^2)} = \frac{h^R(\lambda,z_3^2,\mu)}{h^R(\lambda_0, z_3^2,\mu)},\ \lambda_0 = P_3^0 z_3,
\eeq{ratio1}
with respect to the matrix element at a fixed momentum $P^0$ is an
RGI quantity that we can determine on the lattice.  We obtain the
leading twist expression for ${\cal M}$ from the $\msbar$ expressions
for $h^{\rm tw2}$ in
\eqn{mope} and \eqn{cope} by making use of its RGI nature as
\beq
{\cal M}^{\rm tw2}(\lambda,z^2, P^0) = \frac{h^{\rm tw2}(\lambda,z^2,\mu)}{h^{\rm tw2}(\lambda_0,z^2,\mu)}.
\eeq{ratiotw2}
The actual lattice data in the range of $z_3$ and $P_3$ that we use
could suffer from lattice corrections and higher-twist corrections
to the continuum leading-twist expressions in \eqn{mope} and
\eqn{cope}.  We model the two corrections using some functions
$L(z,P,a)$ and $H(z,P)$ respectively. With such corrections, we use
expressions of the type,
\beqa
&&{\cal M}^{\rm tw2,corr}(\lambda,z^2, P^0) =\cr &&\qquad\frac{h^{\rm tw2}(\lambda,z^2,\mu) + L(z,P,a) + H(z,P) }{h^{\rm tw2}(\lambda_0,z^2,\mu)+ L(z,P^0,a) + H(z,P^0)},
\eeqa{ratiotw2corr}
to perform our fits to the lattice QCD data for ${\cal M}(\lambda,z^2,
P^0)$ to obtain information on the Mellin or Gegenbauer moments of
the DA depending on whether M-OPE or C-OPE is used for $h^{\rm
tw2}$, respectively. Equivalently, by modeling the functional form
of $\phi(x,\mu)$, we also reconstruct the $x$ dependence of the
pion DA.  Alternatively, from such analyses, we can also infer the
$\msbar$ light-front Ioffe-time distribution (ITD) as
\beq
{\cal I}(\lambda,\mu) = \sum_{n=0} \frac{(-i\lambda/2)^n}{n!} \langle x^n\rangle(\mu).
\eeq{msbaritd}
We discuss
the implementation of the above set of steps further in the following
sections presenting our results.

\section{Analytical results for conformal and Mellin OPE}\label{sec:matching}
\subsection{Short distance factorization of the quasi-DA matrix element}
\label{sec:sdf}

In $x$-space, the quasi-DA can be perturbatively matched onto the
light-cone DA at large momentum, where the matching coefficient has
been derived in the $\MS$ scheme at one-loop order~\cite{Liu:2018tox}.
In coordinate space, the quasi-DA matrix element $h(\lambda,
z^2,\mu^2)$ can also be perturbatively matched onto the light-cone
correlation ${\cal I}(\lambda,\mu)$ through a short-distance
factorization formula, which has been derived in QCD at one-loop
order~\cite{Radyushkin:2019owq} as
\begin{align}
	h^R(\lambda, z^2,\mu^2)
	&=\int_0^1 dw\ C(w,\lambda, z^2\mu^2){\cal I}(w\lambda,\mu) \nn\\
		& \qquad  + {\cal O}(z^2\Lambda_{\rm QCD}^2)\,,
\end{align}
where the matching kernel
\begin{align}
	&C(w,\lambda, z^2\mu^2) \nn\\
	&=\delta(\bar{w}) + {\alpha_sC_F\over 2\pi} \Bigg\{ 2\left({\mathbf L}+1\right) \delta(\bar{w}) + \big({\mathbf L}+1\big)\nn\\
	& \quad \times \left[- \left({2w\over \bar{w}}\right)_+ \cos({\bar{w}\lambda \over 2})  - {\sin(\bar{w}\lambda /2)\over \lambda/2}\right]  \nn\\
	& - 4\left({\ln\bar{w}\over \bar{w}}\right)_+\cos({\bar{w}\lambda\over2}) + {(2+2\delta_{\rho3})\sin(\bar{w}\lambda/2)\over \lambda/2}\Bigg\}\nn\\
	&\quad  + {\cal O}(\alpha_s^2)\,,
\end{align}
with $C_F=4/3$, ${\mathbf L}=\ln(z^2\mu^2e^{2\gamma_E}/4)$, and
$\bar{w}=1-w$.  The $2\delta_{\rho3}$ term in the curly bracket can
be inferred from the factorization of the forward quasi-PDF matrix
elements~\cite{Izubuchi:2018srq}.

\subsection{Conformal OPE}
\label{sec:cope}

The LCDA  can be expressed as the sum of Gegenbauer moments,
\begin{align}
	\phi(x, \mu) &= {3\over4}(1-x^2) \sum_{\substack{n=0,\\\rm even}}^\infty {\cal C}_n^{3\over2}(x)a_n(\mu) \,,
\end{align}
where ${\cal C}_n^{3\over2}(x)$ is a Gegenbauer polynomial
(refer~\cite[Table~18.3.1]{NIST:DLMF}). The Gegenbauer moment
$a_n(\mu)$ can also be projected from $\phi(x,\mu)$ as
\begin{align}
a_n(\mu) = {4(n+3/2)\over 3(n+1)(n+2)}\int_{-1}^1 dx \ \phi(x,\mu) {\cal C}_n^{3\over2}(x)\,.
\end{align}

At leading logarithmic (LL) accuracy, QCD is conformal, and
$\phi_n(\mu)$ evolves multiplicatively with the anomalous dimension
\begin{align}
	\gamma_n(\alpha_s) &=  {\alpha_sC_F\over 4\pi} \gamma_n^{(0)} + {\cal O}(\alpha_s^2)\\
	&= {\alpha_sC_F\over 4\pi}\left[4H_{n+1}-{2\over (n+1)(n+2)}-3\right] + {\cal O}(\alpha_s^2),\nn
\end{align}
with $H_n=\sum_{i=1}^n 1/i$.  The value of $\gamma_n$ is different
from that for the Mellin moments of PDFs by a minus sign.  Therefore,
the Gegenbauer moments should be the basis of OPE under the conformal
approximation to QCD.

The conformal OPE of the quasi DA matrix element in the $\MS$ scheme
can be inferred from that for the current-current
correlator~\cite{Braun:2007wv} as
\begin{align}
	h^{\rm tw2}_{\rm cf}(\lambda, z^2, \mu^2) &= \sum_{\substack{n=0,\\\rm even}}^\infty  {\cal F}_n({\lambda\over2},z^2\mu^2;\alpha_s)a_n(\mu)\,,
\end{align}
where $\lambda =zP^z$, and the LL resummed coefficient
\begin{align}
{\cal F}_n(\lambda,z^2\mu^2;\alpha_s) &=  c_n(\alpha_s)(\mu^2 z^2)^{{\gamma_n  }+ \gamma_O} {\Gamma(2-{\gamma_n})\Gamma(1+n) \over \Gamma(1+n+{\gamma_n})}\nn\\
	&\quad \times {3\over 4}\ i^n \sqrt{\pi}{(n+1)(n+2)\over 2} {\Gamma(n+\gamma_n+{5\over2}) \over \Gamma(n+{5\over2})} \nn\\
	&\quad \times \Big({\lambda\over2}  \Big)^{-{3\over2}-\gamma_n} J_{n+\gamma_n+{3\over2}}(\lambda) \,.
\label{partwavedef}
\end{align}
with $\Gamma$ and $J_n$ being the standard
gamma function and Bessel function of first-kind 
respectively, and
\begin{align}
c_n &= 1+ {\alpha_sC_F\over 2\pi}\Bigg[ {5+2n\over 2+3n+n^2} + {2 \delta_{\rho3}\over 2+3n+n^2}\nn\\
&\qquad  + 2(1-H_n)H_n -2 H_n^{(2)}\Bigg] + {\cal O}(\alpha_s^2)\,,
\end{align}
which is the same as the Wilson coefficients in the OPE of the
helicity quasi PDF matrix elements~\cite{Izubuchi:2018srq}, and
\begin{align}
	\gamma_O&=\gamma_O^{(0)}+ {\cal O}(\alpha_s^2)={\alpha_sC_F\over 4\pi}\cdot {3}+ {\cal O}(\alpha_s^2)\,,
\end{align}
which is the anomalous dimension of the nonlocal operator $O_\rho
(z,\mu)$.  Note that in Eq.~(\ref{partwavedef}) the running of
strong coupling is turned off because we assumed conformal symmetry.

\subsection{OPE in terms of Mellin moments}
\label{sec:mellin}

If we do not include the scale evolution in the OPE, then we can
consider expansion in terms of the Mellin moments in Eq.~(\ref{mope}).
The coefficient functions
\begin{align}
	C_{n,m} &= C^{(0)}_{n,m}+{\alpha_sC_F\over 2\pi} C^{(1)}_{n,m} + {\cal O}(\alpha_s^2)
\end{align}
can be obtained by the relation
\begin{align}
	& C^{(0)}_{n,m} = \delta_{n, m}\,,\\
	& \sum_{m= 0}^n C^{(1)}_{n,m}(z^2 \mu^2) x^m  \nn\\
	& = 2 \big({\mathbf L}+1\big)x^n  +   \big({\mathbf L}+1\big)  \int_0^1 dw \nn\\
		&\qquad \times\left[ \left({-2w\over \bar{w}}\right)_+  {(xw - \bar{w})^n + (xw + \bar{w})^n \over2} \right.\nn\\
		&\qquad\qquad \left. - {-(xw - \bar{w})^{n+1} + (xw + \bar{w})^{n+1} \over2(n+1)}\right] \nn\\
		& +   \int_0^1 dw\left[ \left(-{4\ln\bar{w}\over \bar{w}}\right)_+{(xw - \bar{w})^n + (xw + \bar{w})^n \over2} \right.\nn\\
		& \quad \left. +(2+2\delta_{\rho3}) {-(xw - \bar{w})^{n+1} + (xw + \bar{w})^{n+1} \over2(n+1)}\right]\,,
\end{align}
where $C^{(1)}_{n,m}$ can be read off from the coefficients of $x^m$. 
The lowest few coefficient functions are
\begin{align}
	C_{0,0}^{(1)} &= {3\over 2}{\mathbf L} + {7\over2}\,,\\
	\sum_{m= 0}^n C^{(1)}_{1,m}(z^2 \mu^2) x^m &= \left({17\over 6}{{\mathbf L}} - {1\over2}\right) x\,,\\
	\sum_{m= 0}^n C^{(1)}_{2,m}(z^2 \mu^2) x^m &= \left(\frac{43}{12}{\mathbf L}\!-\!\frac{37}{12}\right) x^2 \!+\! \frac{11}{12}\!-\!\frac{5}{12}{\mathbf L}\,,\\
	\sum_{m= 0}^n C^{(1)}_{3,m}(z^2 \mu^2) x^m &= \left(\frac{247}{60}{\mathbf L}-\frac{923}{180}\right) x^3 \nn\\
	&\qquad\qquad + \left(\frac{79}{60}-\frac{11}{20}{\mathbf L}\right) x\,, \\
	\sum_{m= 0}^n C^{(1)}_{4,m}(z^2 \mu^2) x^m &= \left(\frac{68}{15}{\mathbf L}-\frac{247}{36}\right) x^4 \\
	&\qquad + \left(\frac{5}{3}-\frac{19}{30}{\mathbf L}\right) x^2 + \frac{1}{4}-\frac{2}{15}{\mathbf L}\,.\nn
\end{align}

\section{Computational setup}\label{sec:setup}

We used a mixed fermion action setup consisting of a 2+1 flavor
HISQ sea quark action and a clover-improved Wilson fermion action for the
valence quarks.  The HISQ ensemble~\cite{Bazavov:2019www} was
generated by the HotQCD collaboration, and consists of $L_s^3 \times
L_t = 64^3 \times 64$ lattice sites at a lattice spacing of $a=0.076$
fm.  The sea quark mass in the setup corresponds to a near physical
pion mass of 140 MeV.  The tadpole improved Wilson clover valence
quarks couple to 1-HYP smeared gauge links~\cite{Hasenfratz:2001hp}.
We tuned the Wilson mass to obtain the valence pion mass of 140
MeV. Therefore, both the sea and valence quarks are tuned to the
physical point.

In order to compute the quasi-DA matrix element of a pion with
momentum $P_3$, the two essential ingredients are the $\pi$-$\pi$
and $\pi$-${\cal O}_3$ correlators . The pion-pion two-point function
at source-sink time separation of $t_s$ is
\beq
C_{\pi\pi}(t_s,P_3) = \left\langle \pi(\mathbf{P},t_s) \pi^\dagger(\mathbf{x}_0, 0) \right\rangle,
\eeq{pipi}
where
\beq
\pi^\dagger(\mathbf{x},t_s) = \overline{u}_s(\mathbf{x},t_s)\gamma_5 d_s(\mathbf{x},t_s), 
\eeq{pisource}
with
$\pi^\dagger(\mathbf{P},t_s)=\sum_{\mathbf{x}}\pi^\dagger(\mathbf{x},t_s)
e^{i \mathbf{P}\cdot \mathbf{x}}$.  The $u_s$ and $d_s$ represent
Coulomb-gauge Gaussian smeared quark operators, with the smearing
radius as 0.59 fm. At nonzero spatial momentum $\mathbf{P}=(0,0,P_3)$,
we implemented the boosted quark smearing~\cite{Bali:2016lva} with
quark boosts, $k_3 = \pm \zeta P_3$, for $u$ and $d$ respectively.
The other ingredient, the pion-quasi DA-operator correlator with
time separation $t_s$, is
\beq
C_{\pi \tilde{O}_3}(t_s; z_3,P_3) = \left \langle \tilde{O}^B_3(z_3; \mathbf{P}, t_s) \pi^\dagger(\mathbf{x}_0,0)\right\rangle,
\eeq{pitoO}
where
\beqa
&
\tilde{O}^B_3(z_3; \mathbf{P}, t_s) = \nonumber\\[2mm]
&
\displaystyle
\quad \sum_{\mathbf{x}} \overline{d}(\mathbf{x},t_s)\gamma_z\gamma_5 W(\mathbf{x},t_s; \mathbf{x}+\mathbf{z}, t_s) u(\mathbf{x}+\mathbf{z},t_s) e^{-i \mathbf{P}\cdot \mathbf{x}}. \nonumber\\
\eeqa{o3tilde}
The spatial part of the quark-antiquark separation $z = (0,0,0,z_3)$
is denoted with $\mathbf{z}$.
The straight Wilson-line along the $z$-direction is
$W(\mathbf{x},t_s; \mathbf{x}+\mathbf{z}, t_s) = \prod_{k=0}^{z_3/a}
U^{\rm HYP}_3(\mathbf{x}+k a \hat z, t_s)$, where $U^{\rm HYP}$ is the
1-HYP smeared gauge link, the same as those used in the Wilson Dirac
operator. The $u$ and $d$ quark operators in \eqn{o3tilde} are {\sl
not} Gaussian smeared. One should note that the
coordinates of the antiquark and quark are at $x$ and $x+z$ which
differs from the one in \eqn{quasidaop}, and hence the tilde on top
of $O_3$ to make this distinction clear.  This is due to the ease
of implementation of the former convention on the lattice, and we
defer the conversion to the analysis-wise convenient convention
in \eqn{quasidaop} at a later stage by multiplying the results with
a phase $\exp\left(-i P_3 z_3/2\right)$.  We computed 
the quark
propagators that occur in the Wick contractions of \eqn{pipi} and
\eqn{pitoO} on GPUs using the multigrid algorithm~\cite{Brannick:2007ue} as
implemented in the QUDA suite~\cite{Clark:2009wm,Babich:2011np, Clark:2016rdz}.

We performed the above set of computations at eight different spatial
momenta $\mathbf{P} = (0,0,P_3)$,
\beq
P_3 = \frac{2\pi}{L_s a} n_3 \approx 0.254\times n_3 {\rm\ GeV},
\eeq{p3vals}
for $n_3 \in [0,7]$. Thus, the highest momentum we use in this work
is $1.78$ GeV which is sufficiently larger than $\Lambda_{\rm QCD}$,
and at the same time corresponds to $P_3 a = 0.69$ which is below
the lattice-like scales.  For these momenta, we decided to choose
the phase parameter $\zeta$ in momentum smearing such that $\zeta
n_3 = 2$ for $n_3 \le 3$ and $\zeta n_3 = 5$ for $n_3 > 3$, so that
we could reuse smeared sources for multiple momenta to balance the
computational cost and the signal-to-noise ratios.

We used 350 statistically independent configurations.  We effectively
increased the statistics many folds using the all-mode averaging
method~\cite{Shintani:2014vja}, implemented using exact inversions
of the Dirac operator at $N_{\rm ex}$ source locations $\mathbf{x}_0$,
and sloppy inversions at $N_{\rm sl}$ source locations. For $n_3
\le 3$, we used $(N_{\rm ex},N_{\rm sl})=(4,80)$, and for $n_3 >
3$, we used $(8,160)$.

\section{Determination of matrix element}\label{sec:extrapol}

In this section, we discuss the details of the determination of
the ground-state matrix elements of the bare quasi-DA operator at
different momenta and quark-antiquark separations, and the RGI
ratios that we construct from them.

\subsection{Bare matrix element}\label{sec:BareM}

\befs
\centering
\includegraphics[scale=0.7]{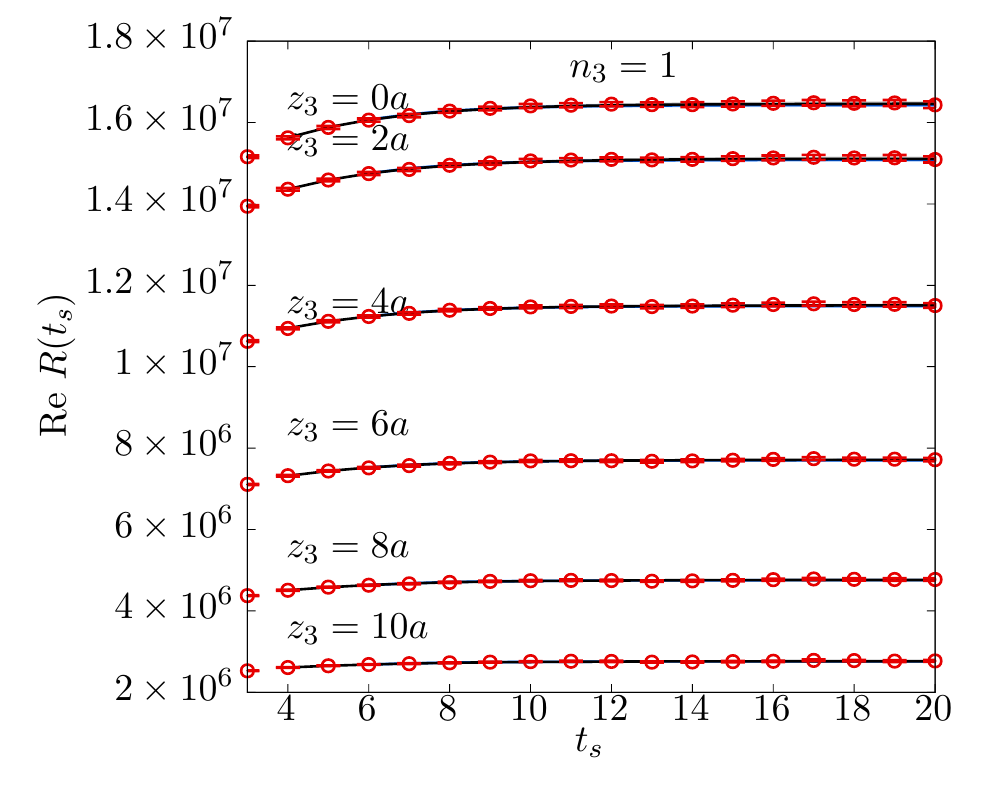}
\includegraphics[scale=0.7]{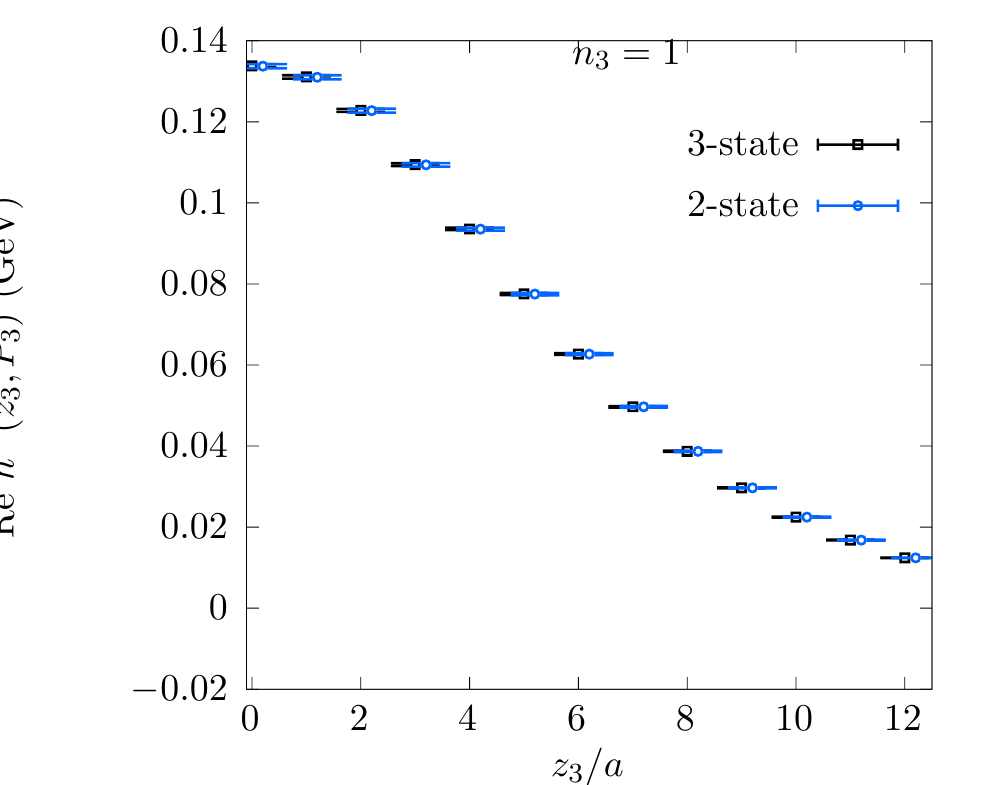}

\includegraphics[scale=0.7]{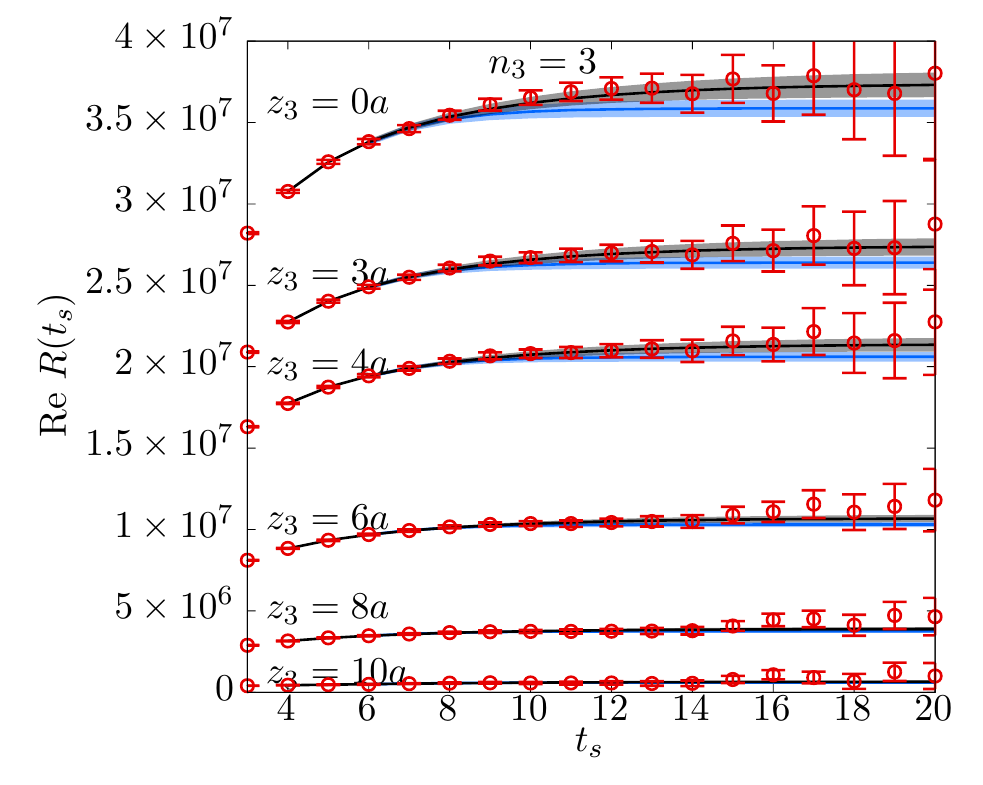}
\includegraphics[scale=0.7]{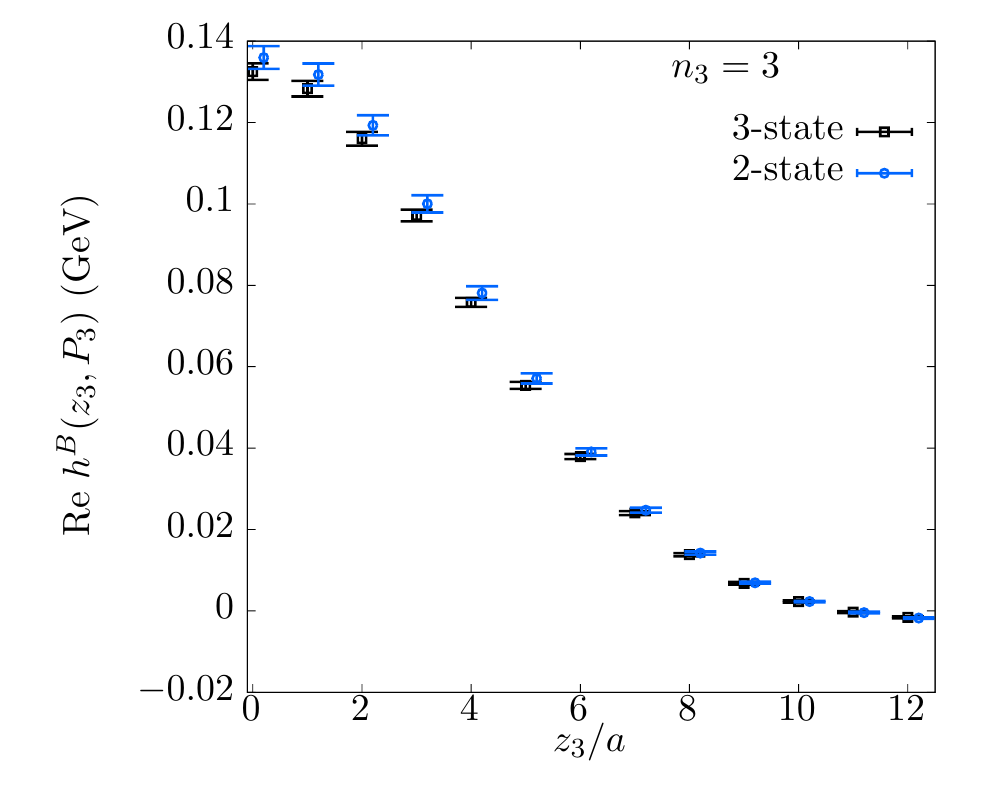}

\includegraphics[scale=0.7]{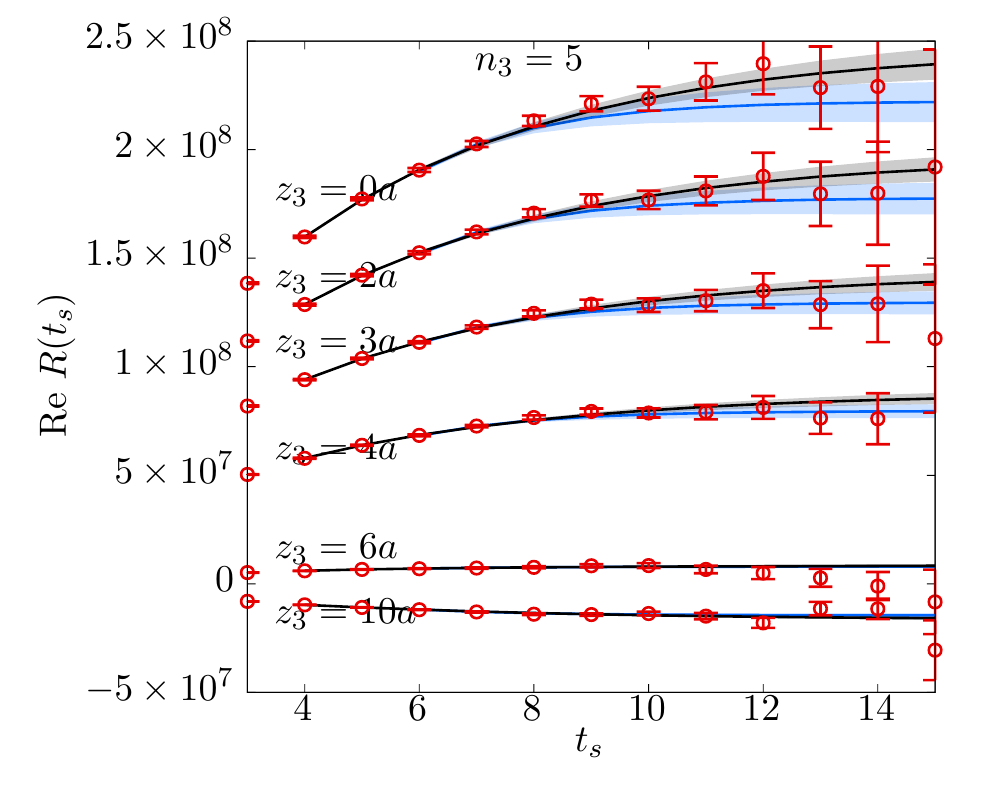}
\includegraphics[scale=0.7]{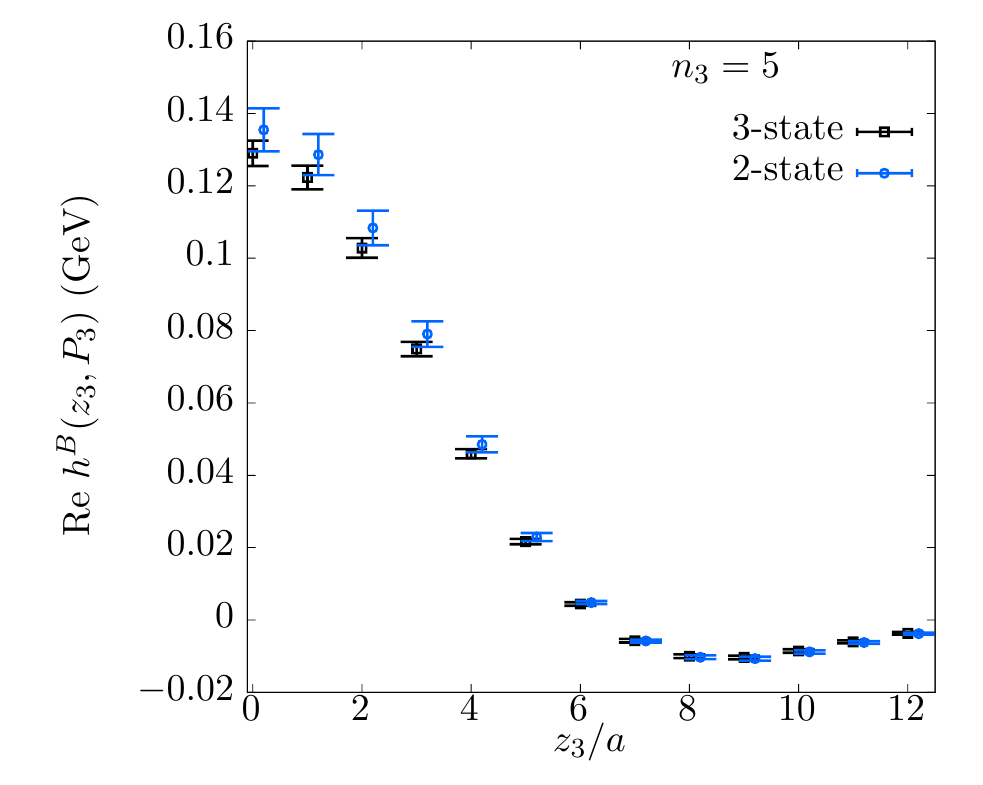}

\includegraphics[scale=0.7]{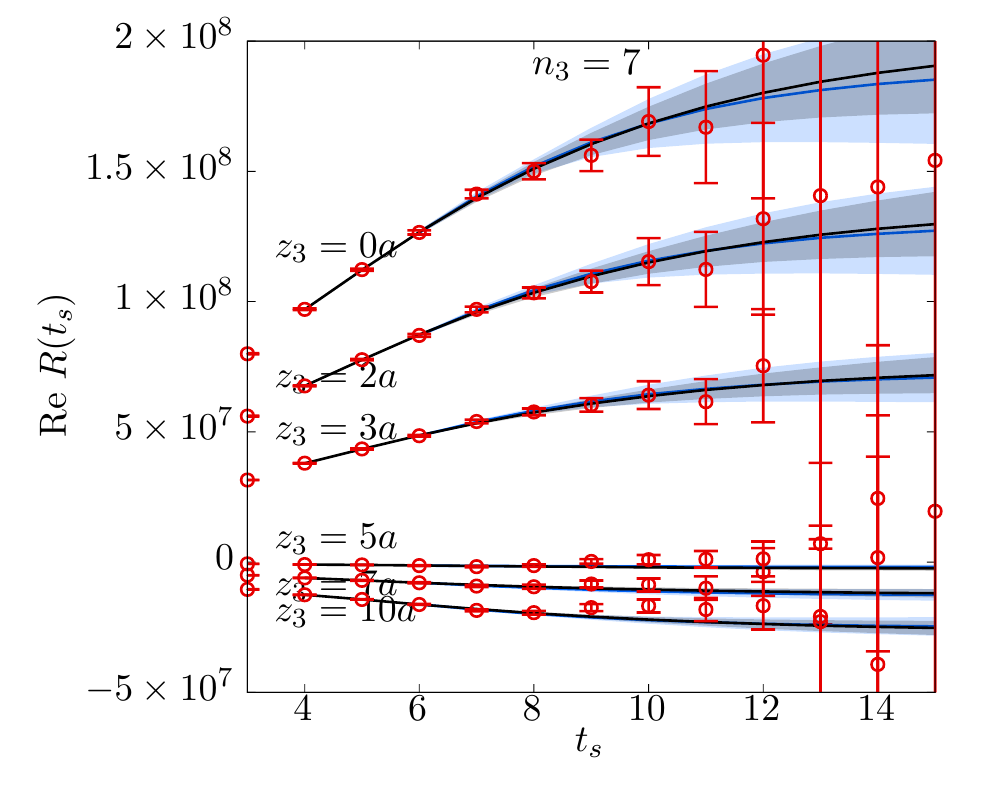}
\includegraphics[scale=0.7]{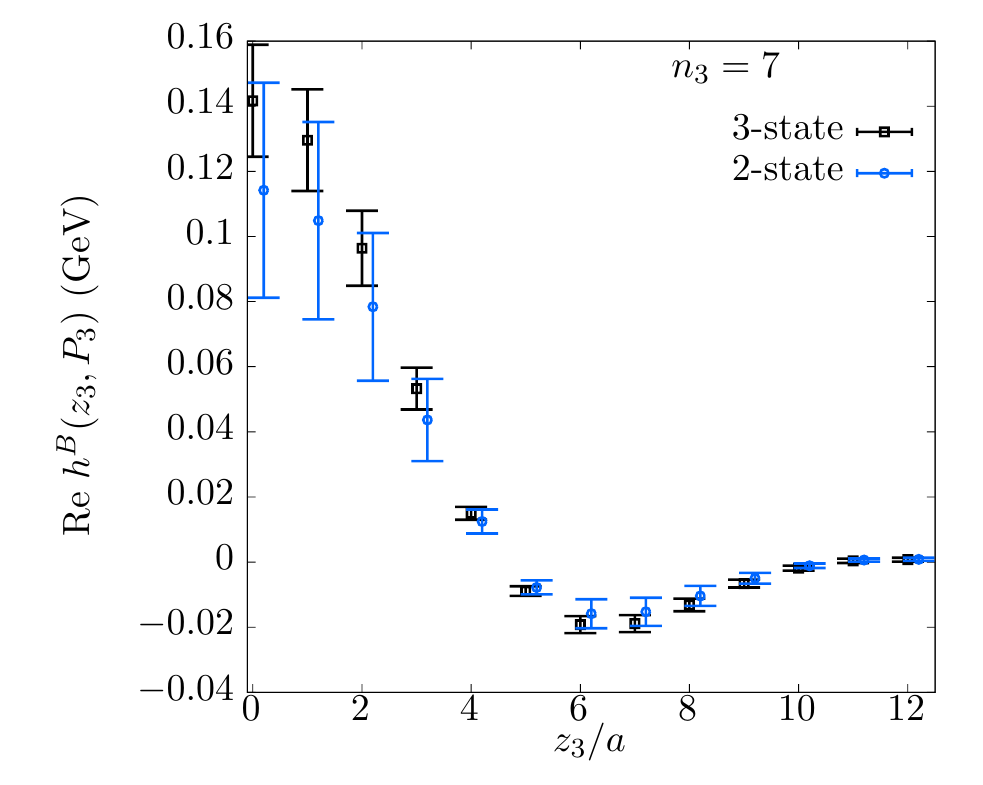}
\caption{The determination of the real part of the bare matrix element
${\rm Re} \: \tilde{h}^B(z_3,P_3)$ from two-state and three-state fits to the
ratio $R(t_s)$ for momenta $P_3=0.254 n_3$ GeV for $n_3=$ 1, 3, 5
and 7 from top to bottom.  The left panels show the extrapolations
in $t_s$ with the two-state and three-state fits shown as the blue
and black bands.  The results at different representative values of
$z_3$ used in this work are also shown together.  The right panels
show the resulting extrapolated values of ${\rm Re} \: h^B(z_3,P_3)$,
in units of GeV, as a function of $z_3/a$.  The results using two-state
and three-state fits are slightly displaced horizontally for clarity.
}
\eefs{realextrapol}

We used the spectral decomposition of $C_{\pi\pi}(t_s; P_3)$ and
$C_{\pi \tilde{O}_3}(t_s; P_3,z_3)$ to extract the bare quasi-DA
matrix element. That is,  the pion-pion correlator,
\beq
C_{\pi\pi}(t_s) = \sum_{i=0}^{N_{\rm st}-1} \frac{|Z_n|^2}{2 E_n} \left (e^{-E_n t_s} + e^{-E_n(L_t-t_s)} \right),
\eeq{pipispect}
and the pion-quasiDA correlator,
\beqa
&&C_{\pi O_3}(t_s) =\cr&&\qquad \sum_{i=0}^{N_{\rm st}-1} \frac{Z_n}{2 E_n} \mel**{0}{\tilde{O}^B_3(z)}{E_n} \left (e^{-E_n t_s} + e^{-E_n(L_t-t_s)} \right), \nonumber\\
\eeqa{piOspect}
where the kets are relativistically normalized, and $Z_n = \mel**{\pi;
P_3}{\pi^\dagger(P_3)}{0}$, which we assume to be real and positive
in this work.  The summations in \eqn{pipispect} and \eqn{piOspect}
run over all the eigenstates at definite momentum $P_3$. However,
for practical considerations, one truncates them including only the
lowest $N_{st}$ states.  We refer to the $N_{\rm st}=2$ truncated
expression as the two-state ansatz, and the $N_{\rm st}=3$ truncation
as the three-state ansatz.  The periodicity of the two correlators
is imposed above; for $C_{\pi \tilde{O}_3}(t_s)$ the periodicity
can be seen by reflection $(x_0,x_1,x_2 , x_3)\to (-x_0, -x_1, -x_2,
x_3)$, along with $q \to \gamma_z q$, $\bar q \to \bar q \gamma_z$
where $q$ is either $u$ or $d$, which is a symmetry of the Euclidean
path integral, and therefore, $C_{\pi \tilde{O}_3}(t_s) = C_{\pi
\tilde{O}_3}(-t_s)$.  We obtained the bare quasi-DA matrix element
from the analysis of the ratio,
\beq
R(t_s) = \frac{-i C_{\pi \tilde{O}_3}(t_s; P_3,z_3)}{C_{\pi\pi}(t_s; P_3)},
\eeq{ratc3c2}
whose spectral decomposition is simply the ratio of \eqn{piOspect}
and \eqn{pipispect}.  It can be seen that the leading term for large
$t_s$ behaves as $R(t_s)\to P_3 \tilde{h}^B(z_3,P_3) / Z_0$.

First, we obtained the best fit values of the spectral parameters
$E_n$ and $Z_n$ from the analysis of $C_{\pi\pi}$ correlators.  In
a previous work~\cite{Gao:2021xsm}, we discussed our fits to the
pion correlator on the same gauge ensemble.  In this work, we used
the two-state and three-state fits with the ground-state energy
fixed to the continuum dispersion, $E_0(P_3) = \sqrt{P_3^2 +
m_\pi^2}$.  We chose the fit ranges $t_s\in[t_{\rm min}, t_{\rm
max}]$ for $C_{\pi\pi}(t_s)$ such that they covered the range used
for the subsequent fits to the ratio $R$ to be discussed next.
Namely, we used $t_s\in [4a,32a]$ for two-state fits and $t_s\in[2a,32a]$
for three-state fits. In this way, we obtained good effective values
of $E_n$ and $Z_n$ that best describe the excited state contribution
to the ratio $R(t_s)$ in the range of $t_s$ we made use of.  However,
as observed in Ref.~\cite{Gao:2021xsm}, the values of $E_0$, $E_1$
and $E_2$ from our final fit choice were within errors of the results
when larger $t_{\rm min}$ were used, albeit with noisier determinations.
Whereas the value of $E_1$ at $P_3=0$ is consistent with the 
pole mass of $\pi(1300)$, the value of $E_2$ at $P_3=0$ from three-state fits is much higher than expected at
3 GeV. Thus, as noted in~\cite{Gao:2021xsm}, it is likely that the three-state fit with $E_2$ capturing the tower of 
excited states above $E_1$ via a single effective state.

In the next step, we used $(Z_n,E_n)$ from two-state and three-state
fits on jackknife samples as inputs in our fits to $R(t_s)$ over
ranges $t_s \in [t_{\rm min}, t_{\rm max}]$ on the same jackknife
samples.  The fits for $R(t_s)$  used the above spectral decomposition
with fit parameters being the amplitudes $\langle 0|O_3|E_n\rangle$,
and therefore, the fits were linear.  We performed these fits to
$R(t_s)$ using two-state and three-state ansatz.  For two-state
fits to $R$, we chose $t_{\rm min}=6a$, whereas for the three-state
fits, we used $t_{\rm min}=4a$. For both two- and three-state fits,
we chose the maximum range of the fits $t_{\rm max}=20a$ for momenta
$n_3 \le 3$ and $t_{\rm max}=15a$ for $n_3 > 3$ to avoid noisier
estimates at larger $t_s$.  In this way, we extrapolated the ratio
to $t_s\to\infty$ to obtain the bare quasi-DA matrix element
$\tilde{h}^B(z_3,P_3)$.

\bef
\centering
\includegraphics[scale=0.8]{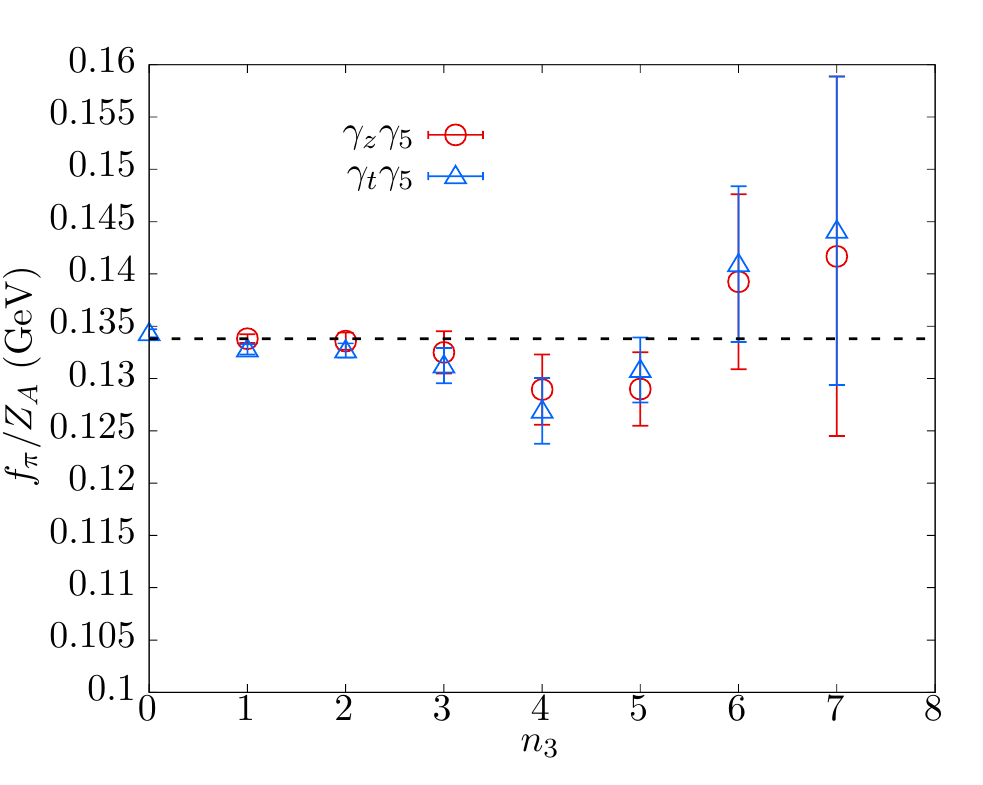}
\caption{The pion decay constant $f_\pi$, modulo the finite
renormalization factor $Z_A$, is shown as a function of momenta
$P_3=0.254 n_3$ that is used in the extraction. The results using
$\Gamma=\gamma_z\gamma_5$ and $\gamma_t\gamma_5$ are shown in the
plot. 
The dashed curve is the value of $f_\pi/Z_A$ from $n_3=1$.
}
\eef{fpi}

In \fgn{realextrapol}, we show some examples from our extrapolations
of the real part of $R$. From top to bottom, the data are from
momenta $n_3=1$, 3, 5 and 7 respectively.  Let us first focus on
the left panels. We show the data for ${\rm Re}(R)$ as a function
of $t_s$ for few sample values of $z_3$ as specified near the data
points.  The magnitude of $R$ is not important as it still depends
on the two-point function amplitude $Z_0$, and only its $t_s$
dependence is important here.  For $n_3=1$ to 3, the variations
with $t_s$ is smaller due to larger energy gap, $E_1-E_0$, which
is about the gap between pion mass and that of $\pi(1300)$. For
larger $n_3$, the variation of $R$ with $t_s$ is significant due
to the states being relativistic. Thus, the extrapolations using
spectral decomposition of $R$ is necessary in our calculation,
especially in the important large momenta data set. The blue and
the black bands show the extrapolations using the two-state and
three-state ansatz respectively. Within the statistical errors, the two
extrapolation bands satisfactorily describe the $t_s$ dependence
of the lattice data. However, the three-state fits have a tendency
to be closer to the central values of the data when compared to the
two-state ones. 
In the right panels of
\fgn{realextrapol}, we compare the resultant values of the bare
quasi-DA matrix element, $\tilde{h}^B(z_3,P_3)$, from the two-state and
three-state fits to $R$ as a function of $z_3$. Note that the
ordering of blue and black points in the right panel is not in
one-to-one correspondence with the left panel as there is also an
additional factor $Z_0$ that is different between the left and right
panels.  Within errors, the two-state and three-state extrapolated
values are consistent, with perhaps a slight tension in the $n_3=3$
momenta. The errors on the three-state fits are comparable or smaller
than in the two-state fits due to the $t_{\rm min}$ being $4a$ for
three-state ones compared to $6a$ for two-state fits.  The relative error
at larger momenta $n_3 > 5$ increases at even shorter $z_3$. Nevertheless,
as we will discuss in the next subsection, the growth in statistical
error in shorter $z_3$ is reduced due to the RGI ratios that we will construct,
and, due to the same reason, even the slightest discrepancies between
the two-state and three-state fits seen in the bare matrix element
in \fgn{realextrapol} will be reduced further.
We found similar consistency between the 
two-state and three-state fits in the 
case of  ${\rm Im}(R)$.

\befs
\centering
\includegraphics[scale=0.8]{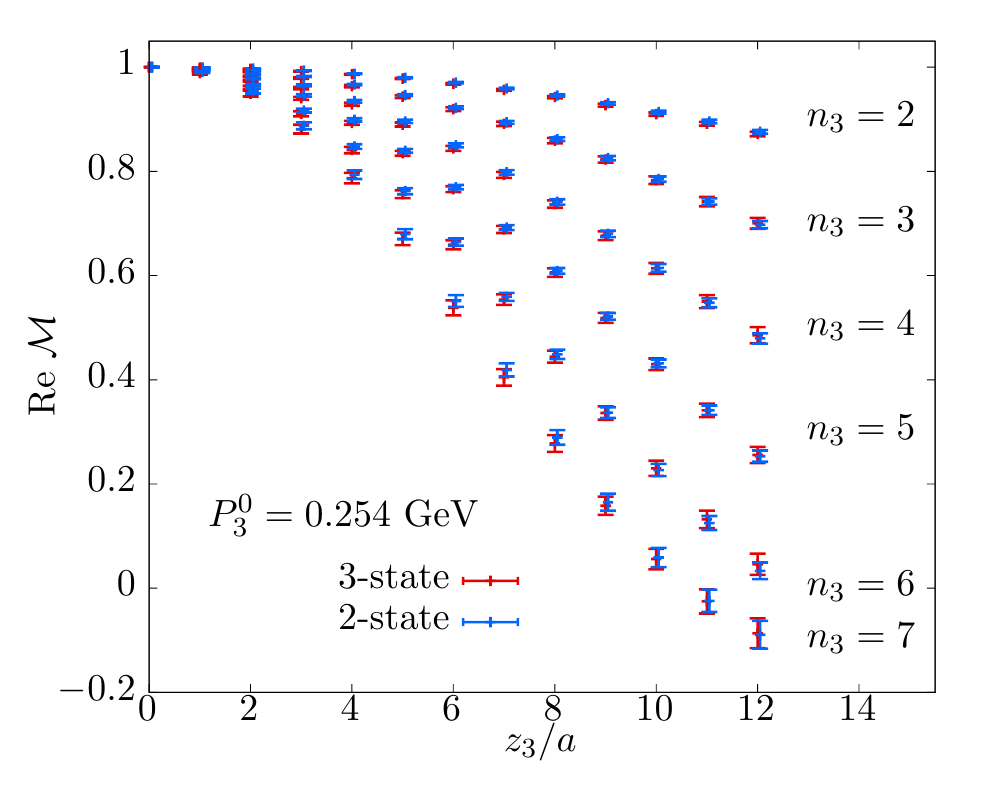}
\includegraphics[scale=0.8]{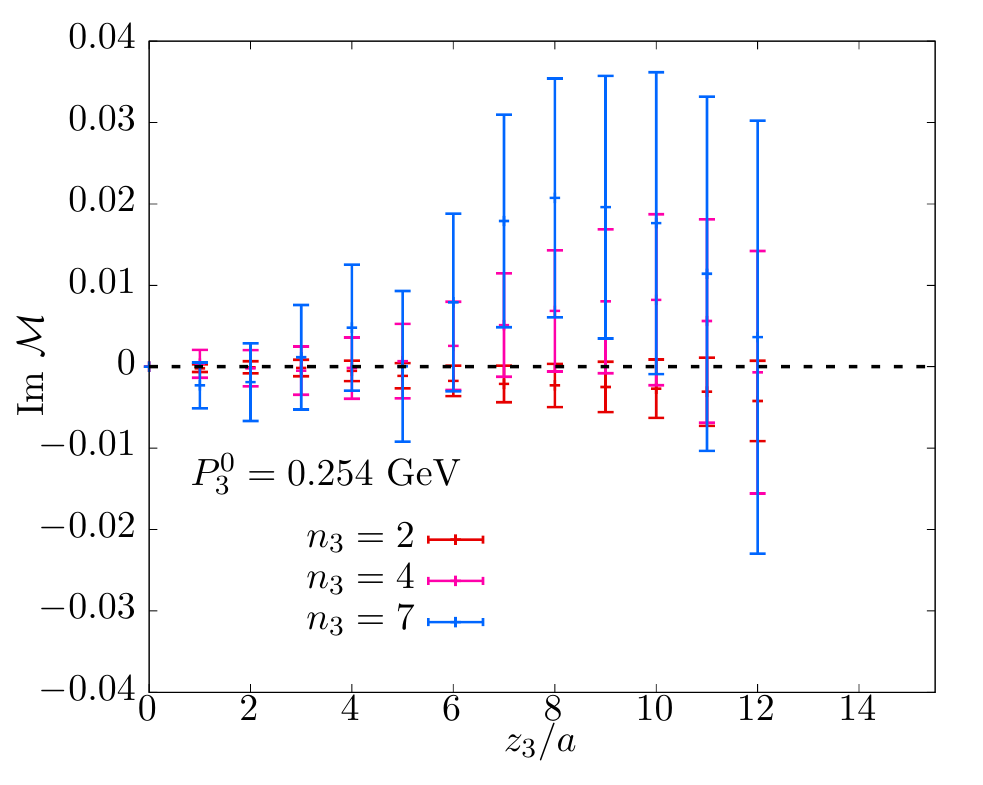}
\caption{Renormalized matrix elements. The left panel shows the
ratio ${\rm Re} \: {\cal M}(z_3P_3, z_3^2, P_3^0)$ as a function of $z_3/a$, for a
specific $P_3^0=0.254$ GeV. The results from the 
two-state and three-state
extrapolations are compared in the panel to demonstrate that the ratio
further reduces any extrapolation uncertainties. The right panel
shows the ratio ${\rm Im} \: {\cal M}(z_3P_3, z_3^2, P_3^0)$ as a function of $z_3$
from three different representative momenta. 
The consistency of ${\rm Im} \: {\cal M}=0$ is demonstrated.
}
\eefs{rencompare}

The $z_3=0$ value of $\tilde{h}^B$ is the bare pion decay constant,
$f_\pi/Z_A$, where $Z_A$ is the finite renormalization constant for
the axial current operator. Thus $\tilde{h}^B(0,P_3)$ has to be
constant with $P_3$ if there were no systematical errors in the
extrapolations and if ${\cal O}(a P_3)$ lattice corrections did not
affect the lattice results, and therefore, provides a cross-check
on our calculation.  In \fgn{fpi}, we show $f_\pi/Z_A$ as a function
of momentum $P_3$ used in $h^B(z_3=0,P_3)$. The red circular data
points are the results using the $O_{3}(z=0)$ operator.  In addition,
we also looked at the local matrix element from $O_0(z=0)$. We show
those values as the blue triangles in \fgn{fpi}. The values of
$f_\pi/Z_A$ are consistent with being constant with respect to
$P_3$, with perhaps a slight dip in the central value around $n_3=3$,
which could be due to statistical fluctuation.  The excited state
contribution and the Lorentz structure of the $O_3$ and $O_0$  matrix
elements are different, and hence, the consistency between the
$f_\pi/Z_A$ determinations from the two observables is reassuring.
Using RI-MOM renormalization procedure, we determined $Z_A=0.969(1)$
for the ensemble used in this paper (see Appendix A).  From the
most precise values of the matrix element at $n_3=0$ for $O_0(z=0)$
and $n_3=1$ for $O_3(z=0)$, the values of $f_\pi$ from the two
observables are 130.0(4) MeV and  129.7(4) MeV respectively.  These
results agree with the FLAG average for 2+1 flavor QCD $f_\pi=130.2(8)$
MeV \cite{Aoki:2021kgd}.  Due to the normalization condition $\langle
x^0\rangle=1$ that we will impose on the matrix elements, $f_\pi$
will not play any further role in this calculation.

\subsection{Renormalized ratios}

We used the RGI ratios of $h^B(z_3,P_3)$ to get
the renormalized quantities. First,
we shifted the location of the operator $\tilde{O}(z)$ by $-z/2$ in
order to conform with the definition in \eqn{quasidaop}. We
did this by multiplying $\tilde{h}^B(z_3,P_3)$ with a phase $\exp(-i
z_3 P_3/2)$ from the translation. Next, we improved the ratio in
\eqn{ratio1} to impose the condition that the ratio should be exactly
unity at $z_3=0$ using the so called double ratio procedure. Thus,
in the end, we determined the RGI ratio~\cite{Orginos:2017kos,Fan:2020nzz,Gao:2020ito} as
\beqa
{\cal M}(\lambda,z^2, P^0) &\equiv& \left(\frac{\tilde{h}^B(z_3,P_3)}{\tilde{h}^B(z_3, P_3^0)}\right)\left( \frac{\tilde{h}^B(0,P_3^0)}{\tilde{h}^B(0, P_3)}\right)\cr
&& \times e^{-i \frac{z_3}{2}(P_3-P_3^0)}.
\eeqa{doublerat}
We have written the arguments in the right-hand side above in terms
of $z_3$ and $P_3$ to make the definition clear.  The factor in the
second parenthesis above should be exactly one, devoid of any
systematical and statistical errors. We refer to the fixed momentum
$P^0$ used to form the ratio as the reference momentum.
We used a non-zero value of $P^0$ for two reasons --- 1) the
leading-twist part of the matrix element of $O_3$ vanishes at zero
spatial momentum. 2) by using larger $P^0$, the higher-twist
corrections, such as $(\Lambda_{\rm QCD}^2 z_3^2)^k$, present in
$h^R(z_3,P_3^0)$ are made relatively smaller compared to the
leading-twist terms containing powers of $P^0_3 z_3$. However, if
we use $P_3 < P_3^0$, the possible advantage of larger $P_3^0$ is
rendered meaningless.  Therefore, we only used $P_3 > P_3^0$.

In \fgn{rencompare}, we show the real and imaginary parts of the
RGI ratio with reference momentum $P_3^0=0.254$ GeV.  In the left
panel, we show ${\rm Re} \: {\cal M}$ as a function of $z_3$ for
easier visibility of data at different $P_3$.  We show 
the results for ${\rm
Re} \: {\cal M}$ obtained using the bare matrix elements from the two-state
and three-state fits together. It is clear that the two
extrapolated results are quite consistent with each other, even
more so after forming the RGI ratios, wherein any correlated
systematical errors could get canceled between the numerator and
denominator.  It is also striking that the errors at larger momenta
at small to moderate range of $z_3$ are statistically well determined,
thanks to the statistical correlation in the data at $P_3$ and
$P_3^0$ at a given $z_3$. By construction, the $z_3=0$ value of
${\cal M}$ is exactly 1. In the rest of the paper, we will use the
data for ${\cal M}$ obtained using three-state fits.

In the right panel of \fgn{rencompare}, we show ${\rm Im} \: {\cal M}$
from three different representative values of $P_3$.  Due to chiral
symmetry, the leading-twist part of $h^R(\lambda,z^2,\mu)$, and hence
${\cal M}$, should be purely real.  This is demonstrated in our data
by the vanishing of ${\rm Im} \: {\cal M}$ well within statistical
errors at different $P_3$ and $z_3$.  Therefore, we only
analyzed ${\rm Re} \: {\cal M}$ and imposed the symmetry of pion DA about
$x=0$ explicitly by setting $\langle x^n\rangle=0$ for odd
$n$.

\section{Results}\label{sec:results}

\subsection{Analysis strategy}
We extracted the leading-twist DA related quantities from $\reM$
by fits to the corrected (as well as the uncorrected) leading-twist
expression $\reMtw$ in \eqn{ratiotw2corr}. Let ${\cal P}$ be the
set of free parameters that enter $\reMtw$; for example, they could
be the set of moments or the parameters of a DA ansatz. We found
the best fit values of ${\cal P}$ by the standard $\chi^2$ fits
using $\chi^2 = \Delta^T \Sigma^{-1} \Delta$ with $\Delta_{z_3,P_3}
= \left(\reM(z_3,P_3) - \reMtw(z_3,P_3;{\cal P})\right)$, including
only the data points with $z_3 \in [z_3^{\rm min}, z_3^{\rm max}]$
and $P_3 \in [P_3^{\rm min}, P_3^{\rm max}]$.  We chose $z_3^{\rm
min}>a$ to reduce the effect of lattice corrections at lattice-like
separations. We used $z_3^{\rm max}$=0.456 fm, 0.608 fm and 0.76
fm to take into account possible variations in the fitted values
due to higher-twist contaminations that we did not capture in
\eqn{ratiotw2corr}, and at the same time remain in moderately small
values of $z_3$ that are allowed given the constraint of the lattice
spacing we are using.  We used the momenta from $P_3^{\rm min} >
P_3^0$, the reference momentum. We used the full covariance matrix
$\Sigma$ to take care of correlations between the data at different
$z_3$ and $P_3$.

The value of the strong-coupling constant $\alpha_s$ enters the
leading-twist OPE expressions.  At NLO, the scale at which it needs
to be determined is ambiguous. In this work, we use the value of
$\alpha_s(\mu)$ at the same scale at which the DA is determined,
namely, at $\mu=2$ GeV. We take the value of $\alpha_s\left(2 {\rm
GeV}\right)=0.303$ determined from the running of $\alpha_s$ taken
from the PDG~\cite{Tanabashi:2018oca}.

The leading-twist expansion approach comes with systematic uncertainties
due to the possible analysis choices, such as, the choices of
$z_3^{\rm min/max}$, $P_3^{\rm min/max}$, and the choice of ansatz
for higher-twist corrections, to list a few. Apart from presenting
a scatter of the fitted results for all possible combination of
analysis choices, it is helpful to summarize a result compactly to
capture its central value, the systematic spread due to analysis
variations, and the statistical error on the central value.  To
achieve this, we used the following procedure. Let ${\cal S}$ be
the set of analysis choices, and let ${\cal P}_a$ be the set of
best fit parameters for a particular choice $a\in {\cal S}$.
Following the approach presented in~\cite{Gao:2020ito,HadStruc:2021qdf}
closely, for some function of parameters, $F({\cal P})$, we first
found the mean value $\bar{F}$ and standard deviation $\sigma_{\rm
width}$ of results for $F$ over all analysis choices in a given
jackknife block as
\beq
\overline{F} = \frac{\sum_{a\in {\cal S}} w_a F({\cal P}_a)}{\sum_{a\in {\cal S}} w_a};\quad \sigma^2_{\rm width} = \overline{F^2} - (\overline F)^2,
\eeq{statsys}
for weights $w_a$ for each analysis choice. From the central value
and width of scatter per jackknife sample, we determined the final
estimate of $F$ as (mean)$\pm$(statistical error)$\pm$(systematic
error) by finding $J_{\rm av}(\overline{F})\pm J_{\rm er}(\overline{F})
\pm J_{\rm av}(\sigma_{\rm width})$, where $J_{\rm av}(\ldots)$ is
the jackknife average and $J_{\rm er}(\ldots)$ is the jackknife
standard error of a quantity.  
One possibility for the $w_a$ is the
Akaike information criterion (AIC) weight given by $e^{-\frac{1}{2}(\chi^2 + 2 f)}$ where $f$ is the number 
fit parameters.
We found that such an estimator for our case had a
tendency to choose only a few of the analysis choices and does not
represent the true scatter present in our analysis. Therefore, we
followed the approach we used in our earlier work~\cite{Gao:2020ito},
which is to set a constant $w_a$ (i.e., unweighted averaging) for
all analysis choices. In this way, we took all the choices on equal
footing.

\subsection{Fixed-$z^2$ analysis: looking for corrections to continuum leading-twist expectation}

The simplest analysis of the data is to study the $\lambda$ dependence
of ${\cal M}(\lambda,z_3^2)$ at fixed values of $z_3$ (refer
~\cite{Karpie:2018zaz} where the idea was first proposed for the
forward matrix element).  In this way, the variation in $\lambda$
comes only from the variation in $P_3$. The output of this analysis
is the set of Mellin or Gegenbauer moments at a fixed scale $\mu$,
based on whether M-OPE or C-OPE is used respectively, as a function
of $z_3$. Using the degree of agreement of the $z_3$-dependent
moments with a plateau in $z_3$ is a nice way to understand whether
the fixed-order leading-twist framework is applicable to the lattice
data in a range of $z_3$ or not, and which type of corrections to
continuum leading-twist expansion are seen. Such a diagnosis of the
lattice data has been performed in the case of the forward matrix
element in the PDF determination~\cite{Gao:2020ito}. Here, we apply
such an analysis for the off-forward matrix element for the first
time.
\bef
\centering
\includegraphics[scale=0.82]{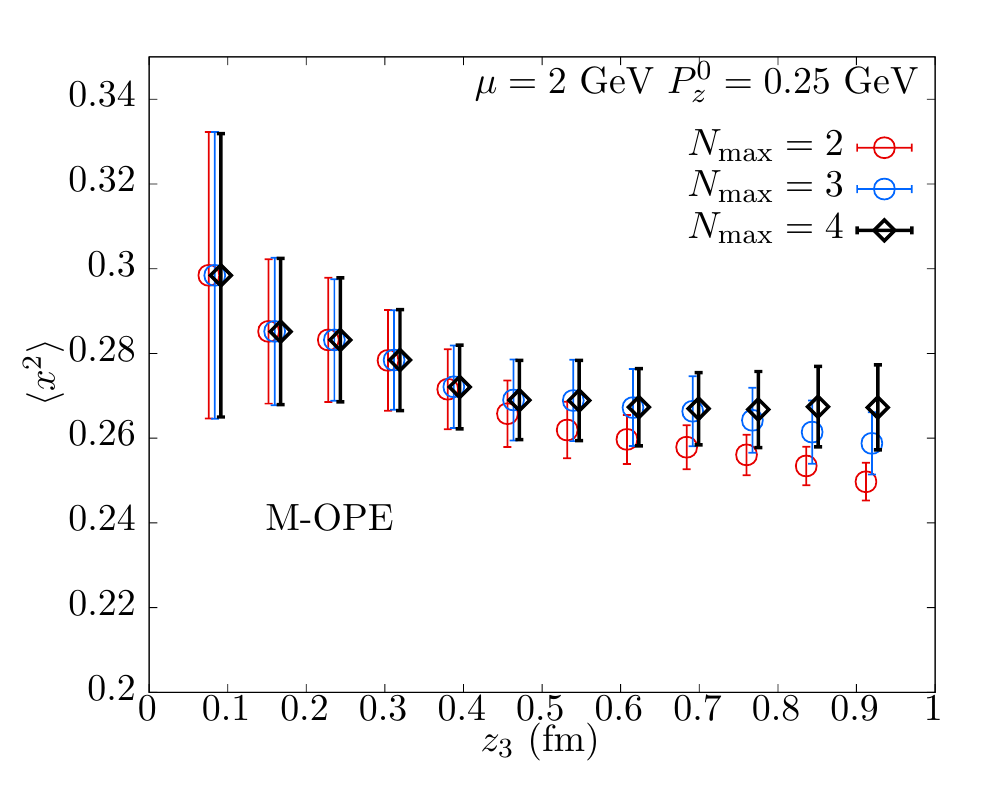}
\includegraphics[scale=0.82]{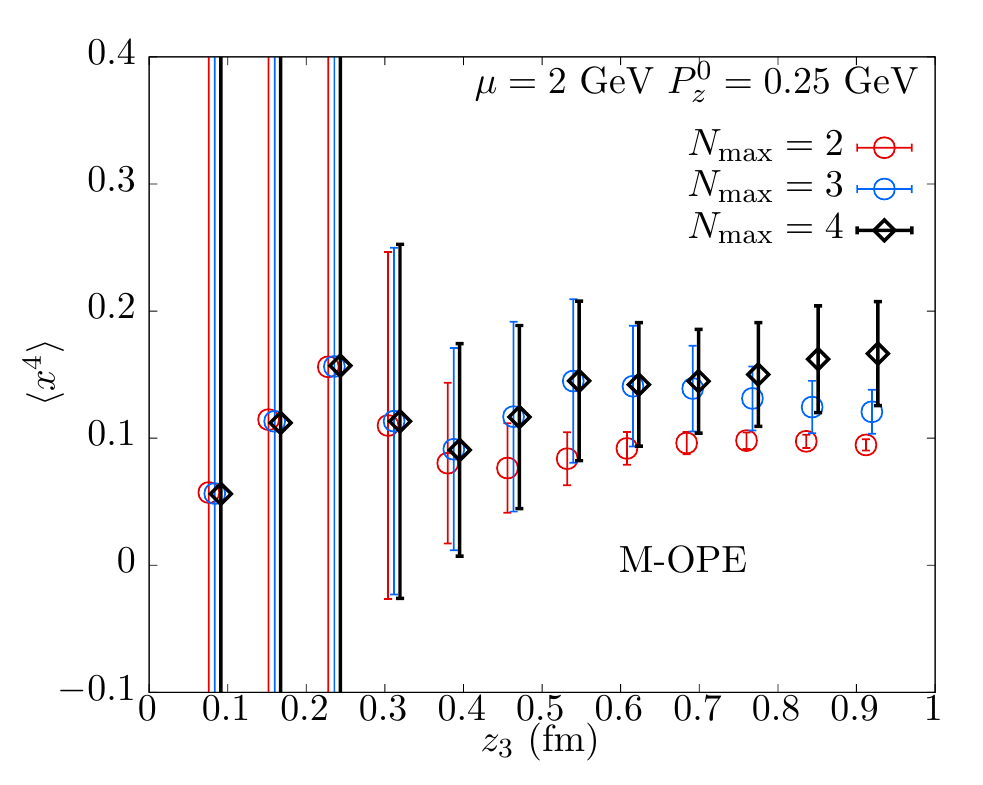}
\includegraphics[scale=0.82]{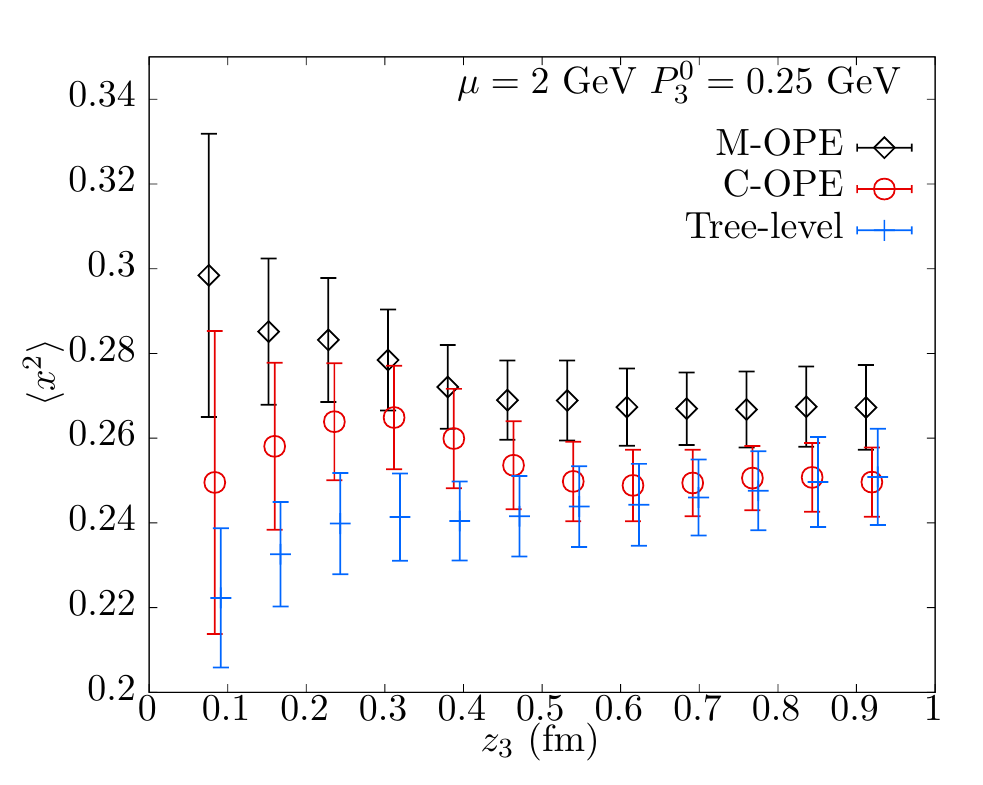}
\caption{The top panel shows the $z_3$-dependence of $\langle
x^2\rangle$ determined from the $\lambda=z_3P_3$ variation of ${\cal
M}(z_3P_3, z_3^2, P_3^0=0.254 {\rm GeV})$ at fixed values of $z_3$.
The convergence with increasing the number $N_{\rm max}$ of even
moments added to the NLO Mellin OPE (M-OPE) is shown.  The middle
panel shows a similar dependence for the determined $\langle
x^4\rangle$. The bottom panel shows a comparison between the results
of $\langle x^2 \rangle$ that are determined from the NLO Mellin
OPE and the conformal OPE. To display the effect of non-zero $\alpha_s$,
we have shown the result using tree-level ($\alpha_s=0$) M-OPE.
}
\eef{fixedz}

We used the purely leading twist expression (i.e., we set the
higher-twist correction $H$ and lattice correction $L$ to zero) in
\eqn{ratiotw2}.  By using the C-OPE for the leading-twist expression,
we obtained the Gegenbauer moments $a_n$ by using them as the fit
parameters. Similarly, by using M-OPE, we obtained the Mellin moments
$\langle x^n \rangle$. Since each Mellin or Gegenbauer moment adds
an additional fit parameter, we needed to truncate the OPE at a
finite order $2 N_{\rm max}$ which contains $N_{\rm max}$ number
of even-$n$ moments; we successively increased $N_{\max}$ from 2
to 4. For C-OPE, the fits became unstable for $N_{\rm max}>3$ as
the dependence on Gegenbauer moments beyond $a_2$ is rather weak.
With M-OPE, we were able to go up to $N_{\rm max}=4$ in this analysis.

In the top panel of \fgn{fixedz}, we show $\langle x^2 \rangle$ as
a function of $z_3$ upto $z_3=0.91$ fm by applying M-OPE to ${\cal
M}$ with $P_3^0=0.254$ GeV.  We show the results using three different
truncations $N_{\rm max}$. For $z_3 < 0.8$ fm, we can see that
$N_{\rm max}=4$ is sufficient given the data precision. The near
plateau in $\langle x^2 \rangle$ around a value of about 0.28 shows
that the leading-twist expansion describes the lattice data for
${\cal M}$ to a good approximation and that any higher-twist or
lattice corrections are subdominant.  This provides a validation
of the leading-twist fixed-order perturbative framework that we are
using to describe the nonpertubative lattice QCD data in the range
of subfermi values of $z_3$ we use.  The near plateau in $\langle
x^2 \rangle$ around a value of about 0.28 shows that our fitting
form based on leading-twist perturbative expansion at NLO, with the
choice of $\mu=2$ GeV, can describe the lattice data within the
current statistical error. However, as shown in
Refs.~\cite{Ji:2020brr,Gao:2021dbh}, perturbation theory may become
unreliable at large $z$ due to the resummation of large
$\ln(z^2\mu^2)$~\cite{Gao:2021hxl}, so a more dedicated study of
the comparison between fixed-order and renormalization-group improved
OPEs needs to be done to understand the results we have observed.
Nevertheless, the central value of $\langle x^2 \rangle$ changes
by about 0.01 ($\approx 4\%$) as $z_3$ is increased from 0.1 fm to
0.6 fm.  Hence, in the subsequent analysis of the data, we will
include correction terms such as $H$ and $L$ to the leading-twist
expansion. Since the correction itself is small, we were not able
to deduce a possible functional form for the functions $H$ and $L$
that might be present, as we did in the case of the pion PDF in
Ref~\cite{Gao:2020ito}.  In the middle panel of \fgn{fixedz}, we
show a similar $z_3$ dependence of $\langle x^4 \rangle$ at $\mu=2$
GeV. Since the analysis only made use of six different data points
at each $z_3$, the errors on $\langle x^4\rangle$ are larger.
Significant information on $\langle x^4 \rangle$ enters ${\cal
M}(\lambda,z^2)$ only beyond $z_3 > 0.5$ fm, wherein we find initial
indication that $\langle x^4 \rangle \approx 0.15$, as we will find
in the subsequent combined analysis of all the data. Within the
larger errors, the data is consistent with $z_3$ independence.

In the bottom panel, we compare the values of $\langle x^2 \rangle$
obtained using M-OPE (black squares) with that from C-OPE (red
circles). For C-OPE, we converted the values of the fitted Gegenbauer
moment $a_2$ to $\langle x^2\rangle$ through the simple linear
relation (e.g.,~\cite{san2012some}), $\langle x^2\rangle=1/5+12/35
a_2$.  While results from both M-OPE and C-OPE are approximately
$z_3$ independent, the values of $\langle x^2 \rangle$ from M-OPE
is about 3\% higher than that from C-OPE.  Since both M-OPE and
C-OPE have converged well with respect to the OPE truncation, this
is likely due to the remnant finite ${\cal O}(\alpha_s)$ corrections
that are missing from C-OPE, but captured correctly in M-OPE. We
also show the result using $\alpha_s=0$ in the M-OPE, which we refer
to as the tree-level result.  Surprisingly, the tree-level result
is also approximately plateaued, showing the effect of perturbative
$\ln(\mu^2 z_3^2)$ in the OPE to be mild in the range of $z_3$ that
we investigated using the choice of scale $\mu=2$ GeV.  Henceforth,
we will primarily show the results using M-OPE, and use C-OPE to
compare those results with and serve as an indirect way to quantify
the perturbative uncertainty by going from leading-log order to
NLO.

\subsection{Determination of Mellin moments}\label{sec:moments}
\befs
\centering
\includegraphics[scale=1.2]{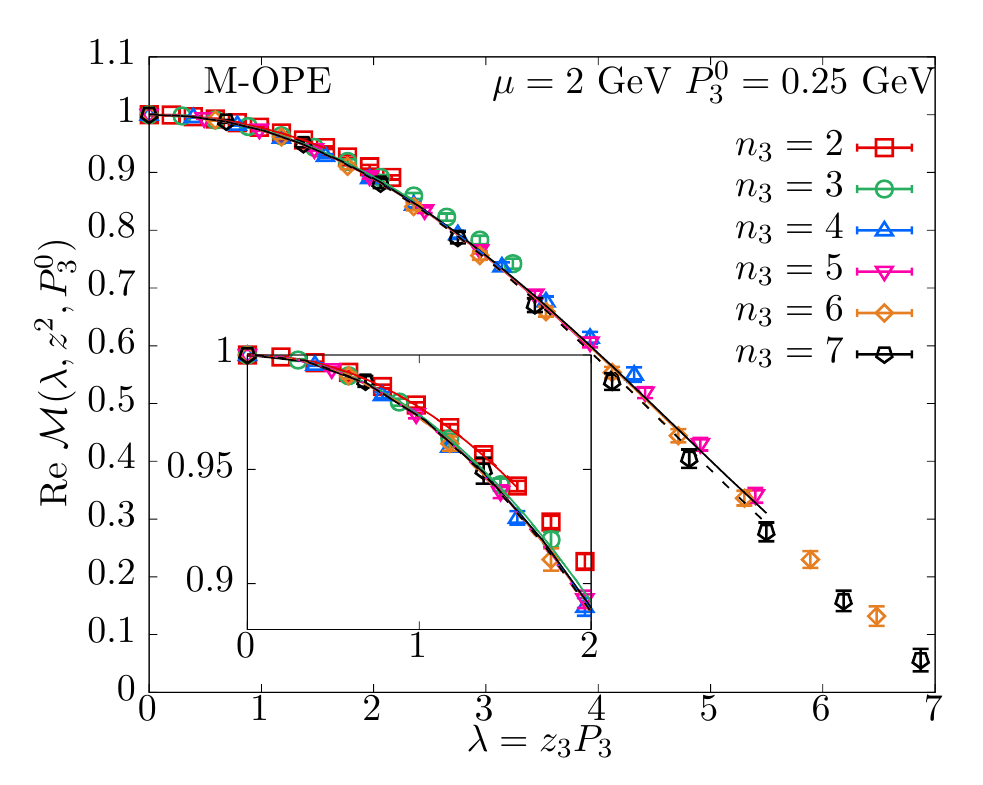}
\includegraphics[scale=1.2]{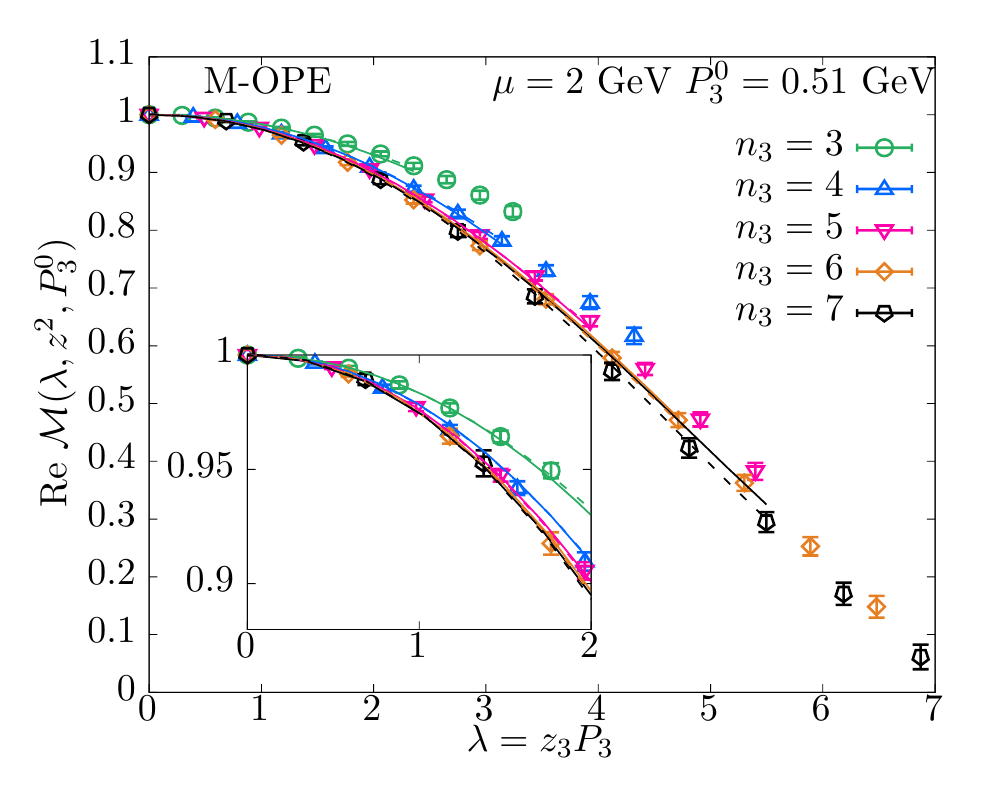}
\caption{
Fits of the leading-twist Mellin OPE to ${\cal M}(\lambda,z^2,P_3^0)$
using $\langle x^{2n}\rangle$ for $n\in[1,5]$ as fit parameters.
The top and the bottom panels are for two-different fixed reference
momenta $P_3^0=0.254$ GeV and 0.508 GeV used to form the ratio. The
central values of the best fit curves from the Mellin OPE are
compared with the lattice data; the solid curves are the results
from using only the leading-twist OPE whereas the dashed curves are
obtained by including a higher-twist term proportional to the
conformal wave ${\cal F}^{(0)}_0(\lambda/2)$ along with $L(z_3)$
lattice correction term.  The data and the curves at different fixed
momenta $P_3$ are distinguished by their colors.
}
\eefs{momentsfit}

Having shown the effectiveness of leading-twist OPE in capturing
the $\lambda$ and $z_3^2$ dependencies of ${\cal M}(\lambda,z_3^2)$,
we now comfortably apply the framework to estimate the lowest two
Mellin moments $\langle x^2\rangle$ and $\langle x^4\rangle$ through
a combined fit to all the data for ${\cal M}(\lambda,z_3^2)$ within
a range of $z_3$. The fits are similar to the ones in the last
section with the moments themselves as the free fit parameters.
Therefore, the method is independent of any modeling of the
$x$-dependence of the DA.

In the absence of any obvious visible evidence in \fgn{fixedz} for
lattice and higher-twist corrections, we simply modeled them.  For
the lattice correction that affects the lattice-like separations
$z_3$, we assumed a possible presence of $(P_3 a)^2$ corrections
in the quasi-DA matrix element similar to the one in the quasi-PDF
matrix element~\cite{Gao:2020ito}. Thus, we chose a functional form,
\beq
L(z_3,P_3) = l_2 (P_3 a)^2,
\eeq{latcorr}
with $l_2$ being a real valued fit parameter in the modeled
correction.  In this way, such a term can effectively affect the
leading twist terms at ${\cal O}(\lambda^2)$ through a $z_3^{-2}$
type correction to the moments. In addition, there could be momentum
independent ${\cal O}(a)$ or ${\cal O}(a^2)$ corrections; since the
continuum leading-twist expressions work quite well in describing
the lattice data at a fixed lattice spacing, it is likely that such
momentum independent corrections can be absorbed as part of the
Mellin moments and the extracted DA themselves at that finite lattice
spacing.  For the higher-twist corrections, we followed a procedure
similar to the one in~\cite{Braun:2007wv}, and assumed that the
corrections resemble the one from twist-4 DA terms captured via a
conformal OPE. For this, we added terms of the form,
\beq
H(z_3,P_3; N_{\rm HT}) =  \sum_{m=0}^{N_{\rm HT}-1} z_3^2 h_m {\cal F}^{(0)}_m(\lambda/2),
\eeq{htmodel}
with $h_m$ being the free parameters.  In the above equation, we
used the tree level conformal partial waves ${\cal F}^{(0)}_m$,
obtained by setting $\alpha_s=0$ in \eqn{partwavedef}, to avoid
modeling the logarithmic $z_3^2$ dependencies using extra fit
parameters. Since we introduce the corrections as a ratio via
\eqn{ratiotw2corr}, the term $H$ can start at ${\cal O}((\lambda)^0)$
as the condition that ${\cal M}^{\rm tw2,corr}(\lambda=\lambda^0,z_3)=1$
is satisfied by construction.  Thus, we included ${\cal F}_0 \sim
(\lambda)^0$ as the leading term to \eqn{htmodel}. We used $N_{\rm
HT}=0,1,2$  in order to keep the number of correction terms required
to be small and at the same time take into account the sensitivity
of the extracted results on the modeled higher-twist effects.
However, one should note that there exists more complex approaches to
model the functional form of $H$
(e.g., see~\cite{Braun:2004bu} that uses renormalon model) than the simpler
functional parametrization that we adopt in this work.

\befs
\centering
\includegraphics[scale=1.05]{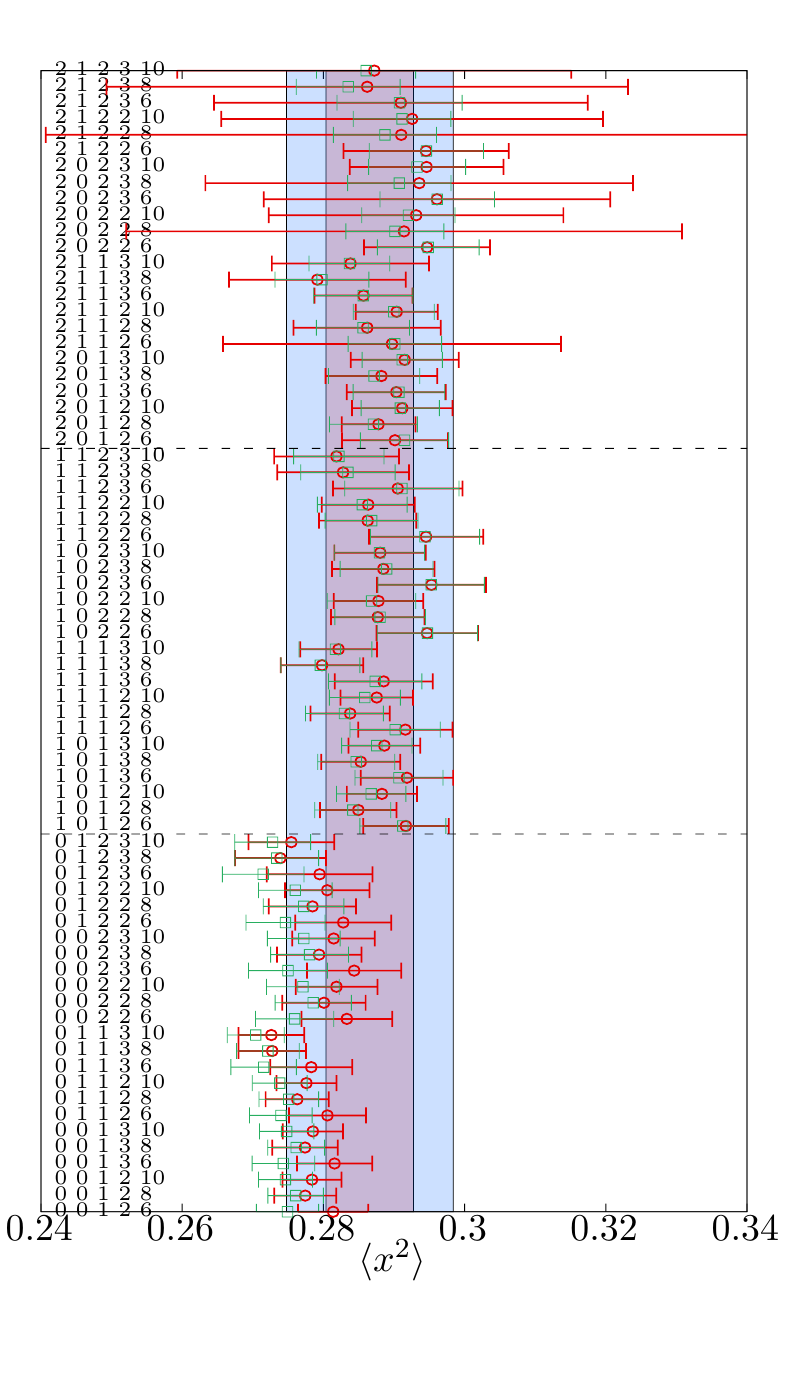}
\includegraphics[scale=1.05]{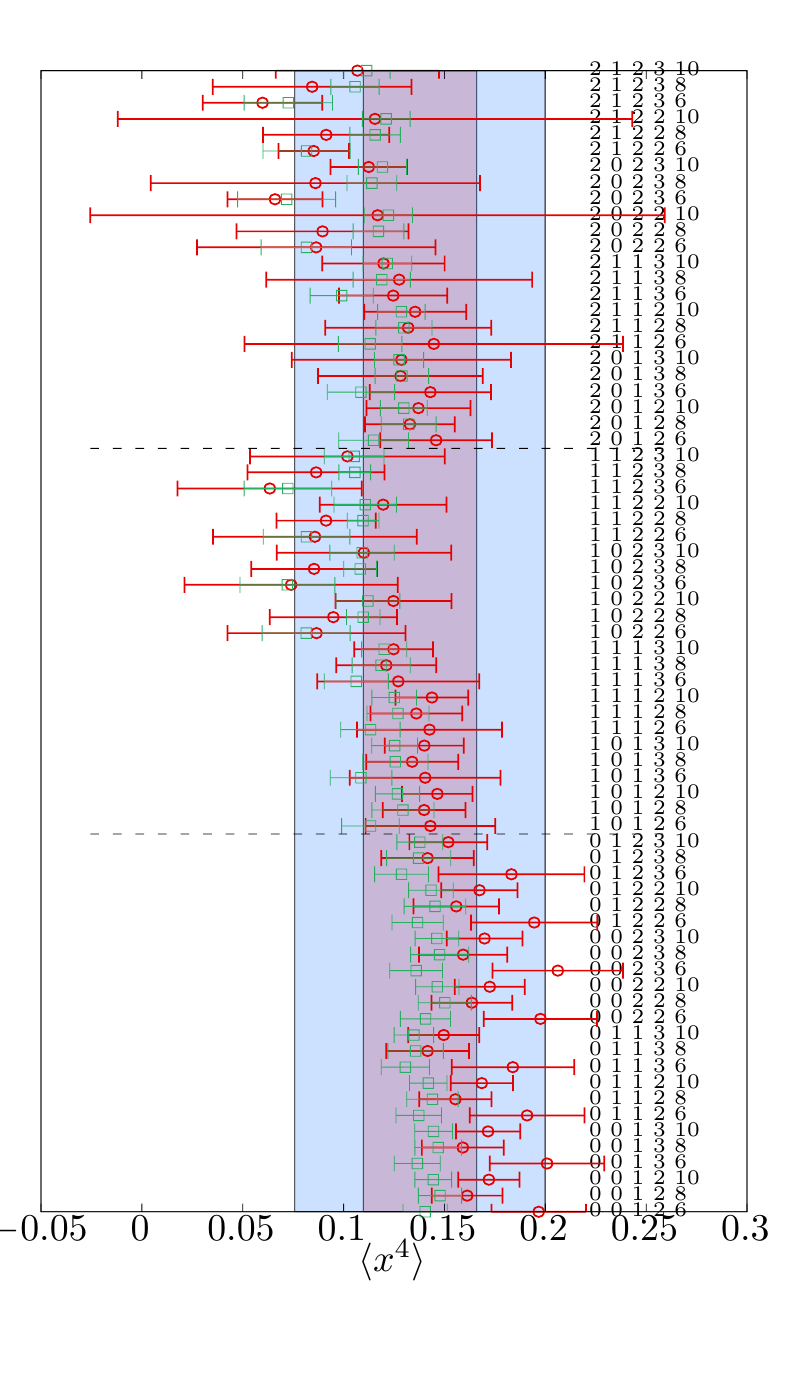}
\caption{
    The scatter of best fit values of $\langle x^2 \rangle$  and
    $\langle x^4 \rangle$ from combined fits to $\lambda$ and $z^2$
    dependencies of ${\cal M}(\lambda,z^2,P_3^0)$ over fit ranges
    $z_3\in[z_3^{\rm min}, z_3^{\rm max}]$ and $P_3\in [P_3^{\rm
    min}, P_3^{\rm max}]$.  The red circles are obtained using
    Mellin moments as fit parameters without constraints, whereas
    the green circles are obtained using a positivity constraint
    on the pion DA.  The variability comes from the number of
    higher-twist correction terms $N_{\rm HT}$, the number of lattice
    correction terms $N_{\rm LC}$, the reference momenta $n_3^0$
    used in the ratio, and the ft ranges. The complete specification
    $(N_{\rm HT}, N_{\rm LC}, n_3^0, z_3^{\rm min}/a, z_3^{\rm
    max}/a)$ is noted on the side of the points. The dashed lines
    separate cases with $N_{\rm HT}=0,1,2$ to see the overall effect
    of adding beyond leading-twist correction terms by hand.  As
    determined from the unconstrained fits, the inner red band is
    the statistical error whereas the outer blue band includes both
    statistical and systematic errors.  To compare, the values in
    the conformal limit are $\langle x^2 \rangle = 0.2$ and $\langle
    x^4 \rangle=0.0857$.
}
\eefs{momsys}

In \fgn{momentsfit}, we show the ratio ${\cal M}(\lambda, z_3^2,
P_z^0)$ as a function of $\lambda=P_3 z_3$, using only the lattice
data with $z_3 < 0.91$ fm.  The top and bottom panels are obtained
using the reference momentum $P_3^0=0.254$ and 0.508 GeV respectively.
We show the lattice data points from different $P_3 > P_3^0$ together
in the two plots, and we differentiate between them by the colors
and symbols used.  The insets in \fgn{momentsfit} simply magnify
the range $\lambda < 2$.  We used the NLO Mellin OPE for the twist-2
contribution in \eqn{ratiotw2corr}, with and without the $H$ and
$L$ correction terms that we discussed above to 
fit the lattice data for ${\cal M}$.
Since the usage of non-zero $P_3^0$ is not common in the literature,
we note that at a given fixed $\lambda$, the $z_3$ dependence
comes from two sources even at leading-twist; namely, the $\ln(z_3)$
dependence due to perturbative evolution and
a polynomial dependence due to terms
such as $P_3^0 z_3$ in the denominator of the ratio. This is the
reason that at $P_3^0=0.508$ GeV, one finds a somewhat larger $z_3$
dependence than one would expect only from the perturbative logarithm.
To be clear, the presence of additional $P_3^0 z_3$ type polynomial dependence 
on $z_3$ is not a disadvantage as such terms are captured within a leading-twist 
framework without any modelling, and the consequent spoiling of near universality with respect to 
$\lambda = z_3 P_3$ is not a practical issue from the point of view of fits.
We truncated the OPE using $N_{\rm max}=4$ number of even-$n$ moments.
The dashed curves in \fgn{momentsfit}
are the central values of the best fit curves when $N_{\rm HT}=1$
and $L(z_3)$ terms are used as corrections, whereas the solid curves
are obtained without any correction terms. In the example fit shown,
we used a range $z_3\in[2a, 0.608 {\rm\ fm}]$. It is clear that the
two types of fits work quite well in describing the lattice data.
We found the fits with higher-twist correction terms to perform
marginally better in terms of $\chi^2/{\rm df}$, and this shows up in the tendency for the dashed
curves in \fgn{momentsfit} to pass closer to the central values of
the lattice data points. Taking the case with $N_{\rm HT}=1$ and
$P_3^0=0.25$ GeV
shown in the top panel as a sample case to discuss the 
typical values of the fit parameters that we obtained in our fits, we find
\beqa
&& \frac{h_0}{{\rm GeV}^2} = 0.0067(29),\quad
l_2 = 0.0008(11),\cr
&&\langle x^2\rangle = 0.2838(56),\quad
\langle x^4\rangle = 0.136(23),\cr
&&\langle x^6\rangle = 0.11(11),\quad
\langle x^8\rangle = 0.32(40),\cr
&&\chi^2/{\rm df} = 45.1/36
\eeqa{sampvals}
We see that our data sufficiently constrains only the lowest two
even Mellin moments. The lattice correction term $l_2$ does not
impact the fits, whereas the higher-twist correction term $h_0$
cannot be neglected. The value of $h_0=(81{\rm\ MeV})^2$ is in the
ball-park of the value of higher-twist correction we empirically
found in the quasi-PDF matrix element in Ref~\cite{Gao:2020ito}.
However, its value is small compared to the expectation for the
twist-4 0-th Gegenbauer moment based on QCD sum
rules~\cite{Ball:2006wn,Braun:2007wv}, namely, $\delta^2_\pi \approx
(300 {\rm\ MeV})^2$.

Apart from the one case shown in \fgn{momentsfit}, we also repeated
the fits over the following 72 combinations of analysis choices:
a) $N_{\rm HT}=0,1,2$, (b) with and without a lattice correction
term, $N_{\rm LC}=0,1$, (c) $z_3^{\rm min}=2a,3a$ to take short-distance
lattice artifacts into account, (d) $z_3^{\rm max}=0.456, 0.608,
0.76$ fm to take variations from higher-twist effects into account,
(e) $P_z^0=0.254, 0.508$ GeV for variation from reference momentum.
In \fgn{momsys}, we have shown the determination of $\langle x^2
\rangle$ and $\langle x^4\rangle$ from each of the choices $(N_{\rm
HT}, N_{\rm LC}, n_3^0, z_3^{\rm min}/a, z_3^{\rm max}/a)$ as a
data point (red circles).  We specify the analysis choice to the
side of each point. We found $\chi^2/{\rm df} < 1.6$ for all the
fits, and hence were acceptable.  To make the trend in the fit
parameters with respect to the added higher-twist terms visible,
we have grouped the points in \fgn{momsys} in three sets with $N_{\rm
HT}=0,1,2$ as indicated by the horizontal dashed lines. From the
systematic shift in the determined values of moments, there appears
to be a small but non-negligible effect of adding a $z_3^2 {\cal
F}^{(0)}_0(\lambda/2)$ term to the leading-twist OPE. The addition
of one more term $z_3^2 {\cal F}^{(0)}_1(\lambda/2)$ seems to only
make the fits noisier. Thus, given the precision of the data, usage
of $N_{\rm HT}=1$ seems to be sufficient. 
We show the scatter of 
the other fit parameters as well as the values of $\chi^2/{\rm df}$
in the various fits in \fgn{allsys} in \apx{fitvals}.

Using the analysis method we discussed earlier, we summarize the
content of the red points in \fgn{momentsfit} as the following
unweighted averages with the statistical and systematic errors:
\beqa
\langle x^2 \rangle &=& 0.2866(62)(56),\cr
\langle x^4 \rangle &=& 0.138(28)(34),
\eeqa{x2x4avg}
at a scale $\mu=2$ GeV. These summary estimates are shown in the
vertical bands of \fgn{momentsfit}; the inner band includes the
statistical error only, whereas the outer one includes both the
statistical and systematic error. It can be seen that the outer
band covers most of the scatter due to the various choices, and
hence is representative of our data.

If we assume that the pion DA is positive at all $x$ at $\mu=2$
GeV, then we can improve the stability of fits by imposing inequalities
on the Mellin moments that follows from $\phi(x,\mu)>0$,  and
subsequent derivatives of $\langle x^n\rangle$ with respect to $n$
at values infinitesimally closer to integer values. Namely, as we
explain in~\cite{Gao:2020ito}, we obtain the inequalities (a)
$\langle x^n\rangle > \langle x^{n+2}\rangle$ and (b) $\langle
x^{n+2}\rangle + \langle x^{n-2}\rangle > 2 \langle x^n\rangle$.
In the analysis, we imposed the two constraints through a change
of variables $\langle x^{2n}\rangle \equiv \sum_{i=n}^{N_{\rm max}}
\sum_{j=i}^{N_{\rm max}} e^{-\lambda_j}$. We have shown the results
of such constrained fits as the green circles in the two panels in
\fgn{momentsfit}. We find the resulting values of the two lowest
moments to be well determined, especially as the number of fit
parameters is increased when we set $N_{\rm HT}=2$.  In the case
of $\langle x^4\rangle$, such a procedure results in more precise
estimates compared to the unconstrained estimates shown as red
circles. We find from this constrained analysis that
\beqa
\langle x^2 \rangle &=& 0.2848(52)(71),\cr
\langle x^4 \rangle &=& 0.124(11)(20).
\eeqa{x2x4avgconstrain}
at $\mu=2$ GeV.

Using the model-independent estimates of the Mellin moments themselves,
we can reach a few conclusions.  The values of these two Mellin
moments in the large $Q^2$ limit of DA, $\phi(x) =4(1-x^2)/3$, are
$\langle x^2 \rangle = 0.2$ and $\langle x^4 \rangle = 0.0857$.
These differ quite significantly from the values we determined, and
hence we can conclude that the pion DA at the physical point and
at a scale of $\mu=2$ GeV differs from the asymptotic DA. In fact,
since $\langle x^2 \rangle > 0.2$, we can expect the $x$-dependent
DA to be flatter compared to the asymptotic DA. The DA cannot be a
completely flat DA~\cite{Radyushkin:2009zg}, $\phi(x)=1/2$ which
is characterized by $\langle x^2 \rangle = 1/3\approx 0.33$, and
$\langle x^4 \rangle=0.2$.  We can consider another extreme case
of a double humped Chernyak-Zhitnitsky (CZ) DA~\cite{Chernyak:1981zz},
$\phi(x) = 15 (1-x^2)x^2/4$, at $\mu=2$ GeV, with Mellin moments
as $\langle x^2 \rangle = 0.4285$ and $\langle x^4\rangle = 0.2381$.
These values are not compatible with the values we find.  Instead,
if we assume a simple one-parameter ansatz, $\phi(x) = {\cal
N}(1-x^2)^\alpha$, and solve for $\alpha$ using the value of $\langle
x^2 \rangle=0.2866$, we find the exponent should be around $\alpha
= 0.244$. In the next subsection, we perform more elaborate fits
to such Ans\"atze.

We performed a similar set of analyses using the C-OPE for the
leading-twist contribution in \eqn{ratiotw2corr}.  However, we were
not able to obtain stable fits with a more complex $N_{\rm HT}=2$
correction term to C-OPE without imposing any constraints on the
Gegenbauer moments, which resulted in a spurious negative-valued
$a_2$ at the cost of a large-valued higher-twist coefficient $h_1$.
Since it is likely due to overfitting of the data, we excluded this
analysis choice.  To summarize, using all other combinations of
analysis choices, we found $a_2 = 0.227(18)(23)$ and $a_4 =
-0.16(13)(30)$.  These values correspond to Mellin moments, $\langle
x^2 \rangle = 0.2779(63)(79)$ and $\langle x^4 \rangle = 0.121(15)(28)$,
using the linear relations between $a_n$ and $\langle x^n\rangle$
(e.g.,~\cite{san2012some}).  It is reassuring that two ways of
incorporating the leading-twist OPE result in similar values of the
first two Mellin moments.  It is also clear that when written using
C-OPE, the essential non-vanishing contribution mainly comes from
the $a_2$ Gegenbauer moment. The non-vanishing value of $\langle
x^4\rangle$ we find using the Mellin OPE, while being non-trivial
information from the perspective of M-OPE, becomes trivial when
expressed in terms of non-vanishing $a_2$ and a vanishing $a_4$
from the C-OPE perspective.

As a way of estimating perturbative uncertainty, we performed the
above set of analyses at scales of $\mu=4$ GeV and $\sqrt{2}$ GeV,
using $\alpha_s(4{\rm\ GeV})=0.227$ and $\alpha_s(\sqrt{2} {\rm\
GeV})=0.3607$. We then perturbatively ran the estimated Mellin
moments to the fixed scale $\mu=2$ GeV using the NLO implementation
that incorporates the mixing among the Mellin moments (as discussed
in Ref.~\cite{Agaev:2010aq}). Through this procedure, we found
$\left[\langle x^2 \rangle, \langle x^4\rangle\right]$ at $\mu=2$
GeV to be $\left[0.2822(85)(47), 0.126(30)(35)\right]$ and
$\left[0.2928(57)(76), 0.148(20)(34)\right]$ through evolution from
$\mu=4$ GeV and $\sqrt{2}$ GeV respectively. These values agree
with the estimates from the analysis performed exactly at $\mu=2$
GeV within the statistical and systematical errors.  Thus, we expect
the perturbative uncertainties to be less important compared to the
combined statistical and systematical errors.

Finally we compare our findings for the Mellin moments with some
recent lattice QCD calculations at $\mu=2$ GeV. The
work~\cite{Arthur:2010xf} using a dynamical QCD simulation and using
the local twist-2 operator approach obtain $\langle x^2 \rangle
=0.28(1)(2)$. Another series of works from RQCD that culminated in
Ref~\cite{RQCD:2019osh} using the local operator approach obtain
$\langle x^2 \rangle = 0.240(6)(2)(3)(2)$ at the physical point and
take into account various kinds of systematical errors. Whereas,
the usage of the leading-twist expansion method using current-current
correlators~\cite{Bali:2018spj} results in a scatter of values
around $\langle x^2 \rangle \approx 0.3$.  Using the quasi-DA matrix
element as used in this work, but using LaMET $x$-space matching,
the work~\cite{Zhang:2020gaj} estimates $\langle x^2 \rangle
=0.244(30)(20)$, and the most recent work~\cite{Hua:2022kcm} using
the hybrid-renormalization method~\cite{Ji:2020brr} estimates
$\langle x^2 \rangle =0.300(41)$. Our result lies in the ball park
value of previous estimates, but it is about 2.4-$\sigma$ (including
statistical and systematic errors in both works naively as the net
error) larger from the estimate using the local operator approach
in Ref~\cite{RQCD:2019osh}. In the future, we need to investigate
the remaining systematical uncertainties in our work that we did
not quantify, such as the effect of finite lattice spacing, and see
if the tension between the values of Mellin moments obtained with
two completely different methods, reduces or persists.

\subsection{Prior-sensitive reconstruction of the pion DA}\label{sec:xdep}

\bef
\centering
\includegraphics[scale=0.8]{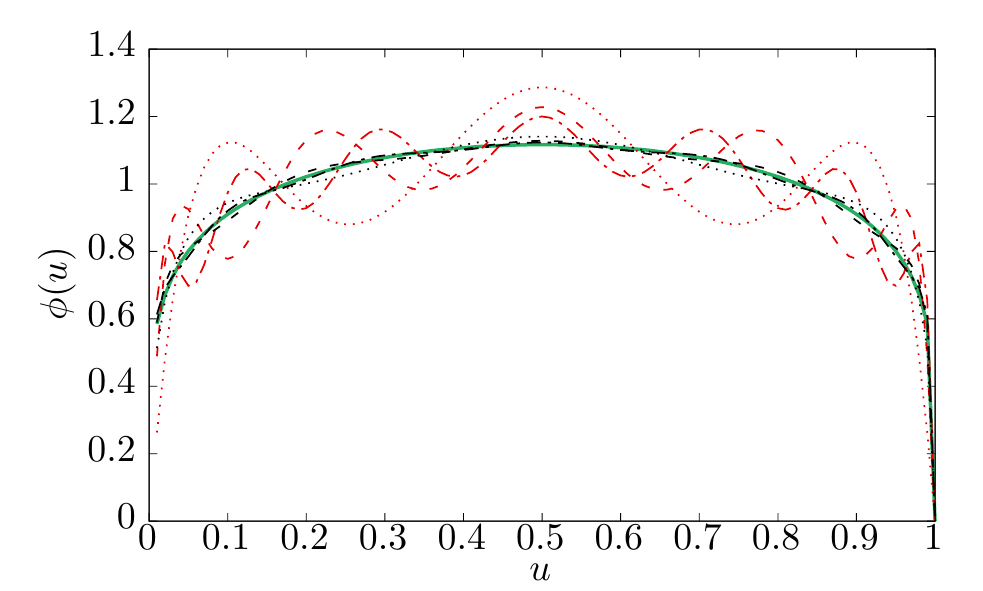}
\caption{The convergence of an example DA, $\phi(x)=1.47
x^{0.2}(1-x)^{0.2}$, shown as the green curve, when expanded in
$C_{2n}^{3/2}$, shown as the red curves, and in another basis
$C_{2n}^{0.9}$, shown as the black curves.  The truncation of the
expansions in $n$ upto $2, 4, 6$ are shown as dotted, dashed and
dot-dashed curves respectively.  The expansion in $C_{2n}^{1/2+\alpha}$
with $\alpha=0.4$ which is close to the actual exponent, 0.2,
converges much faster than with $C_{2n}^{3/2}$.
}
\eef{jacobieg}

Our determination of the lowest two Mellin moments in a model-independent
manner is the important result in this paper.  However, within the
framework of fitting phenomenology motivated Ans\"atze to the lattice
data, we can reconstruct the $x$-dependence of the DA, $\phi(x)$.
For convenience, we define the variable $u$ via
\beq
x = 2u-1,
\eeq{udef}
so that the DA has support from $0$ to $1$.
In principle, once we know all the Gegenbauer moments from 
fits to C-OPE, or inferred from M-OPE, we can perform a 
model-independent reconstruction using
\beq
\phi(u) = 6 u (1-u) \sum_{n=0} a_{2n} C_{2n}^{3/2}(1-2u).
\eeq{gegrecon}
The caveat that all the moments $a_{2n}$ need to be known makes
such an approach not usable in practice;  as we saw, the 
real-space quantity ${\cal M}(\lambda,z^2)$ converges in the 
accessible range of $\lambda < 6$ rapidly with respect to the 
number of Gegenbauer moments $a_n$ (as the main content of ${\cal M}$ can be 
summarized approximately with a value of $a_2$), whereas the corresponding
convergence in $u$ (or $x$)-space is rather slow. 
The problem is easy to understand by considering a behavior
$\phi(u)={\cal N}u^\alpha (1-u)^\alpha$. In the last section,
from the value of $\langle x^2\rangle$, we expected $\alpha\approx0.25$,
which differs significantly from the leading term with $\alpha=1$
in \eqn{gegrecon}. 

We can improve the convergence by using a complete basis that is
orthonormal with respect to a weight function, $w(u)=u^\alpha(1-u)^\alpha$,
rather than the weight function $w(u)=u(1-u)$ that the Gegenbauer
polynomials $C_n^{3/2}$ are orthonormal with respect to. Such an
idea was pursued in~\cite{Chang:2013pq} using the Gegenbauer
polynomial basis $C_n^{\alpha+1/2}(1-2u)$, which we follow in this
paper.  To impose the evenness of $\phi(u)$ around $u=1/2$, we
restrict the functions to even $n$.  That is, we expand,
\beq
\phi(u) = {\cal N} u^\alpha (1-u)^\alpha \sum_{n=0}^{N_G+1} s_n C_{2n}^{\frac{1}{2}+\alpha}(1-2u),
\eeq{jacobiexp}
with $s_0=1$.  The value of $\alpha$ describing the family of
complete functions is arbitrary, but a usage of $\alpha$ that is
close enough to the large/small-$x$ exponent leads to a better
convergence with respect to the truncation order $N_G$.  In
\fgn{jacobieg}, we show a specific example of the better convergence
of an example DA, $\phi(u)=1.47u^{0.2}(1-u)^{0.2}$, when expanded
in a {\sl nearby} $C_n^{0.9}$ polynomial basis as compared to an
expansion in $C_n^{3/2}$ polynomials.  We note that the polynomials
$C_n^{\alpha+1/2}(1-2u)$ are proportional to another complete basis,
the Jacobi polynomials, $P_n^{\alpha,\beta}(1-2u)$ for $\alpha=\beta$,
that have been proposed~\cite{Karpie:2021pap} as a good choice in
the analysis of PDFs even when $\alpha\ne \beta$.
\bef
\centering
\includegraphics[scale=0.8]{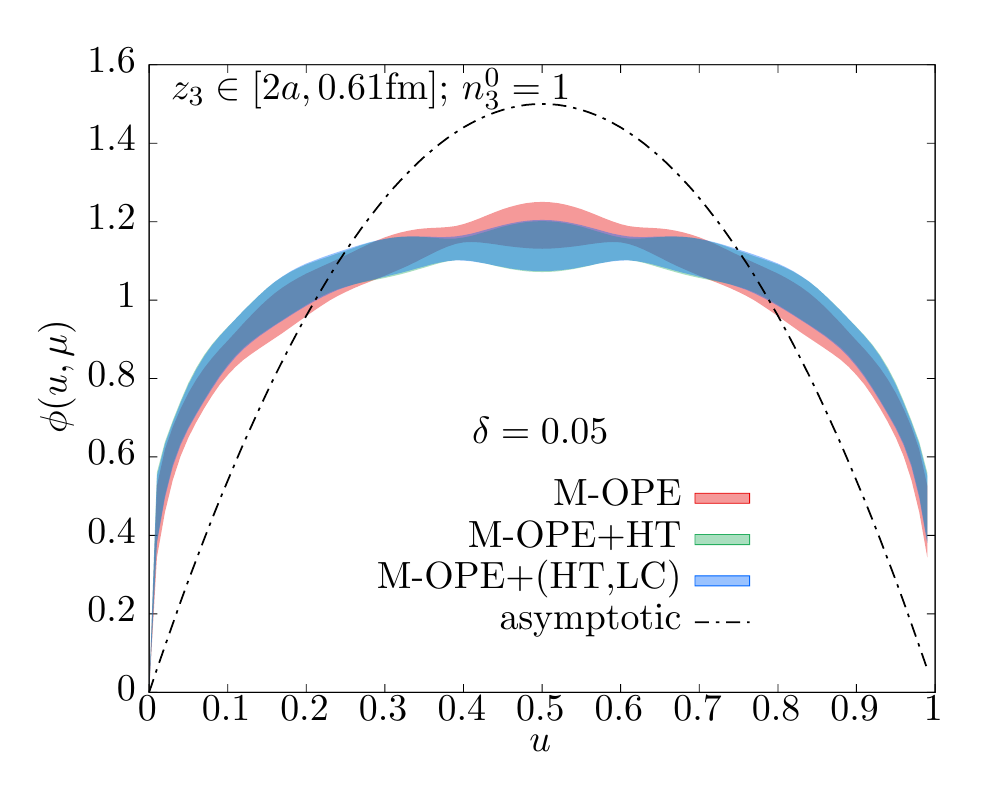}

\includegraphics[scale=0.8]{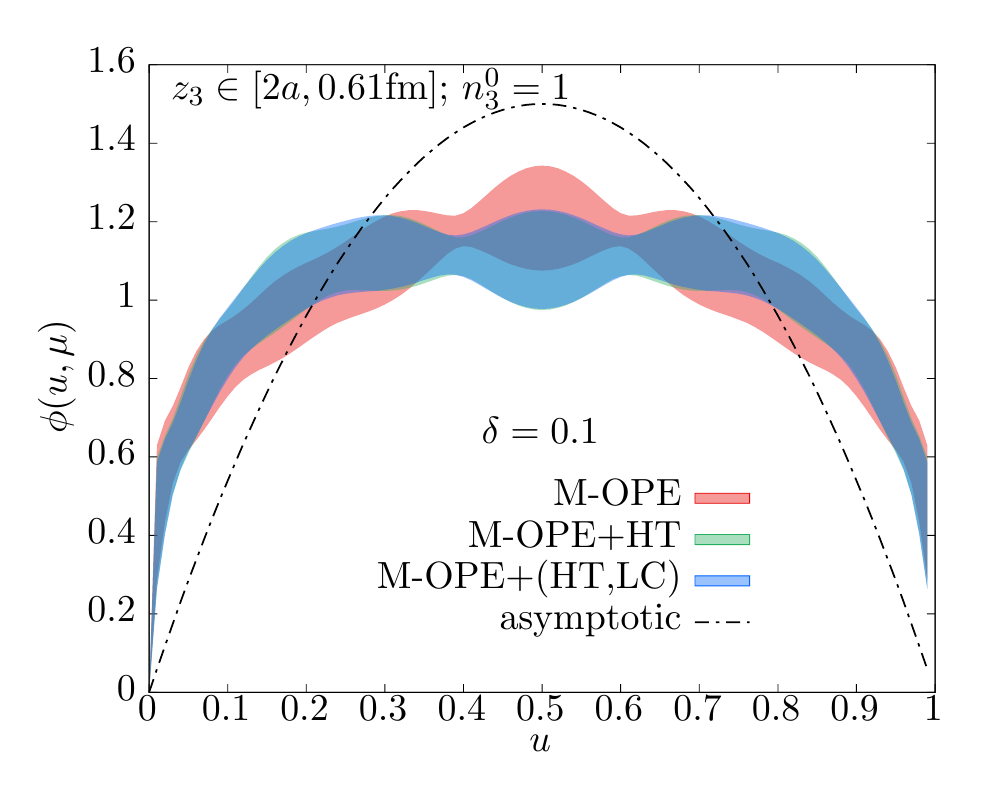}

\includegraphics[scale=0.8]{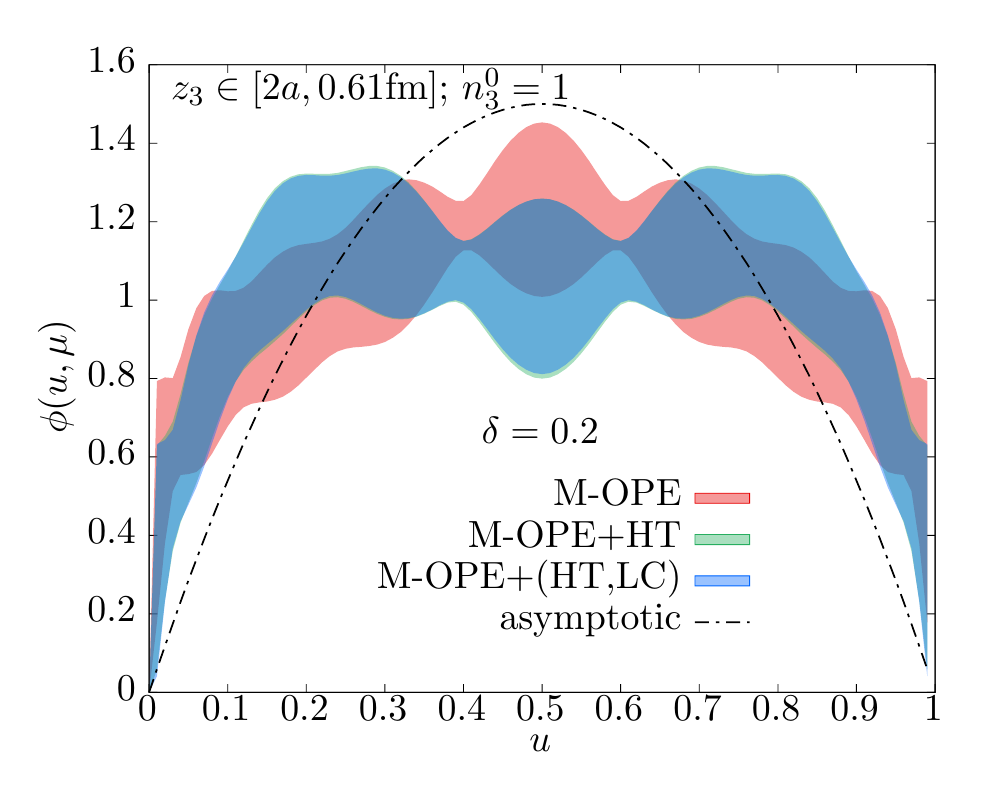}
\caption{The reconstructed $u=2x-1$ dependent pion distribution
amplitude $\phi(u,\mu)$ at $\mu=2$ GeV using the
$C_{2n}^{\alpha+1/2}$ basis from $n=1$ to $n=N_G=4$.  For all the cases shown,
the fit range is $z_3 \in [2a,0.61 {\rm fm}]$, and $P_3^0=0.254$ GeV.  The
top, middle and bottom panels are obtained using prior widths 
$\delta=0.05$, 0.1 and 0.2 on the coefficients of 
$C_n^{\alpha+1/2}$ respectively. The choices of $\alpha$
determining the Gegenbauer polynomial family were obtained 
from a one-parameter fit as explained in the text.}
\eef{dasys}

First, we determined the best fit values of the exponent $\alpha$
of the one-parameter Ansatz,
\beq
\phi_{\rm 1-param}(u)={\cal N}u^\alpha (1-u)^\alpha,
\eeq{1paramans}
that best describes the lattice data via \eqn{ratiotw2corr} for
each analysis choice that we described earlier. Essentially, the
parameter $\alpha$ enters the fits through the $\alpha$-dependent
Mellin moments. Therefore, unlike the model-independent analysis
of moments that we presented in the previous subsection, all the
moments are now related through a single unknown parameter $\alpha$.
We truncated the Mellin OPE at order $N_{\rm max}=6$ as before.
For different analysis choices, we found the values of the $\alpha$
to lie in a range between $0.2$ to $0.32$.

\befs
\centering
\includegraphics[scale=1.1]{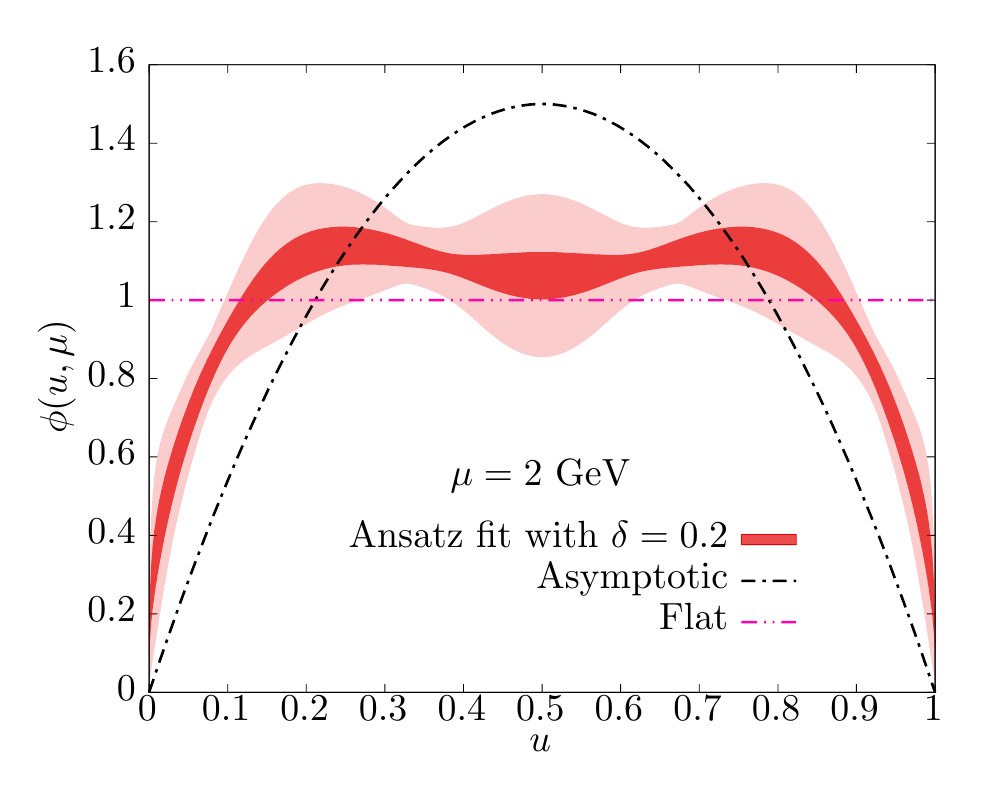}
\includegraphics[scale=1.1]{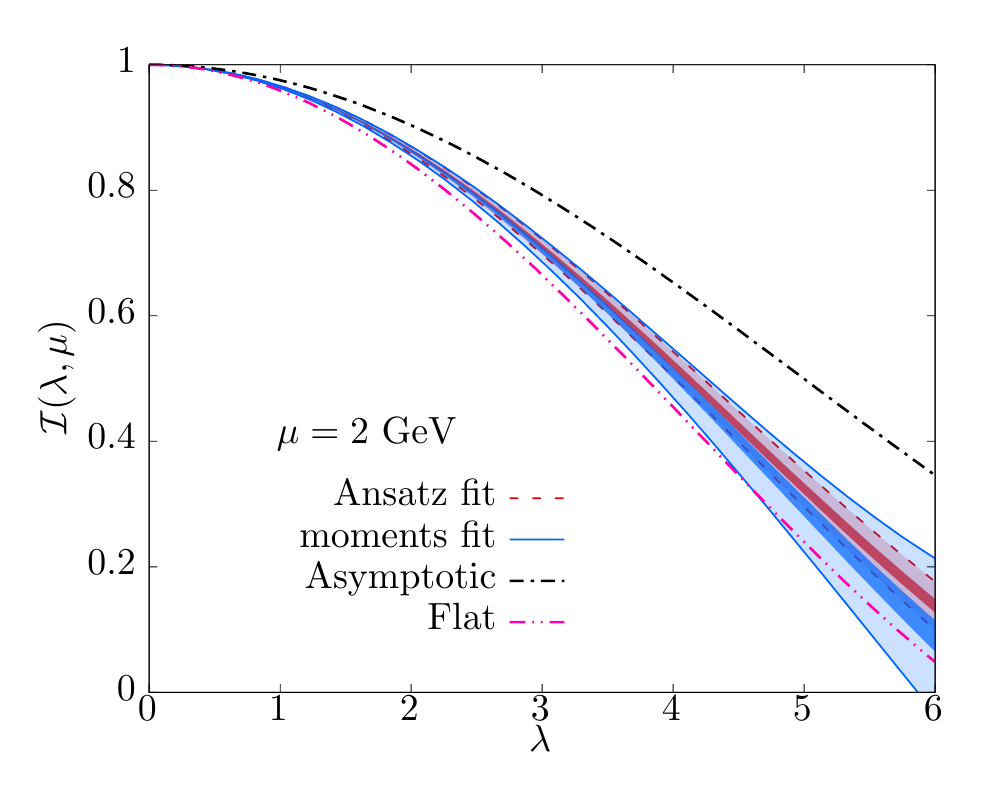}
\caption{ (Top panel) 
The pion DA reconstructed using the $C_n^{1/2+\alpha}$ basis with
the constraint $\delta=0.2$. The inner dark band is the statistical
error band. The outer light band is the combined statistical and
systematic error band. Variations in the fitted range of $z_3$,
reference momentum $P_z^0$, type of lattice correction and higher-twist
corrections added were taken into account in summarizing the result
in the figure. The asymptotic limit of DA is shown as the black
curve.  (Bottom panel) The plot shows the light-front $\msbar$ pion
ITD corresponding to the pion DA in the panel above, as the red
band. The ITD expected from the fits to Mellin moments is shown as
the blue band. In both cases, statistical and combined
statistical-systematical error bands are shown.  For comparison,
the ITDs corresponding to the asymptotic DA (black dot-dashed curve)
and flat DA (magenta dot-dashed curve) are also shown.
}
\eefs{dafinal}

In the next step, we generalized the parametrization for the DA by
an expansion in $C_n^{1/2+\alpha}$ as given in \eqn{jacobiexp}. We
followed the approach in Ref~\cite{HadStruc:2021qdf} to slowly
generalize from the one-parameter Ansatz above to more flexible
ones using a complete basis of functions to capture the corrections
to \eqn{1paramans}.  We used the best fit values of $\alpha$ from
the one-parameter fits from the previous step to choose the basis,
$C_n^{1/2+\alpha}$.  Even though the values of $\alpha$ did not
change much based on the analysis choices, we took care of using
the corresponding $\alpha$ values for a given analysis choice.  The
fit parameters are the coefficients $s_n$ in \eqn{jacobiexp}, which
enter via the Mellin moments or Gegenbauer moments (the implementation
of fits can be made computationally faster by pre-evaluating the
moments of Gegenbauer Polynomials $\int_0^1 (2u-1)^n u^\alpha
(1-u)^\alpha C^{1/2+\alpha}_n du$, from which the Mellin moments
are obtained as linear combinations.) We imposed the external
constraint on the allowed amount of fluctuations about $\phi_{\rm
1-param}$ through constraints on the expansion coefficients that
$|s_{n}| \lesssim \delta$. We realized this via Gaussian priors on
$s_n$ added to the $\chi^2$ using the central values and widths of
the priors of all $s_n$ being $0$ and $\delta$ respectively.  One
should however note that there is a priori no expected value for
$\delta$ simply from general considerations.  From practical
considerations, we will present the reconstructions by imposing
successively weaker constraints from $\delta=0.05$ to $\delta=0.2$.
For even larger values of $\delta$, we found the reconstruction to
be very noisy and oscillatory.

In the panels of \fgn{dasys}, we show the reconstructions of DA at
$\mu=2$ GeV using prior widths $\delta=0.05, 0.1$ and 0.2  from top
to bottom respectively.  For the cases we show, we performed the
fits over a range $z_3\in[2a,0.61{\rm\ fm}]$ and using a reference
momentum $P_3^0=0.254$ GeV. In each panel, we show the reconstructions
based on M-OPE without any correction terms, with $N_{\rm HT}=1$,
and with $(N_{\rm HT},N_{\rm LC})=(1,1)$. The changes due to such
variations are small, especially the effect of $N_{\rm LC}$ being
negligible.  From $\delta=0.05$ to 0.1, the effect of relaxing the
prior of $s_n$ is primarily to increase the statistical error on
the bands while closely agreeing with the one-parameter reconstruction.
The reconstructed DA starts becoming slightly oscillatory and with
larger error band when $\delta$ is relaxed to 0.2.  Unlike the case
of PDFs, which typically show a subdued prior and model dependence,
we found the reconstructed DA to be sensitive to the prior that is
applied. This is not surprising given that the essential content
in our quasi-DA matrix element in the range of $\lambda$ we used
is $a_2$, and the problem posed by such limited information in the
DA reconstruction is well known in the literature. However, the use
of the $C_n^{1/2+\alpha}$ basis was useful to quantitatively and
systematically reconstruct the pion DA that depends on the extent
to which one allows the DA to deviate from the default one-parameter
model.  Thus, the panels of \fgn{dasys} together convey this prior
dependent knowledge of DA from our quasi-DA matrix element.

We repeated the above fits for all analysis choices, which now
includes the truncation order $N_G=2,3$ and 4 in \eqn{jacobiexp}.
In the top panel of \fgn{dafinal}, we show our estimate of the pion
DA as a function of $u$, after taking into account all the analysis
variations, and summarize them with the statistical and systematic
error bands.  To be cautious, we present the reconstruction using
a relatively broad prior width $\delta=0.2$ on the expansion
coefficients. Nevertheless, the reconstruction in the case of DA
is sensitive to the value of $\delta$, however large it is, and
hence, one should interpret the reconstruction of DA in \fgn{dafinal}
as a specific $u$-dependence, given a somewhat broad prior.  We
compare our result with the asymptotic DA shown as the black dashed
curve. Within the precision allowed at $\delta=0.2$, we can only
resolve an overall flat DA over a range of $u\in[0.2,0.8]$ with
sharp fall offs, $u^{\alpha}$ and $(1-u)^\alpha$ with $\alpha\approx
0.3$, to 0 on either side. 
If one focuses only on the central value of the reconstructed
DA, one sees a tendency for a platykurtic DA as noted in
Refs~\cite{Stefanis:2014nla,Stefanis:2015qha,Stefanis:2020rnd}.
The lattice data does not have the
sensitivity to further resolve the concavity or convexity within
the flatter regions, unless one is willing to impose a more stringent
prior width $\delta$.  Apart from providing a reconstruction of the
DA, the ansatz based analysis also provides a way to estimate the
moments of DA.  The usage of ansatz can be thought of as a way to
regulate the values of moments at larger-$n$ for which the lattice
data is less constraining, and therefore, provides robust values
for smaller-$n$ moments.  From the Ansatz based analysis above with
$\delta=0.2$, we estimate the Mellin moments as
\beqa
\langle x^2 \rangle &=& 0.2845(44)(58),\cr
\langle x^4 \rangle &=& 0.1497(50)(38).
\eeqa{x2x4avgjac}
By comparing the values with \eqn{x2x4avg}, we see that the ansatz
based reconstruction for $\langle x^2 \rangle$ agrees quite well
with the completely model-independent reconstruction.  The estimates
of $\langle x^4 \rangle$ also agree with each other, however, the
usage of ansatz has substantially reduced the error.  Thus, from
both the model-independent moments analysis and the model-dependent
reconstruction analysis, we find the values of $\langle x^2 \rangle$
and $\langle x^4\rangle$ to be the quantities that we could reliably
extract from our lattice data.

As another way to summarize our results with less modeling artifacts,
we present the $\msbar$ light-front ITD corresponding to the pion
DA in the bottom panel of \fgn{dafinal} in the range of $\lambda$
that we have lattice data for and performed our analysis on. To
infer the $\msbar$ ITD, we used \eqn{msbaritd}. Since we need only
the information on the Mellin moments to construct the light-front
ITD, we show the resultant ITD based on the above Ansatz-based
analysis as the red band enclosed between the dot-dashed lines, and
the result based on the Mellin moments analysis in the previous
subsection as the blue band enclosed between solid lines. In both
cases, the darker inner bands are the statistical error bands whereas
the lighter outer bands include both statistical and systematical
errors. We see that both the model-independent and ansatz-dependent
reconstructions have similar behavior in the range of $\lambda$
that is constrained by the lattice data, with the latter being a
more precise determination. We show the light-front ITD corresponding
the asymptotic DA as the black dot-dashed curve. We also compare
our result with the expected ITD for a flat pion DA shown as the
magenta dot-dashed curve. Our result is clearly below the asymptotic
DA expectation, and closer to the expectation from a flatter DA.
This expectation is a less model-dependent manifestation of the DA
reconstruction seen in the top panel.

\subsection{Relation between form factors and DA at high momentum
transfer from perturbative factorization}\label{sec:DAtoFF}

The key quantities that characterize exclusive QCD processes, such
as such as the photon-pion transition form
factor~\cite{CELLO:1990klc,CLEO:1997fho,BaBar:2009rrj,Belle:2012wwz},
electromagnetic form factors and GPDs can be factorized into
convolutions of DAs and the perturbatively-calculable partonic
hard-scattering amplitudes if the momentum transfer is sufficiently
large.  For electromagnetic form factor this factorization was
introduced long time ago \cite{Efremov:1979qk,Farrar:1979aw,Lepage:1980fj}.
For photon-pion transition it was discussed in Refs. \cite{Melic:2002ij,
Gao:2021iqq}, and for gravitational form factors it was discussed
in Refs. ~\cite{Tong:2021ctu,Tong:2022zax}, while for GPDs it was
discussed in Refs. \cite{Hoodbhoy:2003uu,Sun:2021pyw}.  The energy
scale where this leading-twist DA-based factorization may work is
unknown at present.  This an important question that can  only be
answered by experiments or through lattice QCD computations. In
this subsection, we put aside this question and simply make predictions
for electromagnetic and gravitational form factors of the pion based
on leading twist factorization and our DA results. These predictions
can be compared to the lattice or experimental results at large
momentum transfer and clarify the range of applicability of the
leading twist factorization for the form-factors.

\bef
\centering
\includegraphics[scale=0.65]{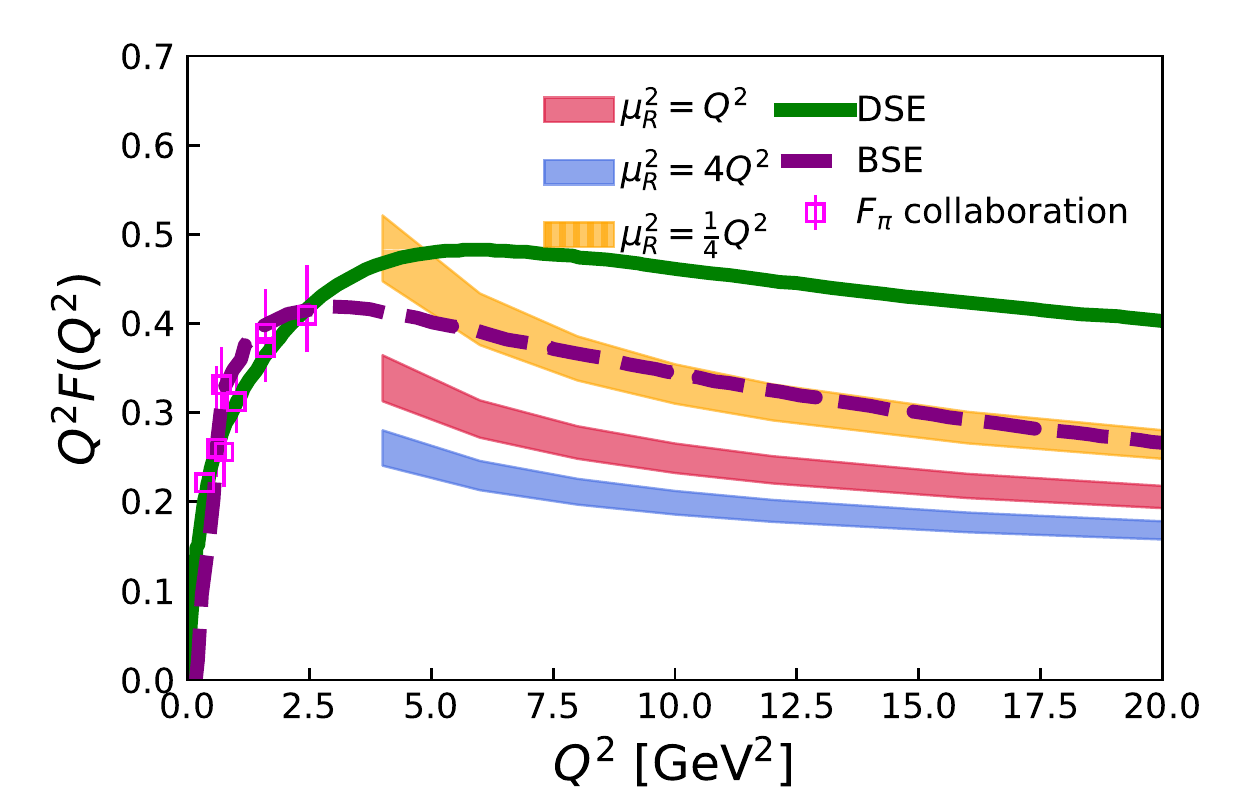}
\caption{The pion electromagnetic form factors reconstructed from
DA using LO matching and evolution are shown; the different bands
capture the variation from using factor 2 variation in scale $\mu_R$
used to determine $\alpha_s$. The experimental data from $F_\pi$
collaboration~\cite{JeffersonLab:2008jve} as well as the calculations
from Dyson-Schwinger equation (DSE)~\cite{Chang:2013nia} and
Minkowski-space Bethe-Salpeter equation (BSE)~\cite{Ydrefors:2021dwa}
are shown.}
\eef{EMFF}

At large $Q^2$, the pion electromagnetic form factors $F_\pi(Q^2)$ can be factorized as
\beqa
	 F_\pi(Q^2)&& =\int^1_0\int^1_0dxdy\ \Phi^*(v,\mu^2_F) \cr &&\qquad \times T_F(u,v,Q^2,\mu^2_R,\mu^2_F)\Phi(u,\mu^2_F)\,,
\eeqa{DAtoFF}
where $Q^2$ is the momentum transfer and $T_F$ is the hard-process
kernel. Though the form factor $F_\pi(Q^2)$ is scale independent,
the fixed-order perturbative factorization introduces dependence
on both renormalization and factorization scales $\mu^2_R$ and
$\mu^2_F$. Here $\Phi(u,\mu^2_F)$ is defined as,
\begin{equation}
	\Phi(u,\mu^2_F)=\frac{f_\pi}{2\sqrt{2N_c}}\phi(u,\mu^2_F),
\end{equation}
where $f_\pi$ is the pion decay constant discussed in \scn{BareM}.
At leading order (LO), the hard kernel reads~\cite{Melic:1998qr},
\begin{equation}\label{eq:THLO}
	T_F^{(0)}(u,v,Q^2)=\alpha_s(\mu^2_R)\frac{4}{3}\frac{16\pi}{Q^2\bar{u}\bar{v}},
\end{equation}
with $\bar{u}=1-u$ and the running coupling constant
\begin{equation}
	\alpha_s(\mu^2_R)=\frac{4\pi}{\beta_0\ln(\mu^2_R/\Lambda^2_{\rm QCD})}\,,
\end{equation}
where $\beta_0=11-\frac{2}{3}n_f$, and we use $n_f=3,\Lambda_{\rm
QCD}=0.2$ GeV in this paper. We take our model fit result at the
initial scale $\mu_0$ = 2 GeV, and evolved it to $\mu_F$ by first
expanding $\phi(u,\mu_0)$ in the Gegenbauer basis in \eqn{gegrecon}
up to a sufficiently large order $n=n_{\rm max}=20$, and then
evolving those Gegenbauer moments from $a_n(\mu_0)$ to $a_n(\mu_F)$
using
\begin{align}\label{gegren_evo}
    a_{n}(\mu_F)&=\left(\frac{\alpha_s(\mu^2_F)}{\alpha_s(\mu^2_0)}\right)^{\gamma_n^{(0)}/\beta_0}a_{n}(\mu_0)\,.
\end{align}
We choose $\mu_R=\mu_F=Q$ as the central value of the scale setup,
and vary the renormalization scale $\mu_R$ by a factor of 2 to
estimate the perturbation uncertainty.  For the ease of implementation,
we used the 1-parameter Ans\"atze $\phi_{\rm 1-param}(x,\mu_R)$
from our analysis using $(N_{\rm HT}, N_{\rm LC}, n_3^0, z_3^{\rm
min}, z_3^{\rm max})=(1,1,1,2a,8a)$.

The results are shown in \fgn{EMFF} with statistical error bands
and compared with the experimental data from the $F_\pi$
collaboration~\cite{JeffersonLab:2008jve} as well as the calculations
from the Dyson-Schwinger equation (DSE)~\cite{Chang:2013nia} and
Minkowski-space Bethe-Salpeter equation (BSE)~\cite{Ydrefors:2021dwa}.
As one can see, our prediction using the LO kernel is systematically
lower than the DSE and BSE calculations. It was found that the
matched form factors could increase with NLO
corrections~\cite{Melic:1998qr}, and the higher-twist corrections
may also make a significant contribution~\cite{Raha:2010kz}. However,
all these arguments can only be tested by the future experimental
results with large momentum transfer $Q^2$ up to 6 $\rm GeV^2$ of
the JLAB E12-09-001 experiment \cite{Dudek:2012vr} and up to 40
$\rm GeV^2$ of the new Electron-Ion Collider (EIC)
facility~\cite{Arrington:2021biu}.

\befs
\centering
\includegraphics[scale=0.6]{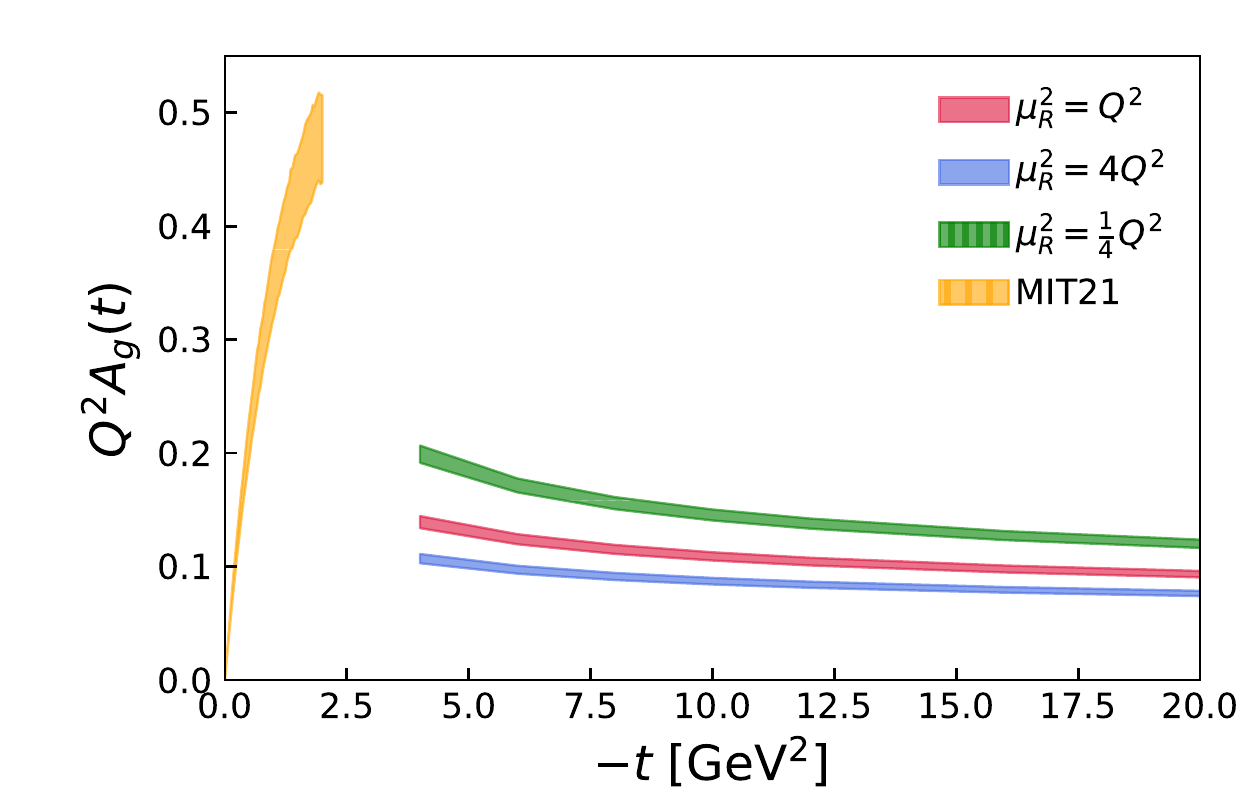}
\includegraphics[scale=0.6]{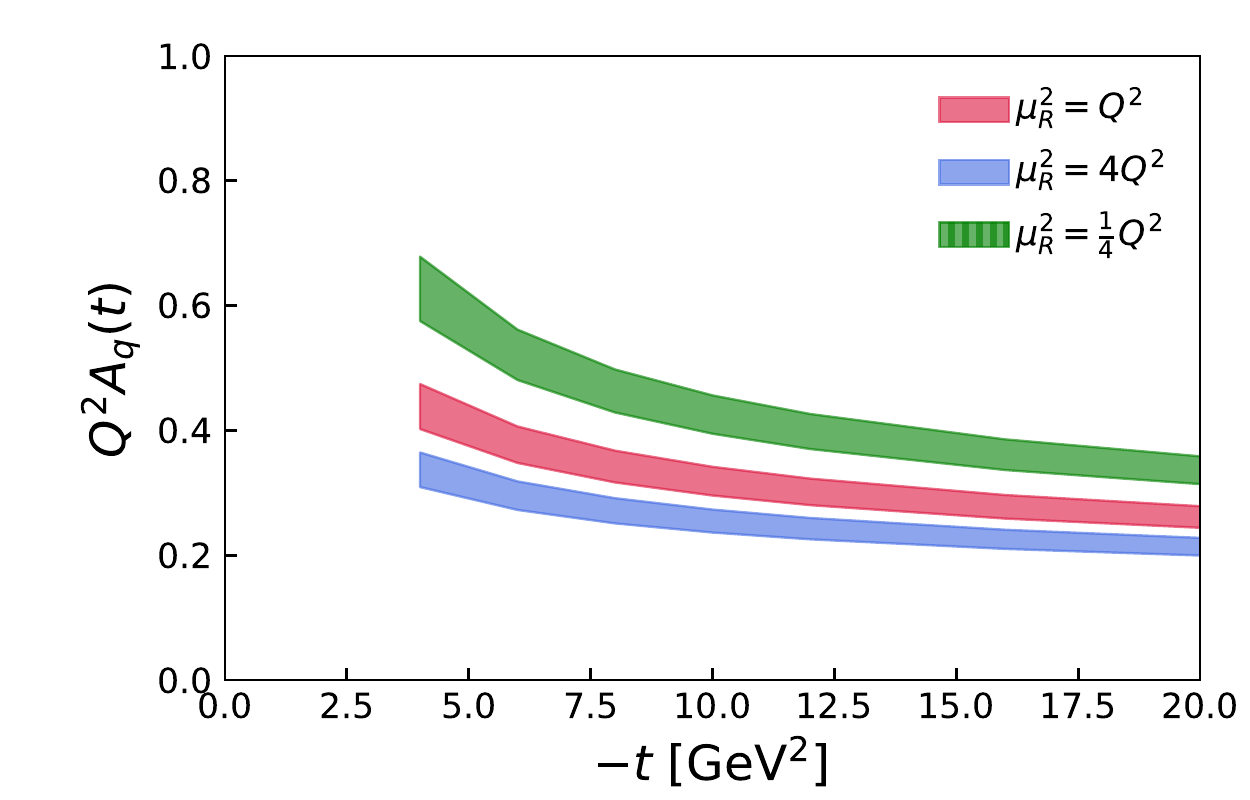}
\includegraphics[scale=0.6]{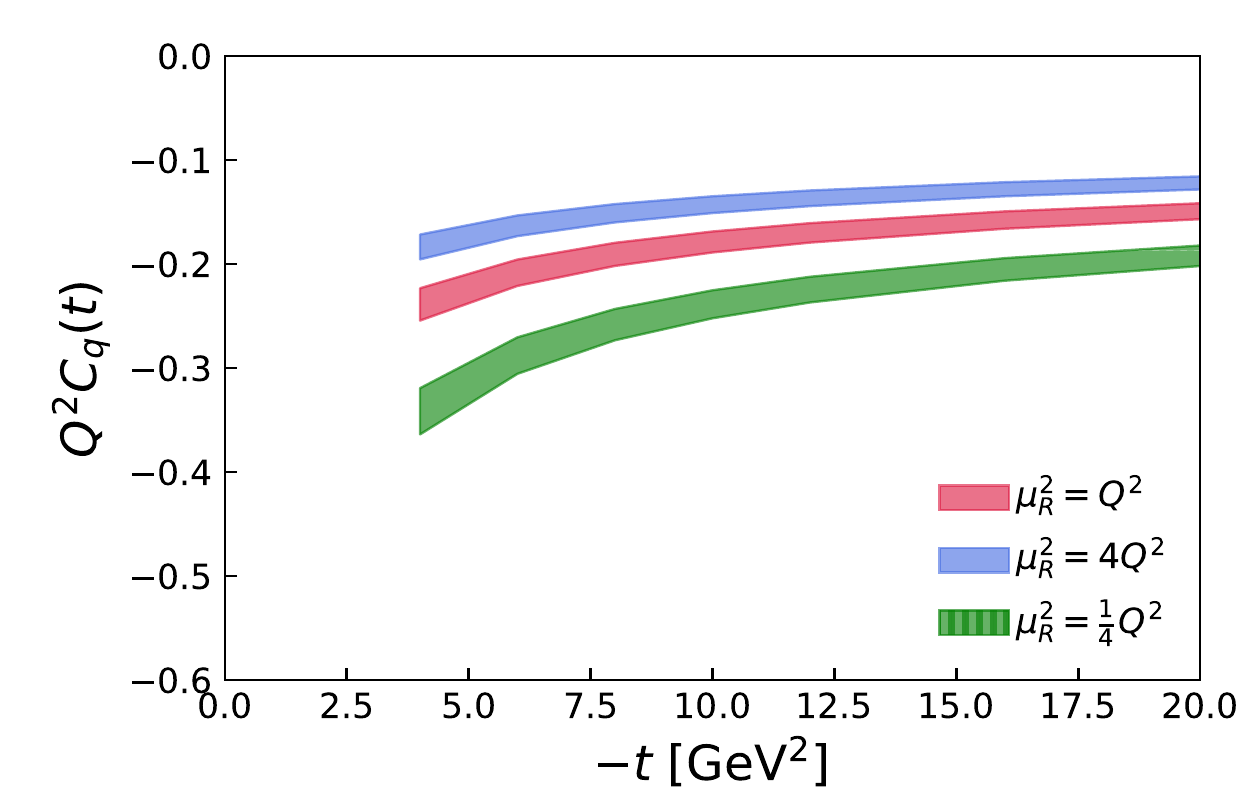}
\includegraphics[scale=0.6]{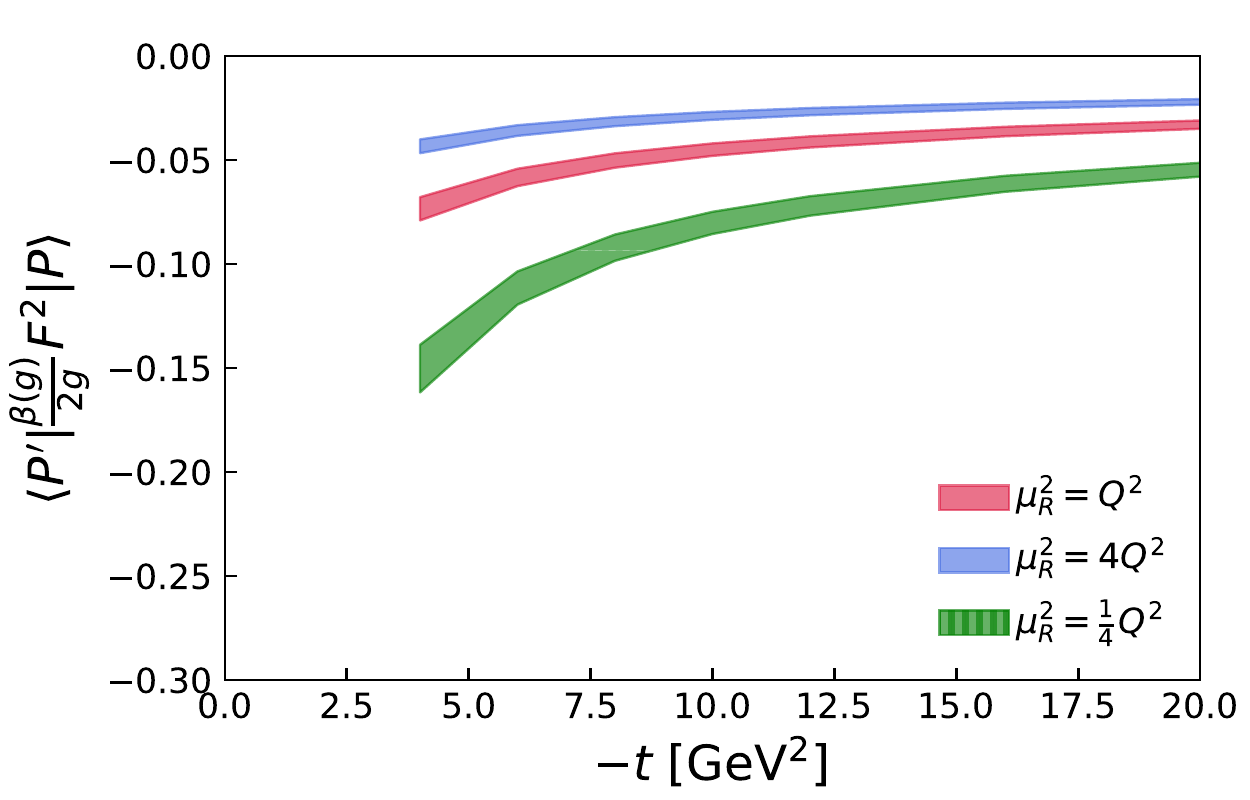}
\caption{The gluon GFFs $Q^2A_g(t)$, quark GFFs $Q^2A_q(t)$ and
$Q^2C_q(t)$ as well as the pion trace anomaly $\langle
P'|\frac{\beta(g)}{2g}F^2|P\rangle$ at large $-t$ predicted by our
determination of pion DA are shown. The bands come from the statistic
errors and we vary the renormalization scale $\mu_R$ by a factor
of 2 to estimate the perturbation uncertainty. The direct lattice
calculation of pion gluon GFF $A_g(-t)$~\cite{Pefkou:2021fni} (the
multipole fit result) using unphysical pion mass $m_\pi=$ 450 MeV
is shown (MIT21, the yellow band in the top-left panel) for
comparison.}
\eefs{GFFs}
The gravitational form factors (GFFs) of the pion are the transition
matrix elements of the QCD energy momentum tensor,
\begin{equation}
	T^{\mu\nu}_{\rm QCD}=T^{\mu\nu}_{q}+T^{\mu\nu}_{g} .
\end{equation}
Though $T^{\mu\nu}_{\rm QCD}$ is conserved, and is therefore UV
finite and scale independent, the quark and gluon contributions,
$T^{\mu\nu}_{q}$ and $T^{\mu\nu}_{g}$, are not and depend on the
renormalization scale $\mu_R$. This dependence is governed by the
corresponding anomalous dimension.  The gluon gravitational form
factors for the pion can be parametrized as,
\begin{align}\label{eq:Tg}
    &\langle P'|T^{\mu\nu}_g(\mu_R)|P\rangle=2\bar{P}^\mu \bar{P}^\nu A^\pi_g(t,\mu_R)
	\\
	&\qquad +\frac{1}{2}(\Delta^\mu\Delta^\nu-g^{\mu\nu}\Delta^2)C^\pi_g(t,\mu_R)
	+2m^2g^{\mu\nu}\overline{C}^\pi_g(t,\mu_R)\,,\nonumber
\end{align}
where $\bar{P}=(P'+P)/2$ is the average momentum, $\Delta=P'-P$ is
the momentum transfer, and $-t=-\Delta^2=Q^2$.  Similar to the case
of the electromagnetic form factor, the leading-twist GFFs perturbative
factorization reads,
\begin{align}\label{eq:facFF}
    A^\pi_g(t,\mu_R)&=\int dudv\ \Phi^*(v,\mu_F)\nonumber\\
    &\qquad\times \mathcal{A}^\pi_g(u,v,t,\mu_R,\mu_F)\Phi(u,\mu_F)\,,
\end{align}
where $\mu_F$-dependence can be introduced in $A^\pi_g$ due to the
use of a fixed-order hard kernel.  At leading order, the kernels
are~\cite{Tong:2021ctu},
\begin{align}
    \mathcal{A}^\pi_g(u,v,t,\mu_R,\mu_F)&=\mathcal{C}^\pi_g(u,v,t,\mu_R,\mu_F)\nonumber\\
    &=\frac{8\pi\alpha_s(\mu_R)C_F}{-t}(\frac{1}{u\bar{u}}+\frac{1}{v\bar{v}})\,.
\end{align}
Due to the traceless feature of \eqn{eq:Tg} one also has the relation
$\overline{\mathcal{C}}^\pi_g=-\mathcal{C}^\pi_{u+d}=\frac{t}{4m^2}\mathcal{C}^\pi_g$.
In terms of the same factorization formula, the hard coefficients
$\mathcal{A}_q$ and $\mathcal{C}_q$ in the quark sector are,
\begin{align}
    \mathcal{A}^\pi_q(u,v,t,\mu_R,\mu_F)&=\frac{8\pi\alpha_s(\mu_R)C_F}{-t}\frac{u+v+1}{\bar{u}\bar{v}},\\
    \mathcal{C}^\pi_q(u,v,t,\mu_R,\mu_F)&=\frac{8\pi\alpha_s(\mu_R)C_F}{-t}\frac{u+v-3}{\bar{u}\bar{v}}.
\end{align}
In addition, the gluon scalar FF defined as,
\begin{equation}
	\langle P'|F^{a,\mu\nu}F^a_{\mu\nu}|P\rangle=m^2_\pi G_\pi(t,\mu_R),
\end{equation}
is closely related to the trace anomaly~\cite{Ji:1995sv},
\begin{equation}
	T_\mu^\mu=\frac{\beta(g_s)}{2g_s}F^{a,\mu\nu}F^a_{\mu\nu}.
\end{equation}
Using the same factorization convention as \eqn{eq:facFF}, the
leading-order hard kernel reads~\cite{Tong:2022zax},
\begin{equation}
	\mathcal{G}^\pi_q(u,v,t,\mu_R,\mu_F)=\frac{16\pi\alpha_s(\mu_R)C_F}{m_\pi^2}(\frac{1}{u\bar{v}}+\frac{1}{\bar{u}v}).
\end{equation}
Comparing to the tensor GFFs $A_{q,g}$ and $C_{q,g}$ shown above,
one can observe that $\mathcal{G}^\pi_q$ does not have a $1/(-t)$
pre-factor and therefore will become flat for large $-t$. In
\fgn{GFFs}, we show the perturbatively determined GFFs and pion
trace anomaly at large $-t$ as expected from our determination of
pion DA. The bands come from the statistical errors and we vary the
renormalization scale $\mu_R$ by a factor of 2 to estimate the
perturbation uncertainty. The direct lattice calculation of the
pion gluon GFF $A_g(-t)$~\cite{Pefkou:2021fni} (the multipole fit
result) using unphysical pion mass $m_\pi=$ 450 MeV is shown (MIT21,
the yellow band in the top-left panel) for comparison.  And it can
be seen our estimate of the perturbative contribution is much
smaller, which could come from sizable higher-order perturbative
or higher-twist contribution. Direct lattice calculations or
experimental results at large $-t$ are needed to clarify the issue.

In order to perform the above perturbative convolutions, we relied
on the  Ansatz-based reconstruction to a large extent.  Instead,
one could ask if there is a way to perform an alternate less model
dependent analysis based only on the ITD in the range of $\lambda$
spanned by the lattice data, or equivalently, only using the moments
that the lattice data is sensitive to.  To address this question,
we can refer to the LO perturbative convolution in \eqn{DAtoFF} for
electromagnetic form-factor that makes use of the integral,
\beq
\langle (1-u)^{-1}\rangle = \langle u^{-1} \rangle = \int_0^1 du \frac{\phi(u,\mu)}{u}.
\eeq{invu}
Using \eqn{gegrecon}, one can see that, $\langle u^{-1}
\rangle=3\sum_{n=0}^\infty a_{2n}(\mu)$.  If one truncates the sum
over Gegenbauer moments up to $n=1$ at $\mu=2$ GeV, then one finds
$\langle u^{-1} \rangle=3.72(45)$, whereas when one sums over 20
Gegenbauer moments using the 1-parameter Ans\"atze, we get $\langle
u^{-1} \rangle=4.87(19)$.  As an alternative, we can write \eqn{invu}
using only the ITD as,
\beq
\langle u^{-1} \rangle = \lim_{\lambda_{\rm max}\to\infty} \int_0^{\rm \lambda_{\rm max}} {\cal I}(\lambda,\mu)\sin\left(\frac{\lambda}{2}\right) d\lambda.
\eeq{invufromitd}
With $\lambda_{\rm max}\approx 6$ as in this work (see bottom panel
of \fgn{dafinal}), we find this region of ${\cal I}(\lambda,\mu)$
to contribute 2.64(2) to $\langle u^{-1}\rangle$, which is only
about 50\% to the Ansatz-based expectation for $\langle u^{-1}\rangle$,
and the rest comes from $\lambda > \lambda_{\rm max}$.  Therefore,
for the two model-independent methods to be reliable, one has to
truncate at much higher Gegenbauer moments or larger $\lambda$,
which poses a challenge for lattice calculations. As a result, one
has to rely on the Ans\"atze for the $x$-dependence of DA, but the
systematic uncertainty from truncation is transformed to the model
dependence of the Ans\"atze. It would be important in the future
to compare and cross-check our current results for the form factors
here with the expectations based on $x$-space LaMET DA matching on
the same ensemble.

\section{Conclusions}\label{sec:concl}

We presented a lattice QCD study of the quasi-DA matrix element in
real-space using the leading-twist OPE method for the first time.
We performed our study at the physical point using clover-improved Wilson
valence quark propagators determined on a physical HISQ ensemble.
The quantities central to this work are the renormalization group
invariant ratios of quasi-DA matrix elements, with non-zero momenta
in both the numerator and denominator; the non-zero momenta were
inevitable as well as helped us remain closer to the leading-twist
approximation.  In the first part of the paper, we adapted the
results from Refs.~\cite{Radyushkin:2019owq,Braun:2007wv} and
presented the analytical perturbative results for the leading-twist
expansion in the forms of the Mellin OPE at next-to-leading order
and the conformal OPE at leading-log order. These expressions formed
the basis for our determination of the pion DA from quasi-DA matrix
elements.

From the leading-twist description of $z\cdot P$ and $z^2$ dependencies
of the ratios of quasi-DA matrix elements, we extracted the Mellin
moments and captured the $x$-dependence of the pion DA based on
fits to various Ans\"atze.  We first checked the validity of
leading-twist dominance in our matrix element using a fixed-$z^2$
analysis.  Then, from a model-independent determination via fits
to the few lowest Mellin moments at a factorization scale $\mu=2$
GeV, we were able to obtain a relatively precise determination of
the second Mellin moment, $\langle x^2 \rangle = 0.287(6)(6)$, and
for the fourth Mellin moment, we obtained  $\langle x^4 \rangle =
0.14(3)(3)$; the first parenthesis gives the statistical error and
the second one specifies the systematical error coming from variations
in various analysis choices, such as fit ranges for $z^2$. Based
on NLO perturbative evolution of our corresponding results at
different $\mu$ evolved to $\mu=2$ GeV, we estimated our perturbative
uncertainty to be within our combined statistical and systematical
errors. We reached a similar conclusion from the differences in the
estimates of the Mellin moments from the Mellin-OPE and Conformal-OPE.
We found the Ansatz based reconstruction of the $x$-dependence of
the pion DA to be sensitive to the model used; by using a complete
set of functions to expand the DA and by imposing constraints on
their expansion coefficients, we systematically reconstructed the
pion DA. Using a weak constraint, we found the DA at $\mu=2$ GeV
to be flatter than the asymptotic DA in the region around $x=0$ (or
equivalently, $u=0.5$).  From the Ans\"atze-based reconstruction
of the pion DA, we found the expected the large $Q^2$ dependence
of electromagnetic and gravitational form factors using the
leading-twist LO convolutions.  It would be interesting in the
future to compare the values of the pion form-factors at these large
$Q^2$ with the perturbative expectations.

The systematical error in this work stemmed only from the choices
of fit ranges, type of higher-twist corrections and other such
analysis choices.  Another source of systematic error could be due
to finite lattice spacing corrections.  Since, we used an ensemble
at a fixed lattice spacing, $a=0.076$ fm, we were unable to quantify
the effect in this work, and we need to revisit this in a future
work. We found the perturbative uncertainties to be about $~3\%$
as estimated through differences in results for $\langle x\rangle$
from Mellin- and Conformal-OPE, and through the effect of evolution
to 2 GeV starting from different initial scales used for fits.
However, in the present work, we were not able to directly address
this issue using NNLO DA matching as is the state-of-art for the
current lattice PDF calculations.  In the immediate future, we plan
to extend the current work using the leading-twist expansion of the
quasi-DA to study the Kaon DA and quantify the effects of explicit
SU(3) flavor symmetry breaking.

\begin{acknowledgments}
We thank V.~Braun and N.~G.~Stefanis for their comments.
This material is based upon work supported by: (i) The U.S. Department
of Energy, Office of Science, Office of Nuclear Physics through
Contract No.~DE-SC0012704; (ii) The U.S. Department of Energy,
Office of Science, Office of Nuclear Physics, from DE-AC02-06CH11357;
(iii) Jefferson Science Associates, LLC under U.S. DOE Contract
No.~DE-AC05-06OR23177 and in part by U.S. DOE grant No.~DE-FG02-04ER41302;
(iv) The U.S. Department of Energy, Office of Science, Office of
Nuclear Physics and Office of Advanced Scientific Computing Research
within the framework of Scientific Discovery through Advance Computing
(SciDAC) award Computing the Properties of Matter with Leadership
Computing Resources; (v) The U.S. Department of Energy, Office of
Science, Office of Nuclear Physics, within the framework of the TMD
Topical Collaboration. (vi) YZ is partially supported by an LDRD
initiative at Argonne National Laboratory under Project~No.~2020-0020.
(vii) SS is supported by the National Science Foundation under
CAREER Award PHY-1847893 and by the RHIC Physics Fellow Program of
the RIKEN BNL Research Center. (vii) This research used awards of
computer time provided by the INCITE and ALCC programs at Oak Ridge
Leadership Computing Facility, a DOE Office of Science User Facility
operated under Contract No. DE-AC05-00OR22725. (viii) Computations
for this work were carried out in part on facilities of the USQCD
Collaboration, which are funded by the Office of Science of the
U.S. Department of Energy.

\end{acknowledgments}

\appendix

\section{Renormalization constants in RI-MOM scheme for $a=0.076$ fm ensemble}
\begin{figure}
\includegraphics[width=7.5cm]{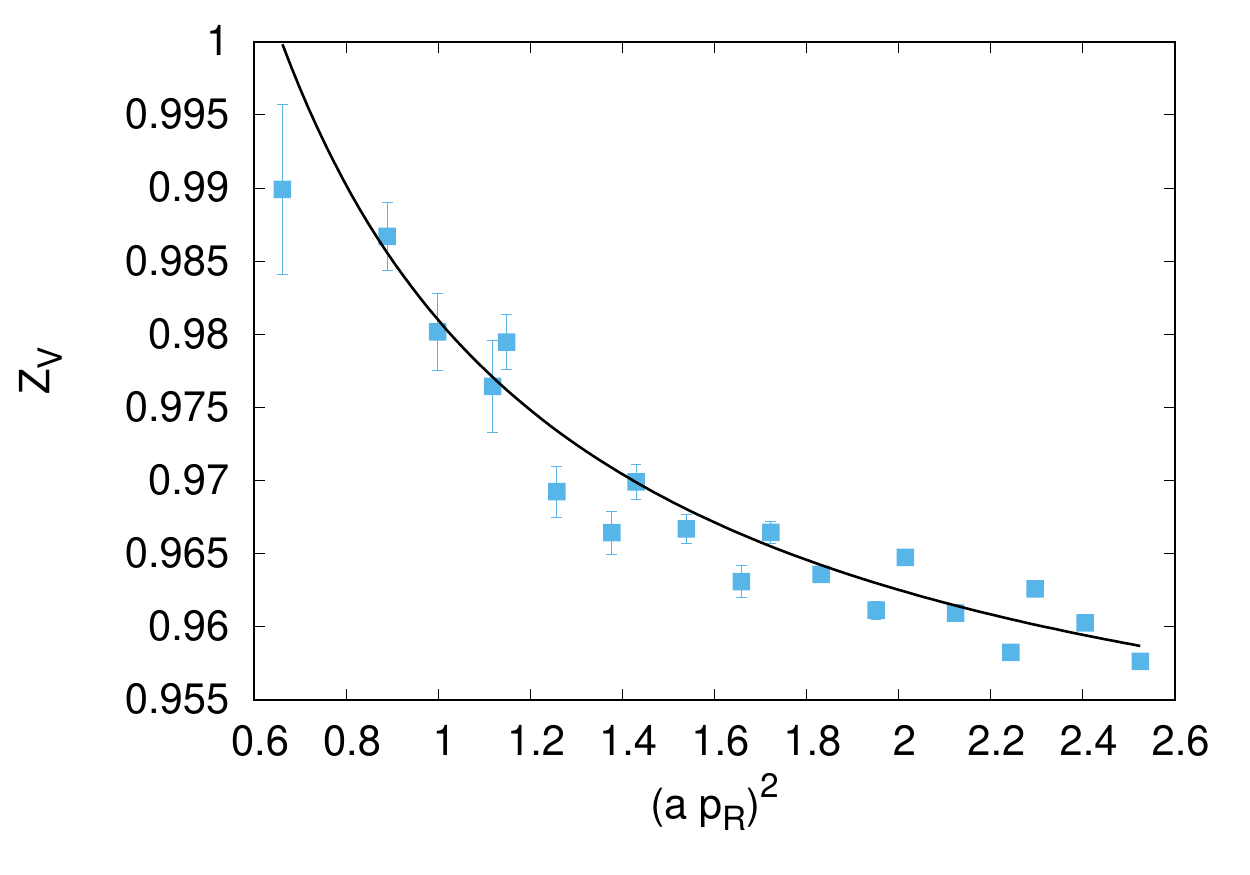}
\includegraphics[width=7.5cm]{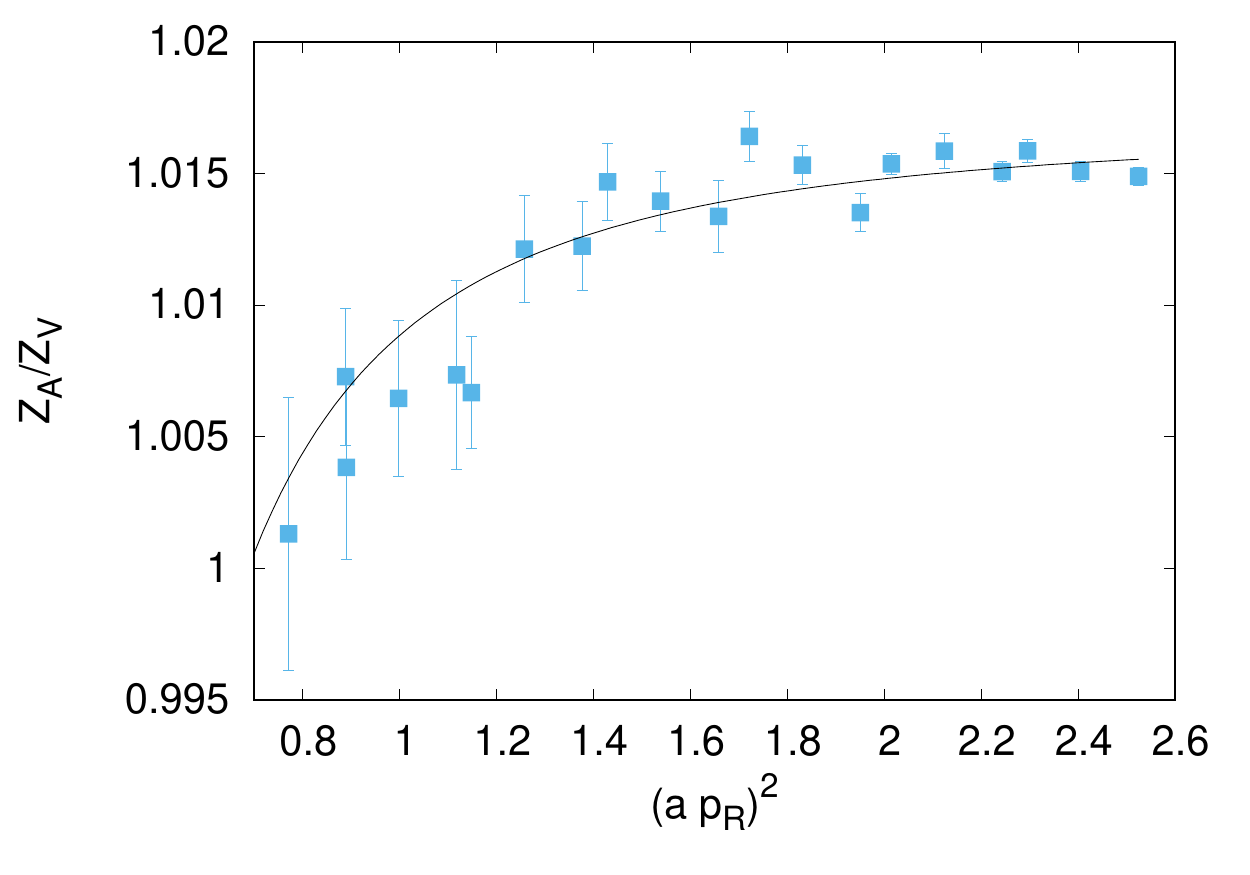}
\caption{
The vector current renormalization factor $Z_V$  (top) and the ratio
$Z_A/Z_V$ (bottom) as function of RI-MOM momentum $p_R$ in lattice
units.
}
\label{fig:ZV}
\end{figure}
In this appendix, we discuss the calculation of the renormalization
of the vector current $Z_V$ and axial-vector current $Z_A$ in RI-MOM
scheme for our setup with $a=0.076$ fm. We use off-shell quark
states in the Landau gauge with different values of lattice momenta
\begin{equation}
a p_{\mu}=\frac{2 \pi}{L_{\mu}} (n_{\mu} + \frac{1}{2} \delta_{\mu,0}).
\end{equation}
To minimize the discretization errors the lattice momenta are
substituted by $a p_{\mu}'=\sin (a p_{\mu})$, so the renormalization
point is $(a p_R)^2=\sum_{\mu=1,4} (a p'_{\mu})^2$.  In \fgn{ZV},
we show our results for $Z_V(p_R)$.  The vector current renormalization
constant should not depend on $p_R$, because in the $a \rightarrow
0$ limit the local current is conserved. Nevertheless, we see a
significant dependence on $p_R$. This dependence is caused by
non-perturbative effects, that for large values of $p_R$ can be
parameterized by local condensates. As we use off-shell quark states
in the Landau gauge in the RI-MOM renormalization procedure, the
lowest dimension local condensate is the dimension-two gluon
condensate $\langle A^2 \rangle$ \cite{Chetyrkin_2010,Blossier:2010ky}.
Lattice artifact shows up as the breaking of the rotational symmetry
on the lattice. We see from \fgn{ZV} that the fish-bone structure
in the lattice data at the level much larger than the statistical
errors on $Z_V$ for large values of $a p_R$. Therefore, to obtain
$Z_V$ we fit our lattice data with the following form:
\beqa
&& Z_V(p_R)=Z_V+B/(ap_R)^2+ C\cdot (ap_R)^k \bigg{(}1+C_4 \Delta^{(4)}
\cr && \quad + C_6 \Delta^{(6)} +C_8 \Delta^{(8)} \bigg{)},
\eeqa{fit}
where 
\begin{equation}
\Delta^{(4)}=\frac{\sum_{\mu}( p_{\mu}' )^4}{p_R^4},~\Delta^{(6)}=\frac{\sum_{\mu}( p_{\mu}' )^6}{p_R^6},
~\Delta^{(6)}=\frac{\sum_{\mu}( p_{\mu}' )^8}{p_R^8}.
\end{equation}
This form is motivated by the 1-loop lattice perturbation theory
\cite{Constantinou:2009tr,Constantinou:2010gr} and the perturbative
analysis with dimension two gluon condensate \cite{Lytle:2018evc}.
For the non-perturbative clover action $k=2$, while for Wilson
action $k=1$. For HYP smeared clover action with tadpole improved
value of $c_{sw}$ we expect ${\cal O}(a)$ discretization errors to
be proportional to $\alpha_s^2$ with a very small coefficient, so
it is reasonable to assume that the dominant cutoff effects scale
like $a^2$.  Nevertheless we also perform fits using $k=1$.  To
limit the size of the lattice artifacts we impose the additional
constraint:  $\Delta^{(4)}<0.4$. We performed different fits varying
the fit interval in $p_R$ as well as  setting some coefficients to
zero in certain cases. Fits with $C=0$ typically have very large
$\chi^2$ but this has almost no effect of the extracted $Z_V$ value.
From the fits we obtain $Z_V=0.947(8)$, where the error is mostly
systematic and corresponds to the scattering of the results from
different fits. We could also estimate $Z_V$ from the matrix element
of the vector charge of the pion calculated in Ref. \cite{Gao:2021xsm}.
Using the result for the matrix element from the two state fit we
obtain $Z_V=0.9534(5)$ \cite{Gao:2021xsm}.  This agrees with the
above result within errors.

In Fig. \ref{fig:ZV} bottom panel, we also show the ratio $Z_A/Z_V$
as function of $p_R$. This ratio too should be independent of $p_R$.
We see some dependence on $p_R$ due to non-perturbative effects,
though it is considerably milder than for $Z_V$. This is likely due
to the fact that the leading non-perturbative contributions cancel
out in the ratio $Z_A/Z_V$.  The lattice discretization effects
also seem to largely cancel in the ratio $Z_A/Z_V$ and no clear
fish-bone structure can be seen in our data. The $p_R$ dependence
of $Z_A/Z_V$ is incompatible with $B/p_R^2$ form. Therefore, we fit
our data with $const+B'/(a p_R)^4$ form. This gives $Z_A/Z_V=1.0168$.
For $(a p_R)^2 > 1.5$ it is also possible to fit the data with
constant, which gives $Z_A/Z_V=1.01514$. Combining these results
with the value of $Z_V$ from the pion matrix element of the vector
charge we obtain $Z_A=0.969(1)$. The error is systematic and is due
to the difference of the two fits of $Z_A/Z_V$.

\befs
\centering
\includegraphics[scale=0.5]{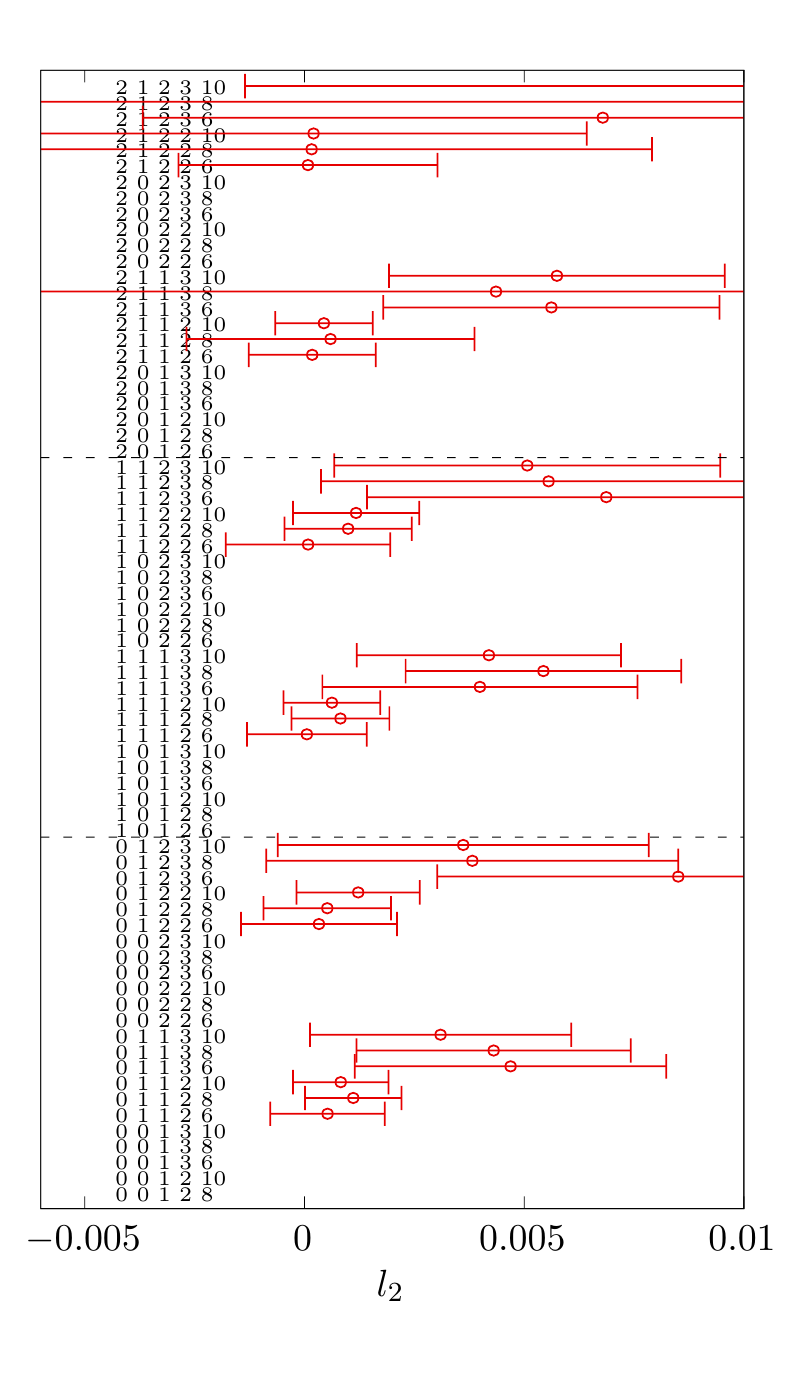}
\includegraphics[scale=0.5]{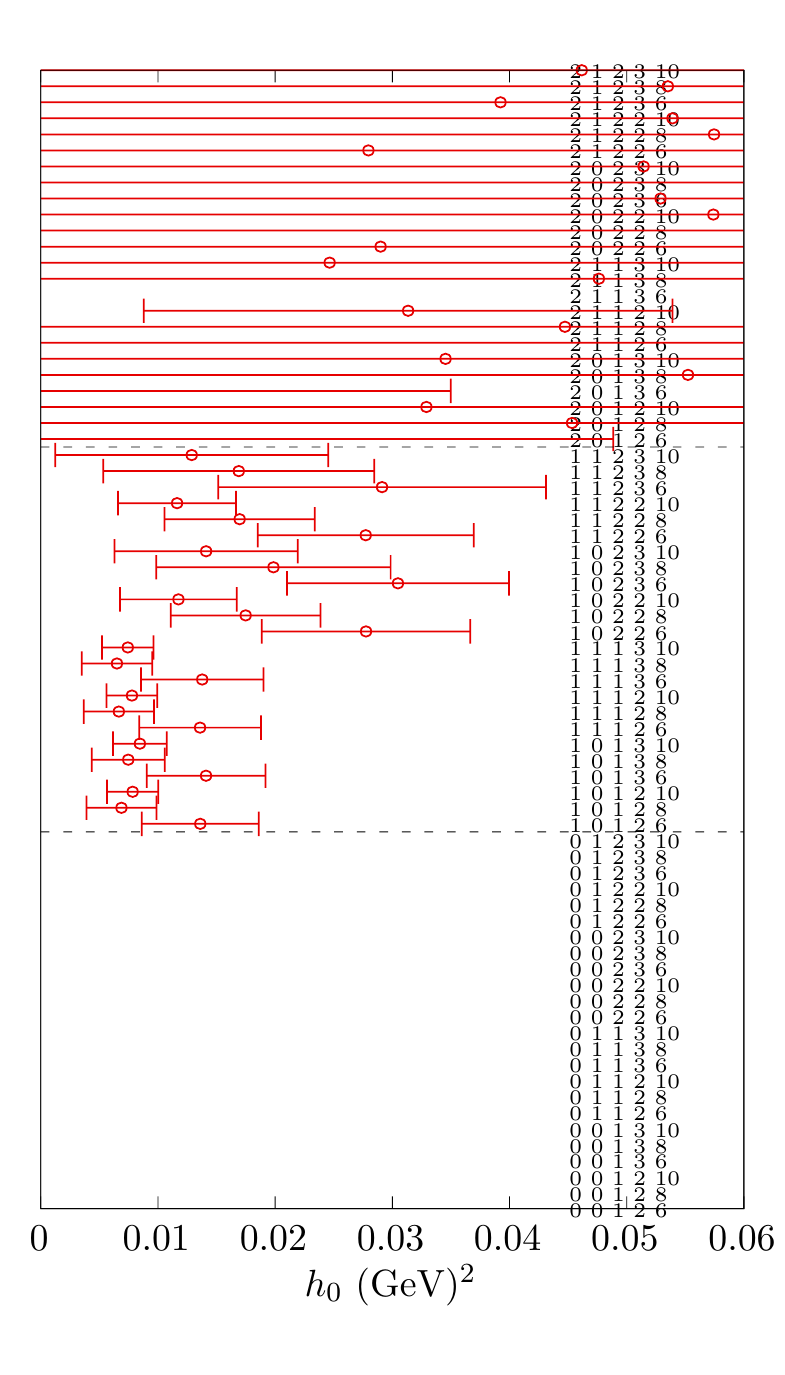}
\includegraphics[scale=0.5]{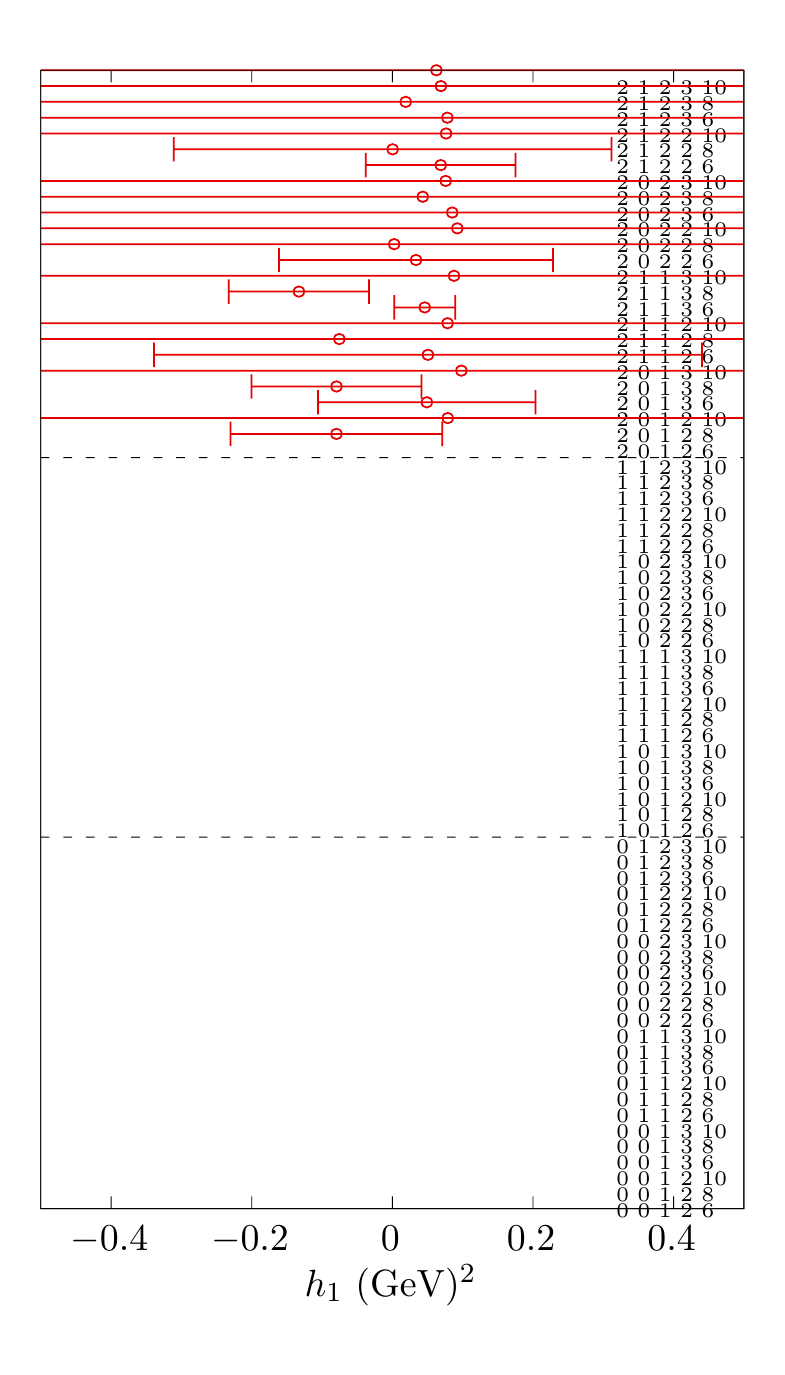}
\includegraphics[scale=0.5]{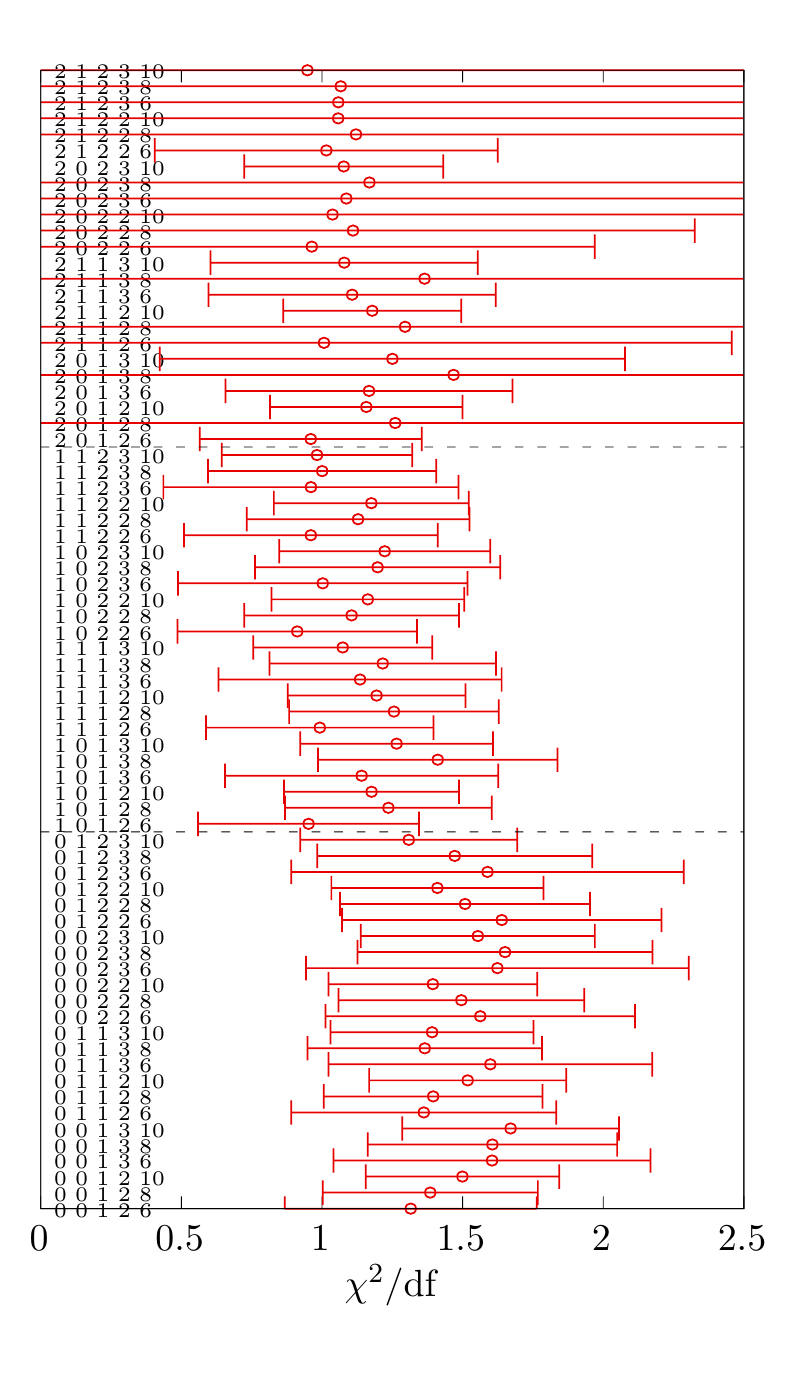}
\caption{
The plot shows additional details of the Mellin moments fit to accompany the 
values of moments shown in \fgn{momsys}. The specification
    $(N_{\rm HT}, N_{\rm LC}, n_3^0, z_3^{\rm min}/a, z_3^{\rm
    max}/a)$ is noted on the side of the points, similar to \fgn{momsys}. 
The first three panels show the best fit values of $l_0$, $h_0$ and $h_1$
respectively -- the cases where a parameter is not included is left without a data point.  
The rightmost panel shows the minimum $\chi^2/{\rm df}$ of the fits. The error bar in this case 
specifies the ranges of $\chi^2/{\rm df}$ occuring in the different Jackknife 
blocks.
}
\eefs{allsys}

\section{Details on the Mellin Moments fit}
\label{sec:fitvals}

In \scn{moments}, we presented the results on the fitted values of the first few Mellin
moments based on fits to leading twist Mellin OPE along with lattice correction of the 
form in \eqn{latcorr} with $l_2$ being a fit parameter, and with higher twist corrections 
in \eqn{htmodel} with $h_0$ and $h_1$ as possible fit parameters. 
In this appendix we discuss the results for these additional fit parameters 
in the Mellin OPE fits with no positivity constraints on the moments.
In the first three panels of 
\fgn{allsys}, we show the scatter of best fit values for $l_2$, $h_0$ and $h_1$ respectively
for various fit types specified by $(N_{\rm HT}, N_{\rm LC}, n_3^0, z_3^{\rm min}/a, z_3^{\rm max}/a)$ beside the data points.
Since certain parameters are not part of all fit types (e.g., $h_1$ does not occur when $N_{\rm HT}=0,1$), data points for 
those cases are left missing in the different panels.  From the figure, the conclusions specified in the main text become 
clearer; namely, the presence of $l_2$ has only a marginal effect, whereas 
the effect of the higher-twist term $h_0$ cannot be neglected.  Also, $N_{\rm HT}=1$ is sufficient for the fits for the 
present data quality, whereas 
inclusion of $h_1$ with $N_{\rm HT}=2$ only makes the fits noisier. 
In the rightmost panel of \fgn{allsys}, we show the scatter of minimum $\chi^2/{\rm df}$ in the various fits. 
The error bars on the minimum $\chi^2/{\rm df}$ is obtained from the Jackknife blocks. The mean values of $\chi^2/{\rm df}$
for different fits all lie approximately between 1 and 1.6.

\bibliography{pap.bib}

\end{document}